\documentclass[twocolumn, journal]{IEEEtran}
\IEEEoverridecommandlockouts

\usepackage{color}
\usepackage{bm}
\usepackage{graphicx}
\usepackage{amsmath}
\usepackage{amssymb}
\usepackage{algorithm}
\usepackage{algorithmic}
\usepackage{multirow}
\usepackage{booktabs}
\usepackage{array}
\usepackage{amsthm}
\usepackage{lipsum}
\usepackage{enumerate}
\usepackage{stfloats}
\usepackage{subfigure}
\usepackage{cases}
\usepackage{diagbox}
\usepackage{threeparttable}
\usepackage{cite}
\usepackage{url}
\usepackage{booktabs}
\usepackage[OT1]{fontenc}

\newcommand{\non}{\nonumber}

\newcommand{\tabitem}{~~\llap{\textbullet}~~}
\allowdisplaybreaks[4]

\title{A Tutorial on Beyond-Diagonal Reconfigurable Intelligent Surfaces: Modeling, Architectures, System Design and Optimization, and Applications}

\author{Hongyu Li,~\IEEEmembership{Member,~IEEE}, Matteo Nerini,~\IEEEmembership{Member,~IEEE},\\ Shanpu Shen,~\IEEEmembership{Senior Member}, and Bruno Clerckx,~\IEEEmembership{Fellow,~IEEE}
\thanks{This work is funded by the National Natural Science Foundation of China (grant no. 62501509), by the Science and Technology Development Fund, Macau SAR (File/Project no. 001/2024/SKL), by University of Macau  (File no. SRG2025-00060-IOTSC), and by UKRI grant EP/Y004086/1, EP/X040569/1, EP/Y037197/1, EP/X04047X/1, EP/Y037243/1 \textit{(Corresponding author: Bruno Clerckx).}}
\thanks{H. Li is with the Internet of Things Thrust, The Hong Kong University of Science and Technology (Guangzhou), Guangzhou 511400, China (e-mail:hongyuli@hkust-gz.edu.cn).}
\thanks{M. Nerini is with the Department of Electrical and Electronic Engineering, Imperial College London, London SW7 2AZ, U.K. (e-mail:m.nerini20,@imperial.ac.uk).}
\thanks{S. Shen is with the State Key Laboratory of Internet of Things for Smart City and Department of Electrical and Computer Engineering, University of Macau, Macau, China (e-mail:shanpushen@um.edu.mo).}
\thanks{B. Clerckx is with the Department of Electrical and Electronic Engineering, Imperial College London, London SW7 2AZ, U.K. and also with Kyung Hee University, Seoul, Korea (e-mail:b.clerckx@imperial.ac.uk).}}

\begin{document}

\maketitle
\thispagestyle{empty}
\begin{abstract}
Written by its inventors, this first tutorial on Beyond-Diagonal Reconfigurable Intelligent Surfaces (BD-RISs) provides the readers with the basics and fundamental tools necessary to appreciate, understand, and contribute to this emerging and disruptive technology. Conventional (Diagonal) RISs (D-RISs) are characterized by a diagonal scattering matrix $\mathbf{\Theta}$ (commonly denoted as phase shift matrix in the literature). Since a very small percentage of the entries of that matrix, namely only the phases of its diagonal entries (in its passive form), are tunable, the wave manipulation flexibility of D-RIS is extremely limited.  In contrast, BD-RIS refers to a novel and general framework for RIS where its scattering matrix is not limited to be diagonal (hence, the ``beyond-diagonal'' terminology) and consequently, all entries of $\mathbf{\Theta}$ can potentially help shaping waves for much higher manipulation flexibility. In its passive form, $\mathbf{\Theta}$ satisfies the unitary property $\mathbf{\Theta}^\mathsf{H}\mathbf{\Theta}=\mathbf{I}$ (for energy conservation in lossless ideal surfaces) and be either symmetric $\mathbf{\Theta}=\mathbf{\Theta}^\mathsf{T}$ or asymmetric $\mathbf{\Theta}\neq\mathbf{\Theta}^\mathsf{T}$ hence leading to reciprocal or non-reciprocal BD-RIS. Such scattering matrix properties correspondingly translate into novel passive (lossless) and reciprocal/non-reciprocal circuitry where each RIS element is not only connected to its own tunable impedance but also to other elements through reconfigurable components. This physically means that BD-RIS can artificially engineer and reconfigure coupling across elements of the surface thanks to inter-element reconfigurable components which allow waves absorbed by one element to flow through other elements. This offers an extra degree of freedom for reconfigurable surfaces that provides new opportunities and flexibility for manipulating, modulating, processing, and computing signals and waves in the analog domain. Consequently, BD-RIS opens the door to more general and versatile intelligent surfaces that subsumes existing RIS architectures as special cases. 
In this tutorial, we share all the secret sauce to model, design, and optimize BD-RIS and make BD-RIS transformative in many different applications. Topics discussed include physics-consistent and multi-port network-aided modeling; transmitting, reflecting, hybrid, and multi-sector mode analysis; reciprocal and non-reciprocal architecture designs and optimal performance-complexity Pareto frontier of BD-RIS; signal processing, optimization, and channel estimation for BD-RIS; hardware impairments (discrete-value impedance and admittance, lossy interconnections and components, wideband effects, mutual coupling) of BD-RIS; benefits and applications of BD-RIS in communications, sensing, power transfer. We also point out challenges of BD-RIS which trigger directions that are promising for future research. 

\end{abstract}


\section{Introduction}
\label{sec:intro}

The sixth generation (6G) networks are driven by three important characteristics linked to human lifestyle and social changes in the near future: High-fidelity holographic society; connectivity for all things; and time sensitive/time engineered applications \cite{tataria20216g,wang2023road,dang2020should}. 
Consequently, 6G networks should be human-centric with the aim of achieving high security and privacy, seamless connectivity between devices, and real-time communications with extremely low latency \cite{dang2020should}. To achieve these stringent requirements, several promising 6G candidate solutions have been proposed, such as advanced multiple access techniques that could better utilize resources \cite{clerckx2024multiple}, extremely large-scale multiple input multiple output (MIMO) that can significantly improve the spectral efficiency \cite{lu2024tutorial}, multifunctional platforms that could simultaneously support wireless communications, sensing, and power transfer \cite{liu2022integrated,clerckx2021wireless}, and reconfigurable intelligent surfaces (RISs) that provide wave manipulation flexibility to reconfigure the wireless propagation environment and enable new processing capabilities in the electromagnetic domain \cite{di2020smart,wu2025intelligent,kaina2014shaping,tang2020wireless}.

\subsection{Background and Motivation}

Among various promising 6G candidate solutions, RIS has gained significant attention in recent years due to its capability to change the way of treating wireless propagation environments: from adapting to them by sophisticated signal processing strategies to manipulating them with low cost, negligible thermal noise, and low power consumption \cite{di2020smart}. 
RIS is a generic term that is also referred to as intelligent reflecting surface (IRS) \cite{wu2025intelligent}, intelligent surface (IS) \cite{wu2024intelligent}, and programmable/dynamic metasurface \cite{di2020smart}.
RIS can be physically fabricated using different implementations, such as using antenna arrays with tunable components and using metasurfaces, while the former way gains more attention in the wireless society since it has a more straightforward and physics-consistent model.
Broadly speaking, an RIS is a planar surface consisting of numerous tunable elements, each of which is able to induce a controllable change of phase shift and/or amplitude to the incident signal. Benefiting from its shape and reconfigurable property, RIS can be flexibly deployed in complex wireless propagation environments, e.g., being attached on high buildings, to bypass obstacles and increase/decrease the directivity of waves according to specific requirements. 
RISs were primarily proposed as nearly passive devices, in the sense that they are not capable of amplifying incident waves and only minimal power is used to control the surface \cite{wu2025intelligent}.
This property brings various practical advantages for RIS implementation, such as being free of power-hungry RF chains, power amplifiers, and thermal noise.

Inspired by the above advantages, the research on RIS has grown exponentially in the past few years, covering a wide range of areas that include but are not limited to beamforming design \cite{guo2020weighted,wu2019intelligent}, channel estimation \cite{swindlehurst2022channel,zheng2022survey}, hardware impairments analysis \cite{abeywickrama2020intelligent,gradoni2021end,li2021intelligent}, and implementation and prototyping \cite{tang2020wireless,araghi2022reconfigurable}.
To fully compensate for the huge path loss induced by RIS, the concept of active RIS, named due to the introduction of reflection-type power amplifiers, has been further proposed to provide significantly enhanced performance gain at the expense of affordable power consumption and cost \cite{zhang2022active}. 
Meanwhile, the industry has also achieved progress on demonstrating and testing RISs. The test mainly originated from NTT DOCOMO, a Japanese network operator, which demonstrated a fifth generation (5G) mobile system using a metasurface reflectarray operating in 28 GHz in 2018 and conducted a trail of a transparent dynamic metasurface for 5G radio signals in 2020, followed by a re-conduction in 2021 \cite{di2020smart}. 
More recently, RIS has been used in a 5G commercial frequency range 2 band network that operates beyond 24 GHz, showing significant performance improvement in the indoor coverage and throughput \cite{yang2023beyond}.
In support of active academia and industrial progress on RIS, key organizations have started to consider the integration of RIS into future commercial networks. 
For example, RIS-related standardization has been kicked off in the China Communications Standards Association and the FuTURE Mobile Communication Forum from China \cite{liu2022path}. 
For another example, within the European Telecommunications Standards Institute, an industry specific group on RISs has been proposed and approved in June 2021, and launched in September 2021 \cite{liu2022path}. The technical details and practical considerations for applying RISs have been thoroughly discussed during 2021-2023 \cite{etsi,wu2024intelligent}. 

\begin{figure}
    \centering 
    \includegraphics[width=0.42\textwidth]{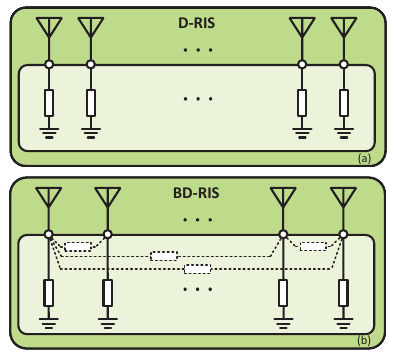}
    \caption{Illustration of (a) D-RIS and (b) BD-RIS.}
    \label{fig:RIS}\vspace{-0.5 cm}
\end{figure}

Despite the comprehensive study and tests of RIS from both academia and industry, one fundamental limitation is that conventional (diagonal) RIS (D-RIS) is characterized by a diagonal scattering matrix $\mathbf{\Theta}$ (commonly denoted as phase shift matrices in the literature). This mathematical structure is realized by a simple architecture where each element of RIS is connected to ground through its own tunable load as illustrated in Fig. \ref{fig:RIS}(a), thereby enabling only the independent control of each diagonal entry of the scattering matrix, as described in the left hand side of equation (\ref{eq:beyond_diagonal}). 
In the passive form (at best lossless) of RIS, only phase shifts of its diagonal entries in the scattering matrix are tunable, such that waves impinging on one element can only be reflected by the same element. Therefore, the wave manipulation capability of D-RIS is limited. 
This limitation in D-RIS motivates a natural question as also mathematically described in equation (\ref{eq:beyond_diagonal}): \textit{Can those zeros in the scattering matrix $\mathbf{\Theta}$ be freely tuned to potentially help shaping waves for more flexible manipulation?} 
\begin{equation}
    \begin{aligned}
        &\mathbf{\Theta} = \left[\begin{matrix}
            ? &  &  & \\
             & ? &  & \\
             & &\ddots  & \\
             &  &  &?
        \end{matrix}\right] \Rightarrow
        \mathbf{\Theta} = \left[\begin{matrix}
            ? & ? & \ldots & ?\\
            ? & ? & \ldots & ?\\
            \vdots & & \ddots & \vdots\\
            ? & ? & \cdots & ?
        \end{matrix}\right].
    \end{aligned}\label{eq:beyond_diagonal}
\end{equation}
The answer is \textit{yes}, with the emerging technology, namely beyond-diagonal (BD) RIS  \cite{li2023reconfigurable}.

\subsection{BD-RIS}

\begin{figure}
    \centering 
    \includegraphics[width=0.48\textwidth]{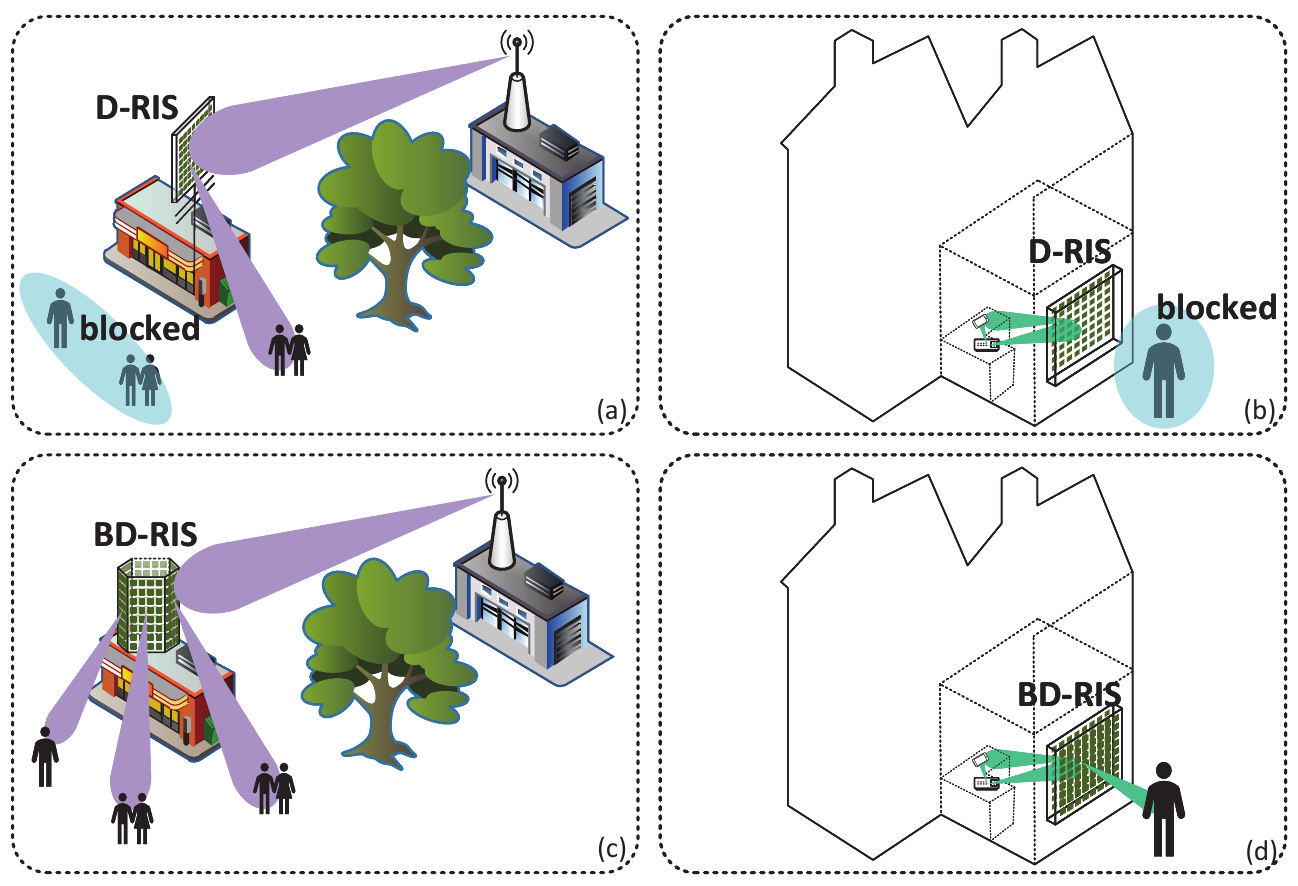}
    \caption{Examples of element arrangements in D-RIS and BD-RIS.}
    \label{fig:arrangement}\vspace{-0.5 cm}
\end{figure}

BD-RIS refers to a novel and general framework for RIS whose scattering matrix $\mathbf{\Theta}$ is not limited to be diagonal, hence not only diagonal entries, but also off-diagonal entries which have been forced to zeros in D-RIS can be flexibly tuned to manipulate waves \cite{li2023reconfigurable,shen2021}. 
This is realized by inter-connecting (part of) elements by additional reconfigurable components as illustrated in Fig. \ref{fig:RIS}(b), hence allowing waves absorbed by one element to flow through other elements and generating nonzero and tunable off-diagonal entries in $\mathbf{\Theta}$.
This brings extra degrees of freedom (DoF) for RIS from the following two perspectives.
\begin{itemize}
    \item \textit{Circuit Design Flexibility:} While D-RIS has individually controllable elements supported by simple architectures, BD-RIS opens the door for diverse architectures by flexibly designing the circuit topology between elements using reciprocal or non-reciprocal devices. This leads to $\mathbf{\Theta}$ with various mathematical structures, e.g., being full matrices, block-diagonal matrices, or permuted matrices, and constraints, e.g., being symmetric $\mathbf{\Theta} = \mathbf{\Theta}^\mathsf{T}$ (realized by reciprocal circuits) or asymmetric $\mathbf{\Theta}\ne\mathbf{\Theta}^\mathsf{T}$ (realized by non-reciprocal circuits). BD-RIS with multiple forms of $\mathbf{\Theta}$ thus provides opportunities to manipulate waves in a smarter way. 
    \item \textit{Element Arrangement Flexibility:} 
    In D-RIS, the arrangement of the elements is constrained by the absence of inter-element interconnections.
    Specifically, since elements in D-RIS are typically designed to cover half of space, they are often arranged as a 2D planar array, leading to practical location constraints for transmitters and receivers, as illustrated in Figs. \ref{fig:arrangement}(a) and \ref{fig:arrangement}(b). 
    By contrast, BD-RIS enables more flexible element arrangements thanks to inter-element connections that allow signals to flow from one direction to another, as illustrated in Figs. \ref{fig:arrangement}(c) and \ref{fig:arrangement}(d). BD-RIS with diverse element arrangements thus provides opportunities to achieve larger wireless coverage and denser connectivity. 
\end{itemize}

\begin{figure}
    \centering
    \includegraphics[width=0.48\textwidth]{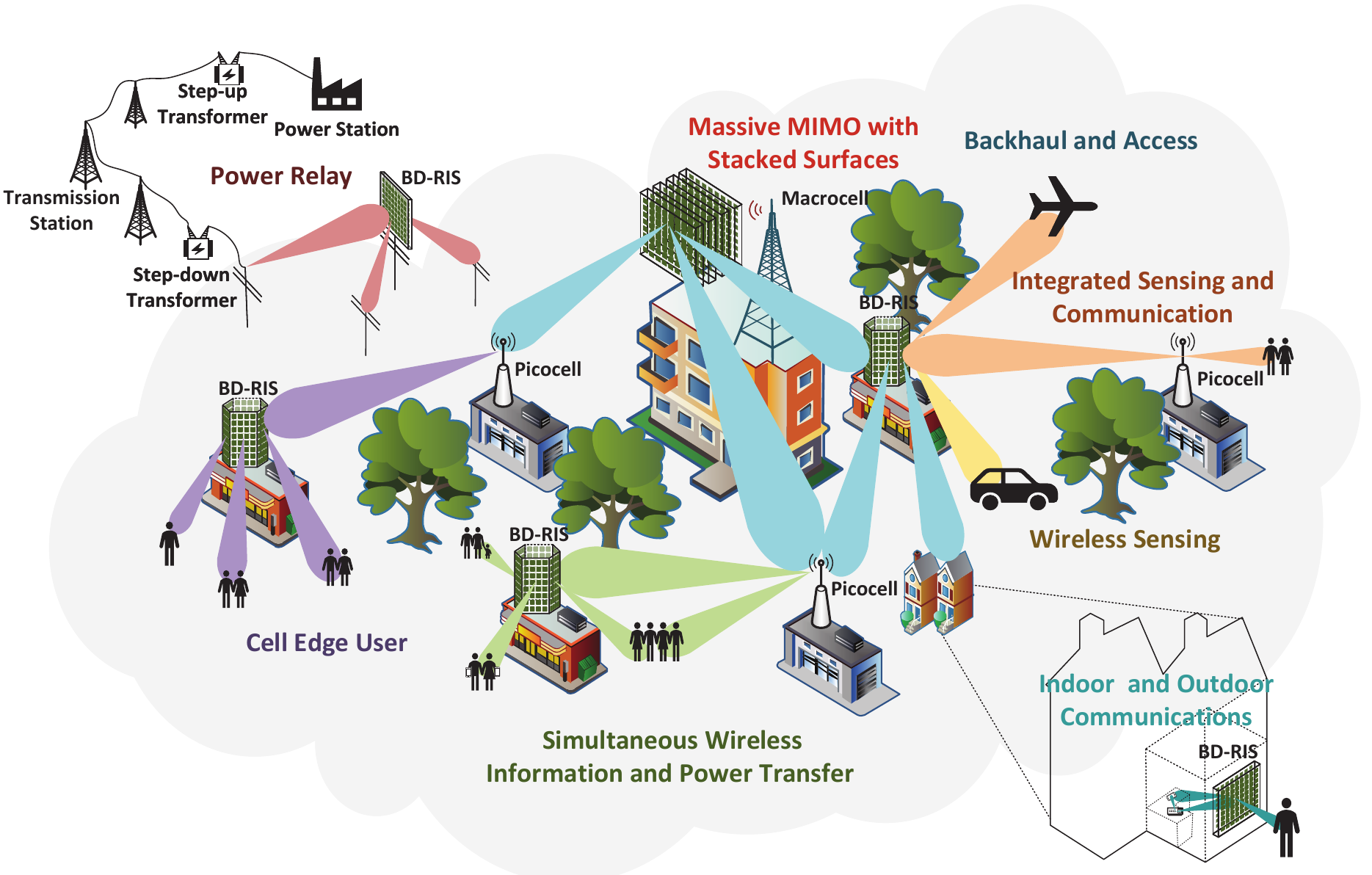}
    \caption{Illustration of future BD-RIS applications.}
    \label{fig:application}\vspace{-0.5 cm}
\end{figure}

Due to the above extra DoF provided by flexible inter-element reactions in BD-RIS, it is expected that BD-RIS will be able to achieve smarter wireless transmissions with higher quality, denser connectivity, and wider coverage. 
Fig. \ref{fig:application} shows an envisioned future application scenario supported by BD-RIS. For instance, BD-RIS can be particularly useful to extend the coverage in joint indoor and outdoor communications in Fig. \ref{fig:application}. Properly deploying BD-RIS with flexible element arrangements can also help serving cell-edge users especially when there are obstacles between the base station and users.
Another interesting application is to enable flexible and scalable integrated  access and backhaul \cite{madapatha2020integrated}. BD-RIS can be flexibly incorporated into real and complex environments to assist and enhance wireless backhauling between macrocells and picocells and wireless access between picocells and users. 
In addition to wireless communications, with proper power levels, suitable deployments, and locations, BD-RIS can be potentially used for higher-efficiency power relay and wireless power transfer (WPT) \cite{clerckx2021wireless}. 
Besides, BD-RIS can also be leveraged to boost wireless sensing, especially in complicated propagation conditions, such as vehicle networks, where strong line-of-sight (LoS) links are not available between the radar and targets.
Not limited to stand-alone functions (wireless communications, power transfer, and sensing), BD-RIS can support more advanced dual-functional systems, such as 
simultaneous wireless information and power transfer (SWIPT) \cite{perera2017simultaneous} and integrated sensing and communications (ISAC) \cite{liu2022integrated,hua2023secure}, or multi-functional integrated communications, sensing, and power transfer systems \cite{zhou2024integrating}.
Not limited to being used to realize smart radio environment, BD-RIS can also be integrated with transmitters/receivers and work as an auxiliary antenna array to enable RIS-aided or stacked surface-aided massive MIMO.
Beyond the above, BD-RIS, as a comprehensive upgrade of D-RIS, is readily applicable to all D-RIS enabled
scenarios while providing more flexible deployment, smarter wave manipulation and better performance, and wider coverage. 

\begin{table}
	\caption{List of Representative Overview/Survey/Tutorial\\ Papers on BD-RIS}
	\label{tab:list}
	\centering
	\begin{tabular}{|c|c|l|}
	   \hline
         Ref. & Type & Highlights\\
         \hline\hline
         \cite{li2023reconfigurable} & \multirow{5}{*}{Overview} & 
        \begin{tabular}[l]{@{}l@{}}The first magazine to provide a high-level \\ overview of the concept of BD-RIS \\ without technical details 
        \end{tabular}\\
        \cline{1-1}\cline{3-3}
        \cite{khan2024beyond,khan2024integration,khan2025beyond} & & \begin{tabular}[l]{@{}l@{}} Briefly revisit the fundamentals of BD-RIS, \\ non-terrestrial networks (NTNs), and Internet\\ 
        of Things (IoT), and outline the application\\ of BD-RIS in NTNs and IoT Networks\end{tabular}\\
        \hline
        This Paper & Tutorial & \begin{tabular}[l]{@{}l@{}}Provide a tutorial of BD-RIS, including\\ fundamentals, signal processing techniques,\\ hardware impairments, and discuss the \\benefits, emerging applications, challenges,\\ and future research directions\end{tabular} \\
        \hline
	\end{tabular}
\end{table}

\subsection{Contributions and Organization}

The appealing benefits and potential of BD-RIS have spurred active research. Specifically, a few overviews \cite{li2023reconfigurable,khan2024beyond,khan2024integration,khan2025beyond} have appeared, focusing either on high-level illustrations to BD-RIS or specific applications of BD-RIS, as summarized in Table \ref{tab:list}. However, there is a lack of a comprehensive tutorial that pedagogically explains the basics and fundamental tools to understand BD-RIS. 
The main goal of this paper is thus to provide the first tutorial on BD-RIS including the fundamentals from a rigorous microwave engineering perspective, the state-of-the-art signal processing techniques, the hardware impairments, and the benefits, applications, technical challenges, and future research directions of BD-RIS. This further leads to the following contributions.

\textit{First}, a toy example is given for BD-RIS-aided single input single output (SISO) systems to provide readers with an intuitive explanation on the rationale behind BD-RIS and its benefits over D-RIS. 

\textit{Second}, the physics-consistent modeling of BD-RIS using multi-port network analysis is explained, followed by a comprehensive summary of the reflecting, hybrid, and multi-sector modes and various reciprocal and non-reciprocal architectures supported by BD-RIS. The unification between architectures and modes is also discussed based on specific examples. 

\textit{Third}, representative optimization methods for various BD-RIS architectures are presented and summarized, taking the BD-RIS-aided SISO system as an example. An exhaustive survey of BD-RIS is provided from the perspective of beamforming design and channel estimation. In addition, the drawback of current channel estimation methods is pointed out with a discussion of possible future research directions. 

\textit{Fourth}, four important hardware impairments of BD-RIS (discrete-value impedance and admittance, lossy interconnections and admittance components, wideband effect, and mutual coupling effect) are modeled, evaluated, and analyzed to capture the practical issues and provide useful guidance for BD-RIS implementation. 

\textit{Fifth}, benefits of BD-RIS are highlighted with supporting simulation results, and applications of BD-RIS in wireless communications, sensing, and power transfer are 
summarized based on a thorough literature review. 

\textit{Sixth}, key technical challenges of BD-RIS are discussed, followed by directions that are promising for future research. 

We hope this tutorial will provide a useful reference on the study and application of BD-RIS and serve as an inspiring resource for future research on BD-RIS.

\textit{Organization:} 
The rest of this paper is organized as follows. Section \ref{sec:example} provides a toy example of BD-RIS-aided SISO systems to briefly explain the rationale and benefits of BD-RIS. Section \ref{sec:model} explains the basic modeling, mode analysis, and various architecture designs for BD-RIS. Section \ref{sec:signal_process} summarizes the key techniques for BD-RIS optimization and channel estimation. Section \ref{sec:hardware} introduces hardware impairments in BD-RIS and illustrates their impacts on system performance.  
Section \ref{sec:benefits} discusses the benefits of BD-RIS supported by analytical and numerical results, and summarizes emerging applications. 
Section \ref{sec:challenges} points out challenges which should be tackled to facilitate the use of BD-RIS in future 6G networks, and sheds light on possible future research directions. 
Finally, Section \ref{sec:conclusion} concludes this paper. The outline of this paper is illustrated in Table \ref{tab:outline}.

\begin{table}[]
	\caption{Outline of the Paper}
	\label{tab:outline}
	\begin{tabular}{|l|}
	   \hline
         Section I. Introduction\\
         \hline
         A. Background and Motivation\\
        B. BD-RIS\\
        C. Contributions and Organization\\
        \hline\hline
        Section II. BD-RIS-Aided SISO: A Toy Example\\
        \hline\hline
        \begin{tabular}[l]{@{}l@{}}Section III. Modeling, Mode Analysis, and Architecture\\ \;\;\;\;\;\;\;\;\;\;\;\;\;\;\;\;\; Design of BD-RIS\end{tabular}\\
        \hline
        A. Preliminaries on Multi-Port Network Analysis\\
        B. Modeling and Classification\\
        C. Mode Analysis\\
        D. Architecture Design: Reciprocal Architectures\\
        E. Architecture Design: Non-Reciprocal Architectures\\
        F. Unified Modes and Architectures\\
        \hline\hline
        Section IV. Signal Processing of BD-RIS\\
        \hline
        A. Optimizing BD-RIS: A Simple SISO Case\\
        B. Survey on BD-RIS Optimization and Performance Analysis\\
        C. Channel Estimation\\
        \hline\hline
        Section V. Benefits of BD-RIS\\
        \hline
        A. Boosting Received Power and Rates\\
        B. Enabling Low-Complexity Architectures with High Performance\\
        C. Enabling Flexible Modes with Highly-Directional Wireless Coverage\\
        D. Providing Orders of Magnitude Gains in Distributed Deployments\\
        E. Enabling Simultaneously Optimal Transmissions for Uplink and\\ \;\;\; Downlink with Non-Reciprocal Architectures\\
        F. Providing Enhanced Gains in Dual-Polarized Systems\\
        \hline\hline
        Section VI. BD-RIS with Hardware Impairments\\
        \hline
        A. Discrete-Value Impedance and Admittance\\
        B. Lossy Interconnections and Admittance Components\\
        C. Wideband Effect\\
        D. Mutual Coupling Effect\\
        \hline\hline
        Section VII. Applications of BD-RIS\\
        \hline
        A. BD-RIS for Communications\\
        B. BD-RIS for Sensing and ISAC\\
        C. BD-RIS for WPT and SWIPT\\
        D. BD-RIS with Other Techniques and Systems\\
        \hline\hline
        Section VIII. Challenges and Future Research Directions of BD-RIS\\
        \hline
        A. BD-RIS Implementation\\
        B. Active BD-RIS\\
        C. AI-Driven Beamforming Solutions\\
        D. CSI-Free Protocols\\
        E. New BD-RIS Architectures\\
        \hline\hline
        Section IX. Conclusion\\
        \hline
	\end{tabular}
\end{table}

\textit{Notations:}
Boldface lower- and upper-case letters indicate column vectors and matrices, respectively. 
$(\cdot)^\mathsf{T}$, $(\cdot)^*$, $(\cdot)^\mathsf{H}$, and $(\cdot)^{-1}$ denote the transpose, conjugate, conjugate-transpose, and inverse operations, respectively.
$\mathbb{C}$, $\mathbb{R}$, and $\mathbb{Z}$ denote the sets of complex, real, and integer numbers, respectively.
The superscript of $(\cdot)^{M\times N}$ identifies the size of a matrix.
$\mathbb{E}\{\cdot\}$ denotes the statistical expectation. 
$\Re\{\cdot\}$ and $\Im\{\cdot\}$ denote the real and imaginary parts of complex numbers, respectively. 
$\angle(\cdot)$ denotes the angle of complex numbers. 
$\mathsf{blkdiag}(\cdot)$ represents a block-diagonal matrix and $\mathsf{diag}(\cdot)$ represents a diagonal matrix.
$|\cdot|$, $\|\cdot\|_2$, and $\|\cdot\|_\mathsf{F}$ denote the absolute value of a scalar, the $\ell_2$ norm of a vector, and the Frobenius norm of a matrix, respectively.
$\jmath=\sqrt{-1}$ denotes the imaginary unit.
$\otimes$ denotes the Kronecker product.
$\mathsf{vec}(\cdot)$ denotes the vectorization operation.
$\mathsf{tr}(\cdot)$ denotes the trace of a matrix.
$\det(\cdot)$ denotes the determinant of a square matrix.
$\mathbf{I}_M$ denotes an $M\times M$ identity matrix.
$\mathbf{0}_{M\times N}$ denotes an $M\times N$ all-zero matrix. 
$\mathbf{A}\preceq\mathbf{B}$ (or $\mathbf{A}\succeq\mathbf{B}$) indicates that $\mathbf{B}-\mathbf{A}$ (or $\mathbf{A}-\mathbf{B}$) is positive semi-definite. 
$a\sim\mathcal{CN}(0,\sigma^2)$ characterizes the circular symmetric complex Gaussian distribution.
$[\mathbf{a}]_i$ denotes the $i$-the entry of $\mathbf{a}$.
$[\mathbf{A}]_{i:i',j:j'}$ extracts the $i$-th to $i'$-th rows and the $j$-th to $j'$-th columns of $\mathbf{A}$.

\section{BD-RIS-Aided SISO: A Toy Example}
\label{sec:example}

To introduce the rationale behind BD-RIS and understand its benefits, in this section, we analyze a BD-RIS-aided SISO system as a toy example.
Consider a SISO wireless system between a single-antenna transmitter and a single-antenna receiver aided by a BD-RIS with $M$ elements.
Denoting the transmitted signal as $x\in\mathbb{C}$, the received signal $y\in\mathbb{C}$ is given by $y=hx+n$, where $h\in\mathbb{C}$ is the wireless channel matrix between the transmitter and receiver and $n\in\mathbb{C}$ is the noise.
The channel $h$ is a function of the phase shift matrix of the RIS $\boldsymbol{\Theta}\in\mathbb{C}^{M\times M}$, given by the well-known cascaded model
\begin{equation}
h=\mathbf{h}_{RI}\boldsymbol{\Theta}\mathbf{h}_{IT},\label{eq:H}
\end{equation}
where $\mathbf{h}_{RI}\in\mathbb{C}^{1\times M}$ is the channel between the RIS and receiver, $\mathbf{h}_{IT}\in\mathbb{C}^{M\times 1}$ is the channel between the transmitter and RIS, and we neglect for simplicity the direct channel between the transmitter and receiver as it is assumed to be obstructed.
Given the channel model in \eqref{eq:H}, the matrix $\boldsymbol{\Theta}$ is commonly optimized to maximize the channel gain $\rho=\vert h\vert^2$ subject to specific constraints depending on the properties of the RIS.

In a D-RIS, $\boldsymbol{\Theta}$ is a diagonal matrix given as
\begin{equation}
\boldsymbol{\Theta}=\mathrm{diag}\left(e^{\jmath\theta_1},e^{\jmath\theta_2},\ldots,e^{\jmath\theta_M}\right),\label{eq:diag}
\end{equation}
where $e^{\jmath\theta_m}$ is the reflection coefficient of the $m$-th RIS element.
Note that we have implicitly assumed the D-RIS to be lossless in \eqref{eq:diag} by considering reflection coefficients with unit magnitude. The detailed derivation for obtaining such a diagonal phase shift matrix at D-RIS will soon be shown in Sections \ref{subsec:model} and \ref{subsec:architecture}.
We can now write the channel gain as $|h|^2 = |\mathbf{h}_{RI}\mathbf{\Theta}\mathbf{h}_{IT}|^2 = |\sum_{m=1}^M[\mathbf{h}_{RI}]_me^{\jmath\theta_m}[\mathbf{h}_{IT}]_m|^2$. This allows us to obtain an optimal phase shift of the $m$-th element of RIS as $\theta_m^\star=-\angle([\mathbf{h}_{RI}]_m[\mathbf{h}_{IT}]_m)$, yielding a maximum channel gain
\begin{equation}
\rho^{\mathsf{D-RIS}}=\left(\sum_{m=1}^M\left\vert\left[\mathbf{h}_{RI}\right]_{m}\left[\mathbf{h}_{IT}\right]_{m}\right\vert\right)^2.\label{eq:G-D}
\end{equation}

The matrix $\boldsymbol{\Theta}$ is commonly referred to as the ``phase shift matrix'' of the RIS, because conventionally it is a diagonal matrix containing phase shifts in its diagonal, as given by \eqref{eq:diag}.
Nevertheless, it has been shown through microwave theory that, more rigorously speaking, $\boldsymbol{\Theta}$ is the so-called ``scattering matrix'' of the RIS.
In microwave theory, the scattering matrix is a matrix that can be used to characterize a linear microwave network linking the reflected waves to the incident waves at its ports.
Intuitively speaking, the scattering matrix can be seen as the multidimensional generalization of the concept of reflection coefficient.
As a lossless 1-port microwave network is characterized by a reflection coefficient having unit magnitude, a lossless multi-port microwave network is characterized by a unitary scattering matrix \cite{pozar2021microwave}.

Since a lossless microwave network has a unitary scattering matrix, a lossless RIS is in principle allowed to have any unitary matrix $\boldsymbol{\Theta}$, i.e., $\boldsymbol{\Theta}^\mathsf{H}\boldsymbol{\Theta}=\mathbf{I}_M$ or, equivalently, $\boldsymbol{\Theta}\boldsymbol{\Theta}^\mathsf{H}=\mathbf{I}_M$, and is not limited to the diagonal constraint in \eqref{eq:diag}. $\mathbf{\Theta}$ in this case can have various expressions based on specific circuit designs, each of which will be explained in detail in Sections \ref{subsec:architecture} and \ref{subsec:architecture_nr}.
We refer to these RIS architectures, which are not limited to having a diagonal scattering matrix $\boldsymbol{\Theta}$, as BD-RIS.
Remarkably, given that the unitary constraint $\boldsymbol{\Theta}^\mathsf{H}\boldsymbol{\Theta}=\mathbf{I}_M$ is a generalization of the constraint in \eqref{eq:diag}, BD-RIS is a generalization of D-RIS and includes D-RIS as a special case.
In this case, 
following the sub-multiplicity property of the spectral norm, we obtain that the channel gain $|h|^2$ is upper bounded by
\begin{equation}
\rho^{\mathsf{BD-RIS}}=\left\Vert\mathbf{h}_{RI}\right\Vert_2^2\left\Vert\mathbf{h}_{IT}\right\Vert_2^2,\label{eq:G-BD}
\end{equation}
based on the Cauchy-Schwarz inequality and the constraint $\mathbf{\Theta}^\mathsf{H}\mathbf{\Theta}=\mathbf{I}_M$.

Comparing the channel gain achievable with D-RIS and BD-RIS, in \eqref{eq:G-BD} and \eqref{eq:G-D}, respectively, we observe that $\rho^{\mathsf{BD-RIS}}\ge \rho^{\mathsf{D-RIS}}$ for any channel realizations $\mathbf{h}_{RI}$ and $\mathbf{h}_{IT}$ because of the Cauchy-Schwarz inequality.
Thus, BD-RIS always achieves a channel gain higher than D-RIS, and this is thanks to its additional flexibility.
Intuitively, while D-RIS is reminiscent of equal gain combining, as it can only adjust the phases of the impinging signal, BD-RIS is reminiscent of maximal ratio combining since it can optimize both phases and amplitude of the impinging signal, increasing flexibility and performance.

To quantify the performance benefits of BD-RIS over D-RIS in a SISO system, we derive the scaling laws of the average channel gains $\rho^{\mathsf{D-RIS}}$ and $\rho^{\mathsf{BD-RIS}}$ under independent and identically distributed (i.i.d.) Rayleigh fading channels, i.e., $\mathbf{h}_{RI}\sim\mathcal{CN}(\mathbf{0}_{1\times M},\mathbf{I}_M)$ and $\mathbf{h}_{IT}\sim\mathcal{CN}(\mathbf{0}_{M\times 1},\mathbf{I}_M)$.
In the case of a D-RIS, the average channel gain is obtained by taking the expectation of \eqref{eq:G-D} as
\begin{equation}
    \begin{aligned}
        \bar{\rho}^{\mathsf{D-RIS}} &=\mathbb{E}\Big\{\Big(\sum_{m=1}^M\left\vert[\mathbf{h}_{RI}]_{m}[\mathbf{h}_{IT}]_{m}\right\vert\Big)^2\Big\}\\
        &= \mathbb{E}\Big\{\sum_{m=1}^{M}\left\vert[\mathbf{h}_{RI}]_{m}\right\vert^2\left\vert[\mathbf{h}_{IT}]_{m}\right\vert^2
        +\sum_{m_1\neq m_2}\left\vert[\mathbf{h}_{RI}]_{m_1}\right\vert\\
        &~~~~\times\left\vert[\mathbf{h}_{IT}]_{m_1}\right\vert\left\vert[\mathbf{h}_{RI}]_{m_2}\right\vert\left\vert[\mathbf{h}_{IT}]_{m_2}\right\vert\Big\},
    \end{aligned}
\end{equation}
by expanding the square term.
By noticing that $[\mathbf{h}_{RI}]_{m_1}$ and $[\mathbf{h}_{IT}]_{m_2}$ are independent $\forall m_1,m_2$, and that $[\mathbf{h}_{RI}]_{m_1}$ and $[\mathbf{h}_{RI}]_{m_2}$ are independent if $m_1\neq m_2$, we obtain
\begin{equation}
\begin{aligned}
\bar{\rho}^{\mathsf{D-RIS}}
&=\sum_{m=1}^{M}\mathbb{E}\left\{\left\vert[\mathbf{h}_{RI}]_{m}\right\vert^2\right\}^2
+\sum_{m_1\neq m_2}\mathbb{E}\left\{\left\vert[\mathbf{h}_{RI}]_{m_1}\right\vert\right\}^4\\
&=\sum_{m=1}^{M}1+\sum_{m_1\neq m_2}\left(\frac{\sqrt{\pi}}{2}\right)^4,
\end{aligned}
\end{equation}
where we exploited that $\mathbb{E}\{\vert[\mathbf{h}_{RI}]_{m}\vert^2\}=1$ and $\mathbb{E}\{\vert[\mathbf{h}_{RI}]_{m}\vert\}=\frac{\sqrt{\pi}}{2}$ in the second equality following the moments of the chi distribution with 2 degrees of freedom, yielding
\begin{equation}
\bar{\rho}^{\mathrm{D-RIS}}=M+\frac{\pi^2}{16}M\left(M-1\right).\label{eq:gain-d-theor}
\end{equation}
For a BD-RIS, the average channel gain is obtained by taking the expectation of \eqref{eq:G-BD}, giving 
\begin{equation}
\begin{aligned}
\bar{\rho}^{\mathrm{BD-RIS}}
&=\mathbb{E}\left\{\left\Vert\mathbf{h}_{RI}\right\Vert_2^2\left\Vert\mathbf{h}_{IT}\right\Vert_2^2\right\}\\
&=\mathbb{E}\left\{\left\Vert\mathbf{h}_{RI}\right\Vert_2^2\right\}\mathbb{E}\left\{\left\Vert\mathbf{h}_{IT}\right\Vert_2^2\right\}.
\end{aligned}
\end{equation}
By noticing that $\mathbb{E}\{\left\Vert\mathbf{h}_{RI}\right\Vert_2^2\}=\mathbb{E}\{\left\Vert\mathbf{h}_{IT}\right\Vert_2^2\}=M$ following the moments of the chi distribution with $2M$ degrees of freedom, we finally have
\begin{equation}
\bar{\rho}^{\mathrm{BD-RIS}}=M^2.\label{eq:gain-bd-theor}
\end{equation}

\begin{figure}
    \centering 
    \includegraphics[width=0.45\textwidth]{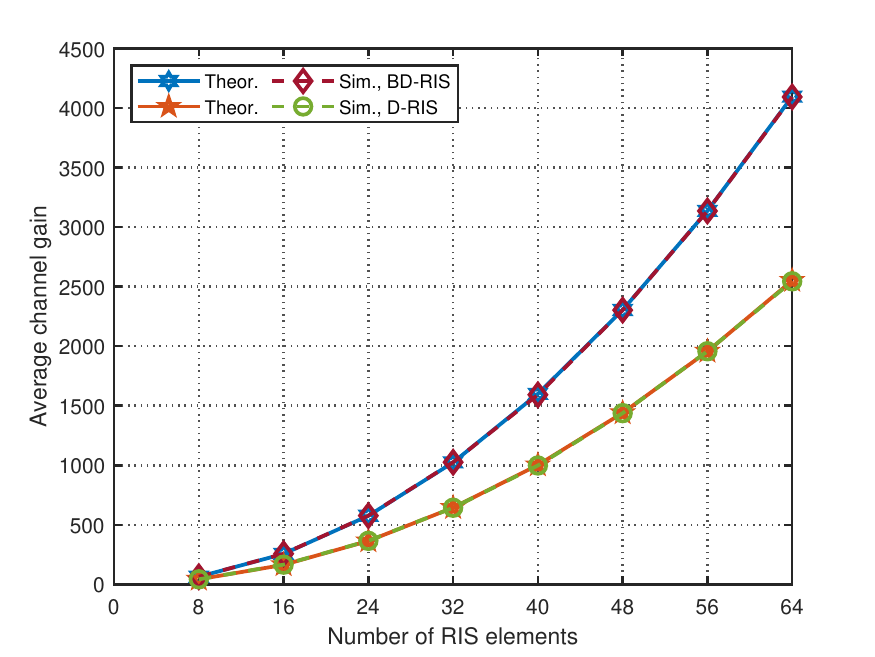}
    \caption{Average channel gain for D-RIS and BD-RIS-aided SISO systems. Both channels through RIS are i.i.d. Rayleigh distributed.}
    \label{fig:gain}
\end{figure}

In Fig.~\ref{fig:gain}, we report the average channel gain offered by D-RIS and BD-RIS under i.i.d. Rayleigh distributed channels.
We compare the simulated channel gains, which are obtained by applying the proposed solutions detailed in Section \ref{subsec:opt_siso} and averaging over multiple channel realizations, with the theoretical channel gains derived in \eqref{eq:gain-bd-theor} and \eqref{eq:gain-d-theor}.
As shown in Fig.~\ref{fig:gain}, the simulation results confirm the accuracy of the closed-form scaling laws in \eqref{eq:gain-bd-theor} and \eqref{eq:gain-d-theor}.
In the asymptotic regime where the number of RIS elements $M\to\infty$, the performance advantage of BD-RIS over D-RIS in a SISO system is quantified by the gain ratio
\begin{equation}
G^\mathsf{BD}
=\lim_{M\to\infty}\frac{\bar{\rho}^{\mathrm{BD-RIS}}}{\bar{\rho}^{\mathrm{D-RIS}}}
=\frac{16}{\pi^2}
\approx1.62,
\end{equation}
implying that BD-RIS achieves approximately a 62\% higher channel gain than D-RIS.
Although this fundamental result highlights the clear advantage of BD-RIS in SISO systems, the benefits of BD-RIS become even more pronounced in multi-user and multi-antenna scenarios. For instance, BD-RIS can provide up to 75\% sum-rate improvement over D-RIS with $M = 128$, as will be shown in Section \ref{sec:benefits}.

\section{Modeling, Mode Analysis, and Architecture Design of BD-RIS}
\label{sec:model}

In this section, we introduce the modeling, architectures, and modes of BD-RIS, including the technical details, graphical illustrations, summaries and discussions. 

\vspace{-0.5 cm}

\subsection{Preliminaries on Multi-Port Network Analysis}

As a foundation on the BD-RIS modeling, we start by concisely reviewing the background knowledge on multi-port network analysis, which is a simplified way to understand transmission properties compared to field analysis using Maxwell's equations, and a powerful and useful technique for modeling and studying wireless systems \cite{pozar2021microwave}.
Specifically, each antenna in a wireless system is regarded as a port, whose behavior is characterized by its terminal voltage, current, incident wave, and reflected wave. This technique has been used to model and analyze MIMO systems \cite{wallace2004mutual,wallace2004termination,morris2005network}, where some hardware impairments, such as mutual coupling and antenna mismatching, can be explicitly captured based on physics-consistent assumptions. 
Due to the fact that RIS is essentially a multi-port network regarding each element as one port, the multi-port network theory is naturally suitable for modeling RIS. Therefore, recently, this technique has also been used to model and study RIS-aided wireless communication systems \cite{shen2021,universal2024,gradoni2021end,nossek2024physically}, by regarding transmit antennas, receive antennas, and RIS elements as a whole multi-port network. In this subsection, we briefly explain the concept of this technique, together with its three important parameters, namely, impedance, admittance, and scattering parameters, to lay the foundation for the modeling of BD-RIS \cite{pozar2021microwave}. 
    
    \begin{figure}
        \centering
        \includegraphics[width=0.43\textwidth]{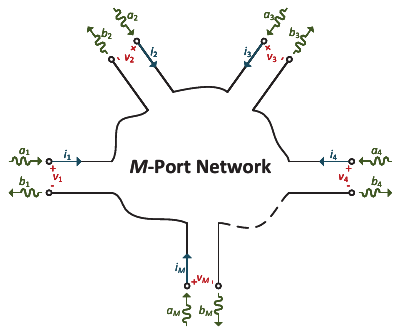}
        \caption{An arbitrary $M$-port microwave network.}
        \label{fig:network}\vspace{-0.5 cm}
    \end{figure}

    \subsubsection{Impedance, Admittance, and Scattering Parameters}
    Consider an arbitrary $M$-port microwave network, as depicted in Fig. \ref{fig:network}. At the $m$-th port, the terminal voltage, current, incident wave, and reflected wave are given by $v_m\in\mathbb{C}$, $i_m\in\mathbb{C}$, $a_m\in\mathbb{C}$, and $b_m\in\mathbb{C}$. Define the terminal voltage vector $\mathbf{v} = [v_1,\ldots,v_M]^\mathsf{T}\in\mathbb{C}^{M\times 1}$, the current vector $\mathbf{i} = [i_1,\ldots,i_M]^\mathsf{T}\in\mathbb{C}^{M\times 1}$, the incident wave vector $\mathbf{a}=[a_1,\ldots,a_M]^\mathsf{T}\in\mathbb{C}^{M\times 1}$, and the reflected wave vector $\mathbf{b}=[b_1,\ldots,b_M]^\mathsf{T}\in\mathbb{C}^{M\times 1}$ for the $M$-port network. These four vectors are related to each other by $\mathbf{v} = \mathbf{a} + \mathbf{b}$, $\mathbf{i} = \frac{\mathbf{a}-\mathbf{b}}{Z_0}$.
    The impedance matrix $\mathbf{Z}\in\mathbb{C}^{M\times M}$ of the $M$-port network then relates the terminal voltage vector and current vector by $\mathbf{v} = \mathbf{Z}\mathbf{i}$. In the impedance matrix $\mathbf{Z}$, each entry $[\mathbf{Z}]_{m,n}$ is the impedance between ports $m$ and $n$ when all other ports are open-circuited, i.e., $[\mathbf{Z}]_{m,n} = \frac{v_m}{i_n}\Big\vert_{i_k=0,\forall k\ne n}$.
    Similarly, the admittance matrix $\mathbf{Y}\in\mathbb{C}^{M\times M}$ of the $M$-port network relates the current vector and voltage vector by $\mathbf{i} = \mathbf{Y}\mathbf{v}$. Therefore, we have $\mathbf{Y}=\mathbf{Z}^{-1}$. In $\mathbf{Y}$, each entry $[\mathbf{Y}]_{m,n}$ is the admittance between ports $m$ and $n$ when all other ports are short-circuited, i.e.,
    $[\mathbf{Y}]_{m,n} = \frac{i_m}{v_n}\Big\vert_{v_k=0,\forall k\ne n}$. 
    The scattering matrix $\mathbf{S}\in\mathbb{C}^{M\times M}$ relates the incident wave vector and reflected wave vector by $\mathbf{b}=\mathbf{S}\mathbf{a}$. In $\mathbf{S}$, each entry $[\mathbf{S}]_{m,n}$ is the scattering coefficient from port $n$ to port $m$ when all other ports are terminated in matched loads, i.e.,
    $[\mathbf{S}]_{m,n} = \frac{b_m}{a_n}\Big\vert_{a_k=0,\forall k\ne n}$.
    Based on the above illustration, the three parameters are related to each other by 
    \begin{equation}
        \begin{aligned}
            \mathbf{S} &= (\mathbf{Z} + Z_0\mathbf{I}_M)^{-1}(\mathbf{Z} - Z_0\mathbf{I}_M)\\
             &= (Y_0\mathbf{I}_M + \mathbf{Y})^{-1}(Y_0\mathbf{I}_M - \mathbf{Y}),
        \end{aligned}
    \end{equation}
    where $Z_0$ denotes the reference impedance and $Y_0=Z_0^{-1}$ denotes the reference admittance. 

    \subsubsection{Reciprocal and Lossless Networks}
    Among all microwave networks, particularly important in practice are networks which are either reciprocal or lossless, or both. These networks also have specific mathematical characteristics. Specifically, for a reciprocal $M$-port network, any current $i_m$ injected into port $m$ produces a voltage $v_n$ at port $n$, $n\ne m$, and $i_m$ injected into port $n$ produces a voltage $v_n$ at port $m$. 
    Note that reciprocal networks are widely adopted in practice since commonly used impedance/admittance components are reciprocal, e.g., resistors, capacitors, and inductors.
    This mathematically means the impedance, admittance, and scattering matrices of a reciprocal network are all symmetric, that is 
    \begin{equation}
        \mathbf{Z} = \mathbf{Z}^\mathsf{T},  \mathbf{Y} = \mathbf{Y}^\mathsf{T},  \mathbf{S} = \mathbf{S}^\mathsf{T}.
    \end{equation}
    Further, for a reciprocal and lossless $M$-port network, the net real power delivered to the network is zero, that is 
    \begin{equation}
        P_\mathsf{net} = \frac{1}{2}\Re\{\mathbf{v}^\mathsf{T}\mathbf{i}^*\} =  \frac{1}{2}\Re\{\mathbf{i}^\mathsf{T}\mathbf{Z}\mathbf{i}^*\} = 0.
    \end{equation}
    This is achieved when all entries of impedance and admittance matrices are purely imaginary, and thus when the scattering matrix is unitary, that is 
    \begin{equation}
        \Re\{\mathbf{Z}\} = \mathbf{0}_{M\times M}, \Re\{\mathbf{Y}\} = \mathbf{0}_{M\times M}, \mathbf{S}^\mathsf{H}\mathbf{S} = \mathbf{I}_M.
    \end{equation}

\subsection{Modeling and Classification}
\label{subsec:model}

\begin{figure}
    \centering
    \includegraphics[width=0.45\textwidth]{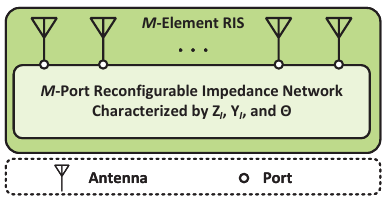}
    \caption{An $M$-element passive RIS modeled as an $M$-antenna array connected an $M$-port reconfigurable impedance network.}
    \label{fig:RIS_general}
\end{figure}

An $M$-element passive RIS is generally modeled as an $M$-antenna array connected to an $M$-port reconfigurable impedance network as illustrated in Fig. \ref{fig:RIS_general}, which can be characterized by its impedance matrix $\mathbf{Z}_I\in\mathbb{C}^{M\times M}$, admittance matrix $\mathbf{Y}_I\in\mathbb{C}^{M\times M}$, or scattering matrix $\mathbf{\Theta}\in\mathbb{C}^{M\times M}$ \cite{shen2021,universal2024}. These matrices are linked to each other by 
\begin{equation}
	\begin{aligned}
		\mathbf{\Theta} &= (\mathbf{Z}_I + Z_0\mathbf{I}_M)^{-1}(\mathbf{Z}_I - Z_0\mathbf{I}_M) \\
		&= (Y_0\mathbf{I}_M + \mathbf{Y}_I)^{-1}(Y_0\mathbf{I}_M - \mathbf{Y}_I), 
	\end{aligned}\label{eq:theta_zy}
\end{equation}
where $\mathbf{Y}_I = \mathbf{Z}_I^{-1}$. 
According to microwave network theory \cite{pozar2021microwave}, for a passive reconfigurable impedance network, we have 
\begin{equation}
    \mathbf{\Theta}^\mathsf{H}\mathbf{\Theta} \preceq \mathbf{I}_M. \label{eq:constraint_general}
\end{equation}
This guarantees that the power of reflected waves is no larger than the power of the waves impinging on the surface.
Specifically when the reconfigurable impedance network is lossless, we have that 
\begin{equation}
    \mathbf{\Theta}^\mathsf{H}\mathbf{\Theta} = \mathbf{I}_M,
    \label{eq:constraint_lossless_theta}
\end{equation}
indicating that it is a unitary matrix.
As per the mathematical properties of $\mathbf{Z}_I$, $\mathbf{Y}_I$, and $\mathbf{\Theta}$, we introduce the following two-layer classifications of RIS.

\subsubsection{D-RIS and BD-RIS}
Based on whether the three matrices, $\mathbf{Z}_I$, $\mathbf{Y}_I$, and $\mathbf{\Theta}$ are diagonal or not, we categorize RIS as D-RIS with diagonal matrices and BD-RIS with matrices not limited to being diagonal \cite{li2023reconfigurable}. 

\textit{Remark 1:} Here we give toy examples for a 4-element D-RIS with a diagonal scattering matrix $\mathbf{\Theta}^\mathsf{D}$ and a 4-element BD-RIS with a full scattering matrix $\mathbf{\Theta}^\mathsf{BD}$. Specifically in the lossless form of D-RIS, (\ref{eq:constraint_lossless_theta}) indicates that $|[\mathbf{\Theta}^\mathsf{D}]_{1,1}|=\ldots=|[\mathbf{\Theta}^\mathsf{D}]_{4,4}|=1$, which is consistent with the constraint commonly considered in existing D-RIS literature \cite{di2020smart,wu2025intelligent}. 
This shows that a lossless D-RIS can only support phase control of signals. 
However, for BD-RIS, (\ref{eq:constraint_lossless_theta}) indicates that $\|[\mathbf{\Theta}^\mathsf{BD}]_{:,1}\|_2 = \ldots = \|[\mathbf{\Theta}^\mathsf{BD}]_{:,4}\|_2 = 1$. Therefore, we have $|[\mathbf{\Theta}^\mathsf{BD}]_{m,n}| \le 1$, $\forall m,n = 1,\ldots,4$, implying that both amplitudes and phases of entries in $\mathbf{\Theta}^\mathsf{BD}$ are tunable even in the lossless case. 

\subsubsection{Reciprocal and Non-Reciprocal BD-RIS}
Further, based on whether the reconfigurable impedance network is reciprocal or not, we categorize BD-RIS as reciprocal BD-RIS with 
\begin{equation}
    \mathbf{Z}_I = \mathbf{Z}_I^\mathsf{T},~\mathbf{Y}_I = \mathbf{Y}_I^\mathsf{T},~\mathbf{\Theta} = \mathbf{\Theta}^\mathsf{T},
\end{equation}
and non-reciprocal BD-RIS with 
\begin{equation}
\mathbf{Z}_I \ne \mathbf{Z}_I^\mathsf{T}, \mathbf{Y}_I \ne \mathbf{Y}_I^\mathsf{T}, \mathbf{\Theta} \ne \mathbf{\Theta}^\mathsf{T}.
\end{equation}

\textit{Remark 2:} Here we reuse the examples for 4-element D-RIS and BD-RIS in Remark 1 to provide more insights. It is obvious that $\mathbf{\Theta}^\mathsf{D}$ is naturally symmetric such that D-RIS is only a special case of the reciprocal BD-RIS. However, $\mathbf{\Theta}^\mathsf{BD}$ can be symmetric or not, hence leading to reciprocal and non-reciprocal BD-RISs. This implies that non-reciprocity is an additional DoF arising from BD-RIS. 

\begin{figure*}
	\centering
	\includegraphics[width=0.92\textwidth]{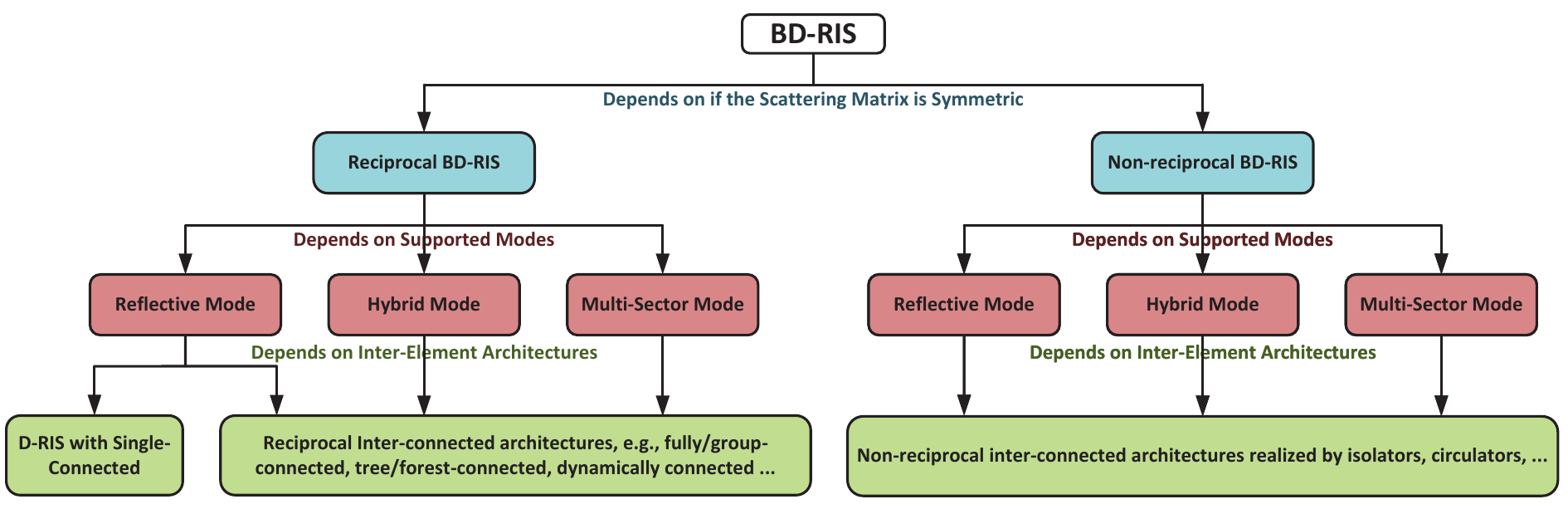}
	\caption{RIS classification tree.}
	\label{fig:tree}
\end{figure*}

Benefiting from the flexible arrangement of elements and the circuit topology design of the reconfigurable impedance network, both reciprocal and non-reciprocal BD-RIS enable various modes and architectures, allowing us to establish an RIS classification tree as illustrated in Fig. \ref{fig:tree}. In the following subsections, we will elaborate on the mode analysis and architecture designs of BD-RIS, providing schematic and mathematical illustrations. 

\subsection{Mode Analysis}

\begin{figure*}
	\centering
	\includegraphics[width=0.92\textwidth]{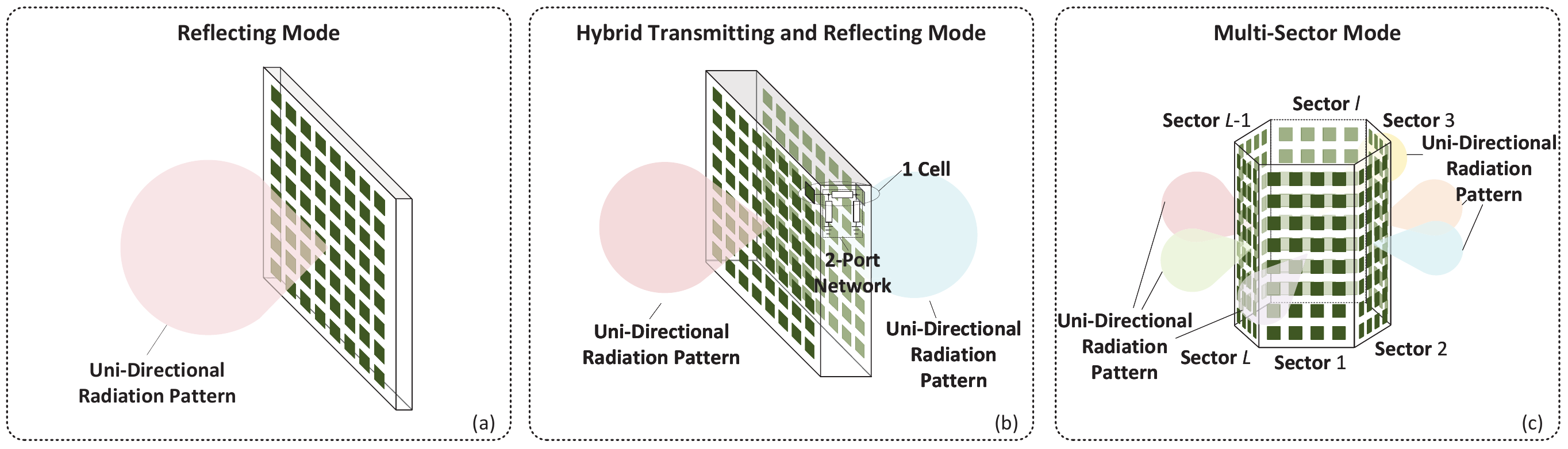}
	\caption{3D views of BD-RIS modes: (a) reflecting mode; (b) hybrid transmitting and reflecting mode; (c) multi-sector mode.}
	\label{fig:mode_3D}
\end{figure*}

According to the element arrangements, RIS has the following three modes. 

\subsubsection{Reflecting Mode}
In this mode, the $M$ elements are uniformly located toward the same direction, as illustrated in Fig. \ref{fig:mode_3D}(a). 
With this arrangement, signals impinging on one side of the RIS are all reflected toward the same side, yielding a half-space coverage. 

\subsubsection{Hybrid Transmitting and Reflecting Mode}
In this mode, every two elements are placed back to back and interconnected with each other via reconfigurable admittance components, as illustrated in Fig. \ref{fig:mode_3D}(b) \cite{li2022beyond}. 
With this arrangement, signals impinging on one sector of the RIS can be partially reflected toward the same sector and partially scattered toward the other sector, yielding a full-space coverage. This is also known as simultaneous transmitting and reflecting surface (STARS), STAR-RIS, or intelligent omni-surface (IOS) \cite{zhang2022intelligent}.
An extreme case of such a hybrid mode refers to the transmitting mode, which means the signals impinging on one sector of the surface are purely scattered toward the other sector. 

\subsubsection{Multi-Sector Mode}
In this mode, every $L$, $L\ge 2$ elements are placed at the edge of an $L$-sided polygon and interconnected with each other via reconfigurable admittance components, as illustrated in Fig. \ref{fig:mode_3D}(c) \cite{li2023beyond}. 
With this arrangement, signals impinging on one sector of RIS can be partially reflected toward the same sector and partially scattered toward the other sectors. 
More importantly, this arrangement allows each antenna to have a uni-directional radiation pattern covering only $\frac{1}{L}$ space to avoid overlapping among sectors, which provides higher gains than the hybrid mode. Consequently, multi-sector mode includes hybrid mode as a special case with $L=2$, while it goes beyond hybrid mode to achieve highly directional full-space coverage. 

\textit{Remark 3:} Note that the D-RIS can only support reflective mode, while hybrid and multi-sector modes can only be supported by BD-RIS. This is because both hybrid and multi-sector modes require interconnections between elements to enable the signal to flow from one element to another, which mathematically leads to admittance, impedance, and scattering matrices with nonzero off-diagonal entries.  

\vspace{-0.3 cm}

\subsection{Architecture Design: Reciprocal Architectures}
\label{subsec:architecture}

In the $M$-port reciprocal reconfigurable impedance network, each port is connected to ground via its own reconfigurable admittance\footnote{We use admittance parameter instead of impedance parameter to describe each reconfigurable component since admittance parameter describes the short-circuited property, that is how easily a circuit allows a current to flow. This allows us to have an admittance component with simply a zero admittance to describe that two elements are not connected. By contrast, the corresponding impedance component between the two elements will have an infinite value, which is not as straightforward as using admittance parameter. This will also simplify the mathematical modeling of various BD-RIS architectures, as will be detailed in the sequel.} component, $Y_m$, and to port $m'$, $m'\ne m$, $\forall m,m'\in\mathcal{M}$ via a reconfigurable admittance component $Y_{m,m'}$, which satisfies $Y_{m,m'} = Y_{m',m}$ for $m'>m$. 
Hence, the admittance matrix\footnote{We use admittance matrix $\mathbf{Y}_I$ instead of impedance matrix $\mathbf{Z}_I$ to describe BD-RIS architectures since, according to (\ref{eq:mapping_y}), there is a linear mapping between the locations of nonzero entries in $\mathbf{Y}_I$ and the circuit topology of a given architecture. For example, when $Y_{m,m'} = 0$, $m\ne m'$, indicating that ports $m$ and $m'$ are not connected, the $(m,m')$-th entry of $\mathbf{Y}_I$ is zero. By contrast, the corresponding impedance parameter-based descriptions have $Z_{m,m'}=Y_{m,m'}^{-1}$, $m\ne m'$, $Z_m = Y_m^{-1}$ and $\mathbf{Z}_I=\mathbf{Y}_I^{-1}$, implying a nonlinear mapping between the structure of $\mathbf{Z}_I$ and certain circuit topologies. This means when $Y_{m,m'} = 0$, equivalently $Z_{m,m'} = \infty$, the $(m,m')$-th entry of $\mathbf{Z}_I$ could be neither zero nor $\infty$.} $\mathbf{Y}_I$ of the reconfigurable impedance network is symmetric and each entry, $[\mathbf{Y}_I]_{m,m'}$ linking ports $m$ and $m'$, can be calculated by making all other ports short-circuited \cite{pozar2021microwave}. Therefore, we have  
\begin{equation}
	[\mathbf{Y}_I]_{m,m'} = \begin{cases} - Y_{m,m'}, &m \ne m',\\
		Y_m + \sum_{k\ne m} Y_{m,k}, &m = m'.
	\end{cases}\label{eq:mapping_y}
\end{equation}
As per the circuit topology of the $M$-port reconfigurable impedance network, the reconfigurable admittance components $Y_{m,m'}$, $m\ne m'$ can be zero (indicating port $m$ is not directly connected to port $m'$) or not (indicating port $m$ is directly connected to port $m'$), leading to the following reciprocal architectures of BD-RIS. 

\subsubsection{Single-Connected (i.e., D-RIS)}
In this architecture, each port $m$, $\forall m\in\mathcal{M}$, in the reconfigurable impedance network is connected to ground via its own reconfigurable admittance component, $Y_{m}$ without interacting with other ports, i.e., $Y_{m,m'}=0$, $m\ne m'$. 
This indicates that the single-connected architecture is realized by in total $M$ reconfigurable admittance components. 
An illustrative example for a 4-element RIS with single-connected architecture is given in Fig. \ref{fig:reciprocal_architecture}(a). 
Hence, $\mathbf{Y}_I$ is a diagonal matrix
\begin{equation}
    \mathbf{Y}_I=\mathsf{diag}(Y_{1},\ldots,Y_{M}).
\end{equation}
By (\ref{eq:theta_zy}) and (\ref{eq:constraint_lossless_theta}), for a lossless network, its impedance matrix $\mathbf{Z}_I$ and scattering matrix $\mathbf{\Theta}$ are also diagonal, and each diagonal entry in the latter has unit modulus, i.e., 
\begin{equation}
	\mathbf{\Theta} = \mathsf{diag}(\varTheta_1,\ldots,\varTheta_M),~|\varTheta_m| = 1, \forall m\in\mathcal{M},
\end{equation} 
where $\varTheta_m = \frac{Y_0 - Y_m}{Y_0+Y_m}$ following (\ref{eq:theta_zy}).
The single-connected RIS is also referred to as D-RIS with a diagonal matrix (\ref{eq:diag}) in Section \ref{sec:example} and has been widely studied and applied in wireless communication systems \cite{di2020smart,wu2025intelligent}. The corresponding scattering matrix $\mathbf{\Theta}$ is very often named as the phase-shift matrix since only the phase shift of each diagonal entry can be tuned.

\subsubsection{Fully-Connected}
\begin{figure}[t]
	\centering
	\includegraphics[width=0.48\textwidth]{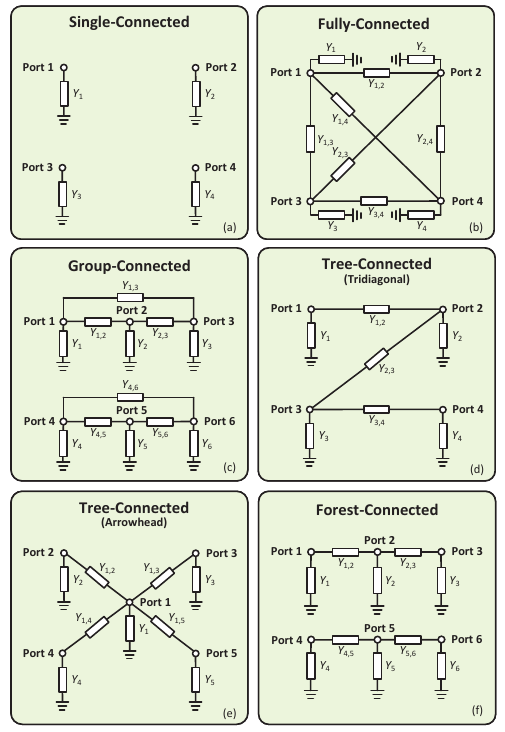}
	\caption{Examples of reciprocal architectures: (a) A 4-element D-RIS with single-connected architectures; (b) a 4-element BD-RIS with fully-connected architectures; (c) a 6-element BD-RIS with group-connected architectures (each group containing 3 interconnected elements); (d) a 4-element tridiagonal BD-RIS with tree-connected architectures; (e) a 5-element arrowhead BD-RIS with tree-connected architectures; (f) a 6-element BD-RIS with forest-connected architectures (each group containing 3 interconnected elements).}
    \label{fig:reciprocal_architecture}
\end{figure}

To break through the strict mathematical constraint of $\mathbf{\Theta}$ in D-RIS and fully make use of the off-diagonal entries in $\mathbf{\Theta}$, a fully-connected architecture which could effectively improve the wave manipulation flexibility of RIS has been proposed in \cite{shen2021}.
In this architecture, each port $m$ in the reconfigurable impedance network is connected to ground via $Y_m$ and to another port $m'$, $m'\ne m$, $\forall m,m'\in\mathcal{M}$ via $Y_{m,m'}$, which satisfies $Y_{m,m'} = Y_{m',m}$ for $m'>m$. 
This indicates that the fully-connected architecture is realized by in total $\frac{M(M+1)}{2}$ reconfigurable admittance components. 
An illustrative example for a 4-element BD-RIS with fully-connected architecture is given in Fig. \ref{fig:reciprocal_architecture}(b).
Hence, $\mathbf{Y}_I$ is a full and symmetric matrix. By (\ref{eq:theta_zy}), for a lossless network, its impedance matrix $\mathbf{Z}_I$ is also a full and symmetric matrix, and the scattering matrix $\mathbf{\Theta}$ is a symmetric and unitary matrix, i.e.,
\begin{equation}
	\mathbf{\Theta} = \mathbf{\Theta}^\mathsf{T}, ~ \mathbf{\Theta}^\mathsf{H}\mathbf{\Theta} = \mathbf{I}_M. \label{eq:fully_connected}
\end{equation} 

\subsubsection{Group-Connected}
The circuit complexity (characterized by the required number of reconfigurable admittance components) of the fully-connected architecture grows quadratically with the number of elements $M$. This makes it difficult to implement fully-connected BD-RISs in practice especially when a large dimension is required. To trade the circuit complexity and wave manipulation flexibility of RIS, a group-connected architecture has been proposed in \cite{shen2021}. In this architecture, the $M$ ports in the reconfigurable impedance network are uniformly divided into $G$ groups, and every $\bar{M} = \frac{M}{G}$ ports within one group are connected to each other to construct a fully-connected architecture. 
Specifically, in group $g$, $\forall g\in\mathcal{G}=\{1,\ldots,G\}$, each port $m_g = (g-1)\bar{M}+m$ is connected to ground via it own reconfigurable admittance component $Y_{m_g}$ and to another port $m'_g = (g-1)\bar{M}+m'$, $m'\ne m$, $\forall m,m'\in\bar{\mathcal{M}}=\{1,\ldots,\bar{M}\}$, via $Y_{m_g,m'_g}$, which satisfies $Y_{m_g,m'_g} = Y_{m'_g,m_g}$ for $m'>m$.
This indicates that the group-connected architecture is realized by in total $G\frac{\bar{M}(\bar{M}+1)}{2}=\frac{M(\bar{M}+1)}{2}$ reconfigurable admittance components. An illustrative example for a 6-element BD-RIS with group-connected architecture is given in Fig. \ref{fig:reciprocal_architecture}(c). Hence, $\mathbf{Y}_I$ is a block-diagonal matrix 
\begin{equation}
    \mathbf{Y}_I = \mathsf{blkdiag}(\mathbf{Y}_{I,1},\ldots,\mathbf{Y}_{I,G}), 
\end{equation}
where each $\mathbf{Y}_{I,g}\in\mathbb{C}^{\bar{M}\times\bar{M}}$ is a full and symmetric matrix. 
Accordingly, $\mathbf{Z}_I$ is a block-diagonal matrix with each block being a full and symmetric matrix. 
By (\ref{eq:theta_zy}), for a lossless network, its scattering matrix $\mathbf{\Theta}$ is a block-diagonal matrix, with each block, $\mathbf{\Theta}_g\in\mathbb{C}^{\bar{M}\times\bar{M}}$, being symmetric and unitary, i.e.,
\begin{equation}
    \begin{aligned}
        \mathbf{\Theta} &= \mathsf{blkdiag}(\mathbf{\Theta}_1,\ldots,\mathbf{\Theta}_G),\\
	\mathbf{\Theta}_g &= \mathbf{\Theta}_g^\mathsf{T}, ~ \mathbf{\Theta}_g^\mathsf{H}\mathbf{\Theta}_g = \mathbf{I}_{\bar{M}}, \forall g\in\mathcal{G}. 
    \end{aligned}\label{eq:group_connected}
\end{equation}   
This architecture is a general illustration, including the fully-connected architecture with $G = 1$ and the single-connected architecture with $G = M$ as two special cases.   

\subsubsection{Tree-Connected}
While fully-connected architecture has the highest wave manipulation flexibility to achieve optimal performance, this is achieved at the cost of complicated circuit topology design requiring numerous reconfigurable admittance components. To reduce the circuit complexity while maintaining the optimal (i.e., same as fully-connected) performance in single-user multiple input single output (MISO) systems, tree-connected architecture where the circuit topology forms a tree based on graph theory \cite{bondy2008graph} has been proposed in \cite{nerini2024beyond}. 
In this architecture, the circuit complexity reduces to in total $2M-1$ admittance components, which can be much less than that of the fully-connected architecture. Two typical tree-connected architectures, leading to respectively tridiagonal and arrowhead admittance matrices, have been introduced in \cite{nerini2024beyond} and are reviewed as follows. 
\begin{itemize}
	\item \textit{Tridiagonal:} In this case, each port $m$ is connected to ground via its own admittance $Y_m$ and to port $m+1$ via an admittance $Y_{m,m+1}$, $\forall m\in\mathcal{M}\backslash \{M\}$. An illustrative example for a 4-element BD-RIS with such architecture is given in Fig. \ref{fig:reciprocal_architecture}(d). This mathematically leads to a tridiagonal admittance matrix 
	\begin{equation}
		\mathbf{Y}_I = \left[\begin{matrix}
			[\mathbf{Y}_{I}]_{1,1} & [\mathbf{Y}_{I}]_{1,2} & \cdots & 0\\
			[\mathbf{Y}_{I}]_{1,2} & \ddots & \ddots & \vdots \\
			\vdots & \ddots & \ddots & [\mathbf{Y}_{I}]_{M-1,M}\\
			0 & \cdots &  [\mathbf{Y}_{I}]_{M-1,M} & [\mathbf{Y}_{I}]_{M,M}
		\end{matrix}\right],\label{eq:tridiagonal}
	\end{equation}
	which is symmetric and contains nonzero entries only on the main diagonal, the lower diagonal, and the upper diagonal.
	\item \textit{Arrowhead:} In this case, each port $m$ is connected to ground via its own admittance $Y_m$ and there is a central port $c$ which connects to all other ports via $Y_{c,m}$, $\forall m \ne c, m\in\mathcal{M}$. An illustrative example for a 5-element BD-RIS with such architecture is given in Fig. \ref{fig:reciprocal_architecture}(e). Assuming $c=1$, we have an arrowhead admittance matrix 
	\begin{equation}
		\mathbf{Y}_I = \left[\begin{matrix}
			[\mathbf{Y}_I]_{1,1} & [\mathbf{Y}_I]_{1,2} &\cdots &[\mathbf{Y}_I]_{1,M}\\
			[\mathbf{Y}_I]_{1,2} & [\mathbf{Y}_I]_{2,2} &\cdots & 0 \\
			\vdots & \vdots &\ddots &\vdots\\
			[\mathbf{Y}_I]_{1,M} & 0 & \cdots & [\mathbf{Y}_I]_{M,M}
		\end{matrix}\right],\label{eq:arrowhead}
	\end{equation}
	which is symmetric and contains nonzero entries only on the main diagonal, the first row, and the first column.
\end{itemize}

\textit{Remark 4:}
Due to the non-linear relationships between $\mathbf{Y}_I$ and $\mathbf{\Theta}$ as in (\ref{eq:theta_zy}), and between $\mathbf{Y}_I$ and $\mathbf{Z}_I$, i.e., $\mathbf{Z}_I = \mathbf{Y}_I^{-1}$, both $\mathbf{Z}_I$ and $\mathbf{\Theta}$ will be full matrices, similar to fully-connected architecture. 
Take Fig. \ref{fig:reciprocal_architecture}(d) as an example. For a 4-element tridiagonal BD-RIS, the resulting admittance matrix writes as 
\begin{equation}
    \mathbf{Y}_I = \left[\begin{matrix}
        [\mathbf{Y}_I]_{1,1} & [\mathbf{Y}_I]_{1,2} & 0 & 0\\
        [\mathbf{Y}_I]_{1,2} & [\mathbf{Y}_I]_{2,2} & [\mathbf{Y}_I]_{2,3} & 0\\
        0 & [\mathbf{Y}_I]_{2,3} & [\mathbf{Y}_I]_{3,3} & [\mathbf{Y}_I]_{3,4}\\
        0 & 0 & [\mathbf{Y}_I]_{3,4} & [\mathbf{Y}_I]_{4,4}
    \end{matrix}\right],
\end{equation}
where the locations of nonzero entries are one-to-one mapped to the circuit topology of a tree-connected architecture. However, the resulting $\mathbf{Z}_I$ and $\mathbf{\Theta}$ will be full matrices, making it difficult to mathematically distinguish tree- and fully-connected architectures. 
This implies that using $\mathbf{\Theta}$ or $\mathbf{Z}_I$ cannot fully reflect the mathematical constraint of the tree-connected architecture.

\subsubsection{Forest-Connected}
Similar to the extension from fully-connected to group-connected architectures, the circuit complexity of the tree-connected architecture can be further reduced by dividing the $M$ ports into $G$ groups and constructing each group as a $\bar{M}$-port tree-connected architecture. This is referred to as the forest-connected architecture \cite{nerini2024beyond} with a circuit complexity $G(2\bar{M}-1)$. An illustrative example for a 6-element RIS with forest-connected architecture is given in Fig. \ref{fig:reciprocal_architecture}(f). Hence, $\mathbf{Y}_I$ is a block-diagonal matrix with each block being a symmetric and tridiagonal (or arrowhead) matrix. Two extreme cases with $G=M$ and $G=1$ respectively correspond to single-connected and tree-connected architectures. For a lossless network with the forest-connected architecture, its impedance matrix $\mathbf{Z}_I$ also has a block-diagonal shape with each block being full and symmetric, and its scattering matrix $\mathbf{\Theta}$ is generally a block-diagonal matrix with each block being symmetric and unitary.
Therefore, using $\mathbf{\Theta}$ or $\mathbf{Z}_I$ is, gain, not sufficient to fully reflect the mathematical constraint of the forest-connected architecture.

\subsubsection{Band- and Stem-Connected}
Inspired by the tree-connected architecture that can perfectly match the performance of fully-connected architecture in single-user MISO systems with significantly reduced circuit complexity, the optimal architectures that can perfectly match the performance of fully-connected architecture in multi-user MIMO systems have been proposed in \cite{wu2025beyond}. In the optimal architectures, each port is connected to ground via its own reconfigurable admittance component, while the number of interconnections between ports is theoretically constrained by the DoF in multi-user MIMO systems, and is generally much less than that of fully-connected architectures. Two representative examples which could construct optimal architectures, namely band-connected \cite{wu2025beyond} and stem-connected \cite{wu2025beyond,zhou2024novel}, are reviewed as follows. 
\begin{itemize}
	\item \textit{Band-Connected:} In this case, each port $m$ is connected to ground via $Y_m$ and to the following $q$ ports, i.e., $m+1,\ldots,m+q$ (if any), via corresponding admittance components, where $q$ denotes the band width. An illustrative example for a 4-element BD-RIS with band-connected architecture ($q=2$) is given in Fig. \ref{fig:band_stem_architecture}(a). This mathematically leads to a band admittance matrix 
	\begin{equation}
		\mathbf{Y}_I = \left[\begin{matrix}
			[\mathbf{Y}_{I}]_{1,1} &\cdots &[\mathbf{Y}_{I}]_{1,q+1} & \cdots~~~~~ 0~~~~~ \\
			\vdots  & \ddots &~ &\ddots~~~~~~~~\\
			[\mathbf{Y}_{I}]_{1,q+1} & \ddots& \ddots & ~~~~~\vdots \\
			\vdots & ~ & \ddots & ~~[\mathbf{Y}_{I}]_{M-q,M}\\
			~ & \ddots & ~ & ~~~~~\vdots\\
			0 & \cdots &  [\mathbf{Y}_{I}]_{M-q,M} & \cdots~[\mathbf{Y}_{I}]_{M,M}
		\end{matrix}\right],\label{eq:band}
	\end{equation}
	which is a symmetric matrix containing nonzero entries only on the main diagonal, the $q$ lower diagonal, and the $q$ upper diagonal. When $q=1$, the band matrix boils down to the tridiagonal matrix, and thus band-connected BD-RIS includes the tridiagonal tree-connected BD-RIS as a special case. 

\begin{figure}[t]
	\centering
	\includegraphics[width=0.48\textwidth]{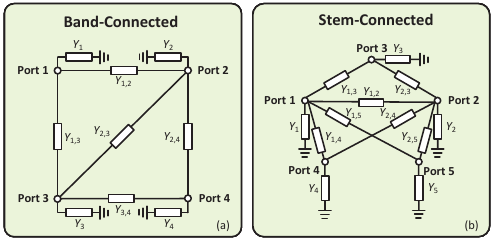}
	\caption{Examples of band- and stem-connected architectures: (a) a 4-element BD-RIS with band-connected architectures (band width $q=2$); (b) a 5-element BD-RIS with stem-connected architectures (stem width $q=2$).}
    \label{fig:band_stem_architecture}
\end{figure}

	\item \textit{Stem-Connected:} In this case, each port is connected to ground via $Y_m$. Meanwhile, there are $q$ ports with $q$ denoting the stem width, marked as $c_i,\forall i\in\{1,\ldots,q\}$, each of which connects to all other ports via corresponding admittance components.  An illustrative example for a 5-element BD-RIS with such architecture ($q=2$, $c_1 = 1$, $c_2=2$) is given in Fig. \ref{fig:band_stem_architecture}(b). Assuming $c_i\in\{1,\ldots,q\}$, we have an admittance matrix with the following structure:
	\begin{equation}
		\mathbf{Y}_I = \left[\begin{matrix}
			\mathbf{A}_q &\mathbf{C}_q\\
			\mathbf{C}_q^\mathsf{T} & \mathbf{D}_q
		\end{matrix}\right],\label{eq:stem}
	\end{equation}
	where
    \begin{subequations}
        \begin{align}
            \mathbf{A}_{q} &= \left[\begin{matrix}
        		[\mathbf{Y}_{I}]_{1,1} & \cdots &[\mathbf{Y}_{I}]_{1,q}\\
        		\vdots &\ddots  & \vdots \\
        		[\mathbf{Y}_{I}]_{1,q} &\cdots & [\mathbf{Y}_{I}]_{q,q}
        	\end{matrix}\right],\\ 
            \mathbf{C}_{q} &= \left[\begin{matrix}
        		[\mathbf{Y}_{I}]_{1,q+1} & \cdots &[\mathbf{Y}_{I}]_{1,M}\\
        		\vdots &\ddots  & \vdots \\
        		[\mathbf{Y}_{I}]_{q,q+1} &\cdots & [\mathbf{Y}_{I}]_{q,M},
        	\end{matrix}\right], \\
            \mathbf{D}_q &= \mathsf{diag}([\mathbf{Y}_I]_{q+1,q+1},\ldots,[\mathbf{Y}_I]_{M,M}). 
        \end{align}
    \end{subequations}
    Similarly, $\mathbf{Y}_I$ is symmetric containing nonzero entries only on the main diagonal, the first $q$ rows, and the first $q$ columns. When $q=1$, the matrix in (\ref{eq:stem}) boils down to the arrowhead matrix, and thus stem-connected BD-RIS includes the arrowhead tree-connected BD-RIS as a special case. 
\end{itemize}
It is worth noting that, similar to tree- and forest-connected architectures, for a lossless network with band- and stem-connected architectures, its impedance matrix $\mathbf{Z}_I$ and scattering matrices $\mathbf{\Theta}$ will both be full matrices\footnote{Note that all the reciprocal architectures we talk about are explicitly described by the admittance parameter, instead of the scattering parameter. For example, the band-connected architecture has a band admittance matrix, where each nonzero entry in the off-diagonal indicates two ports are connected. However, the resulting scattering matrix will not be a band matrix, but instead a full matrix. This is also physically correct since a band-connected architecture essentially makes all the ports connected to each other, such that all entries in the scattering matrix can contribute to the wave scattering.}. Therefore, $\mathbf{Y}_I$ should be used to accurately reflect the constraint of these two architectures.

\subsubsection{Dynamically Connected}

\begin{figure}
	\centering
	\includegraphics[width=0.48\textwidth]{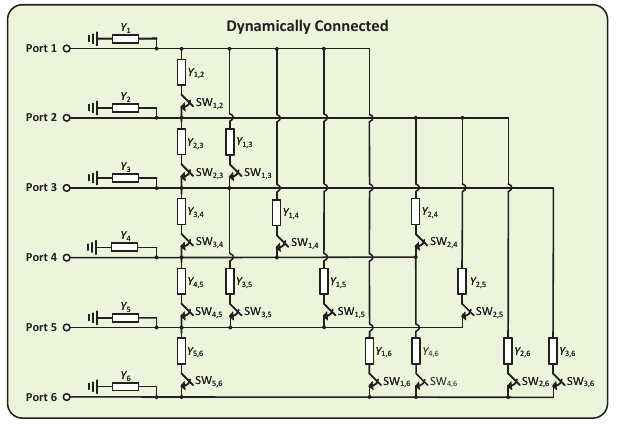}
	\caption{An example of a reciprocal dynamically connected architecture where every two ports are connected by the series of a reconfigurable admittance and a switch.}
	\label{fig:reciprocal_architecture_dynamic}
\end{figure}

\begin{figure}
    \centering
    \includegraphics[width=0.48\textwidth]{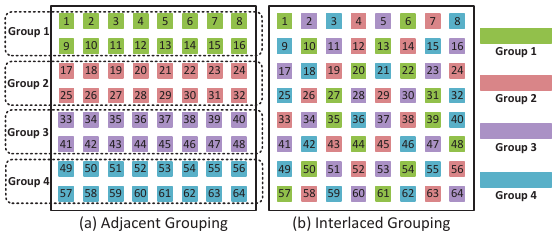}
    \caption{Examples of a 64-element group-connected BD-RIS with (a) adjacent grouping and (b) interlaced grouping.}
    \label{fig:grouping}
\end{figure}

In addition to the aforementioned fixed architectures which do not adapt to the channel environment (i.e., the admittance components are tuned but the circuit topology dictated by the interconnections is fixed), a dynamically connected architecture whose circuit topology can be changed during the transmission has also been proposed in \cite{li2023dynamic}. This can be implemented by an admittance-switch network which enables a joint control of the values of admittance components and the ON/OFF states of switches. In this architecture, every port $m$ is connected to ground via its own admittance $Y_m$ and connected to port $m'$, $m'> m$, $\forall m,m'\in\mathcal{M}$ by the series of an admittance $Y_{m,m'}$ and a switch $\mathsf{SW}_{m,m'}$. This indicates that the dynamically connected architecture is realized by in total $\frac{M(M+1)}{2}$ reconfigurable admittance components and $\frac{M(M-1)}{2}$ switches. An illustrative example for a 6-element RIS with dynamically connected architecture is given in Fig. \ref{fig:reciprocal_architecture_dynamic}. 
It is important to highlight that the admittance-switch network can be regarded as a general framework, which enables single/fully/group/tree/forest/band/stem-connected architectures by activating the corresponding interconnection links between ports. 
For example, the fully-connected architecture is realized when all switches are turned ON, and the single-connected architecture is realized when all switches are turned OFF. 
Beyond that, this admittance-switch network can also support various grouping strategies, different from the group/forest-connected architectures where the ports are uniformly and adjacently grouped. 
For example, one can control the switches to realize group-connected architectures with various group size of each group \cite{li2023dynamic}, or to realize group-connected architectures where each group contains interlaced elements \cite{nerini2024static}, as illustrated in Fig. \ref{fig:grouping}.

\textit{Remark 5:} Note that for D-RIS, its impedance, admittance, and scattering matrices are all diagonal. This friendly mathematical structure has simplified a lot the modeling and signal processing of D-RIS-aided wireless systems. However, for BD-RIS, its three matrices do not always convey equivalent information. For example, for group-connected architecture, its three matrices are all block-diagonal. In this sense, one can use each of them for further studies. For another example, for tree/forest/band/stem-connected architectures, the mathematical structures are only captured in their admittance matrices. This means one should use the admittance matrix $\mathbf{Y}_I$ and cannot use the scattering matrix $\mathbf{\Theta}$ or the impedance matrix $\mathbf{Z}_I$ of these architectures as objective variables for possible optimization and performance analysis.

\begin{figure}
	\centering
	\includegraphics[width=0.48\textwidth]{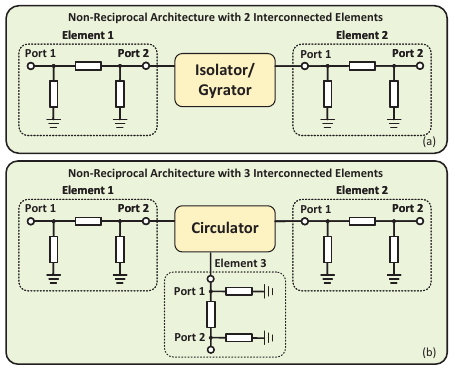}
	\caption{Examples of non-reciprocal architectures: (a) 2 elements interconnect with each other via an isolator or gyrator; (b) 3 elements interconnect with each other via a circulator.}
	\label{fig:nonreciprocal_architecture}
\end{figure}

\subsection{Architecture Design: Non-Reciprocal Architectures}
\label{subsec:architecture_nr}

In addition to the reciprocal architectures \cite{shen2021,nerini2024beyond,li2023dynamic,nerini2024static,wu2025beyond} summarized in the previous subsection, there are also non-reciprocal architectures of BD-RIS realized by non-reciprocal circuits, e.g., isolators, gyrators, circulators \cite{li2022reconfigurable,li2024coordinated,xu2024non}.  
Specifically, \cite{xu2024non} proposes a physics-consistent device model, where all the $M$ elements are uniformly divided into $G$ groups with each containing $\bar{M}$ elements interconnecting with a non-reciprocal device. In each group, each element is modeled as a 2-port reciprocal network where one port interacts with free space and the other is connected to a non-reciprocal $\bar{M}$-port circuit. 
Two illustrative examples\footnote{Although here we only give examples for 2-element and 3-element interconnected non-reciprical architectures for the purpose of easy illustration and simple hardware implementation, we would like to clarify that large-scale non-reciprocal architectures are practically feasible by using the combinations of multiple isolators and circulators.} for a 2-element interconnected architecture realized by the isolator/gyrator and a 3-element interconnected architecture realized by the circulator are respectively given in Fig. \ref{fig:nonreciprocal_architecture}(a) and Fig. \ref{fig:nonreciprocal_architecture}(b). 
Due to the introduction of non-reciprocal circuits in each interconnected group, the resulting scattering matrix for a lossless network is a block-diagonal matrix, i.e., $\mathbf{\Theta} = \mathsf{blkdiag}(\mathbf{\Theta}_1,\ldots,\mathbf{\Theta}_G)$ with each block being asymmetric and unitary, that is 
\begin{equation}
	\mathbf{\Theta}_g \ne \mathbf{\Theta}_g^\mathsf{T}, ~ \mathbf{\Theta}_g^\mathsf{H}\mathbf{\Theta}_g = \mathbf{I}_{\bar{M}}, \forall g\in\mathcal{G}. 
\end{equation}
One extreme case corresponds to the most flexible non-reciprocal BD-RIS whose scattering matrix is simply a full and unitary matrix.
Another extreme case corresponds to the non-diagonal RIS \cite{li2022reconfigurable}, whose scattering matrix is asymmetric and contains only $M$ nonzero entries. In this case, $\mathbf{\Theta}$ has the following form
\begin{equation}
    \mathbf{\Theta} = \mathbf{\Gamma}_\mathsf{r}\bar{\mathbf{\Theta}}\mathbf{\Gamma}_\mathsf{t},~ \bar{\mathbf{\Theta}} = \mathsf{diag}(e^{\jmath\theta_1},\ldots,e^{\jmath\theta_M}), 
\end{equation}
where $\theta_m\in[0,2\pi), \forall m\in\mathcal{M}$, $\mathbf{\Gamma}_{\mathsf{r}}\in\{0,1\}^{M\times M}$ and $\mathbf{\Gamma}_{\mathsf{t}}\in\{0,1\}^{M\times M}$ are two permutation matrices. This implies that signals impinging on one element are purely reflected by another element. 
Here we provide a toy example for a 4-element non-diagonal BD-RIS having 
\begin{equation}
    \mathbf{\Theta}^\mathsf{ND} = \left[\begin{matrix} 
    0 & [\mathbf{\Theta}]_{1,2} & 0 & 0 \\
    0 & 0 & 0 & [\mathbf{\Theta}]_{2,4}\\
    [\mathbf{\Theta}]_{3,1} & 0 & 0 & 0\\
    0 & 0 & [\mathbf{\Theta}]_{4,3} & 0
    \end{matrix}\right],
\end{equation}
where, for instance, $[\mathbf{\Theta}]_{1,2}\ne 0$ controls the signal flow from element 2 to element 1, that is, the signal impinging on element 2 is purely reflected by element 1.

\textit{Remark 6:} Based on the above illustration, we notice that both reciprocal and non-reciprocal architectures have their pros and cons. On the one hand, reciprocal architectures have advantages over non-reciprocal ones from the following two perspectives: 1) The reciprocal architectures have mathematical constraints whose locations of nonzero entries are directly reflected in the circuit topology, and this can potentially facilitate the development of various optimization methods; 2) the reciprocal architectures have simpler circuit designs without embedding non-reciprocal devices and can be more cost-effective in practical implementations. On the other hand, non-reciprocal architectures have the advantage over reciprocal ones in providing more wave manipulation flexibility due to the relaxation of the symmetric constraint in the impedance/admittance/scattering matrix of BD-RIS. 

\subsection{Unified Modes and Architectures}
\label{subsec:unified}

BD-RIS goes beyond D-RIS by introducing interconnections between elements, which provides the possibility to enable various circuit topologies offering more flexible beam manipulation and to enable various modes with enhanced coverage. To clearly show how the circuit topologies can support different
modes, we provide the following two examples.

\begin{figure}
	\centering
	\includegraphics[width=0.48\textwidth]{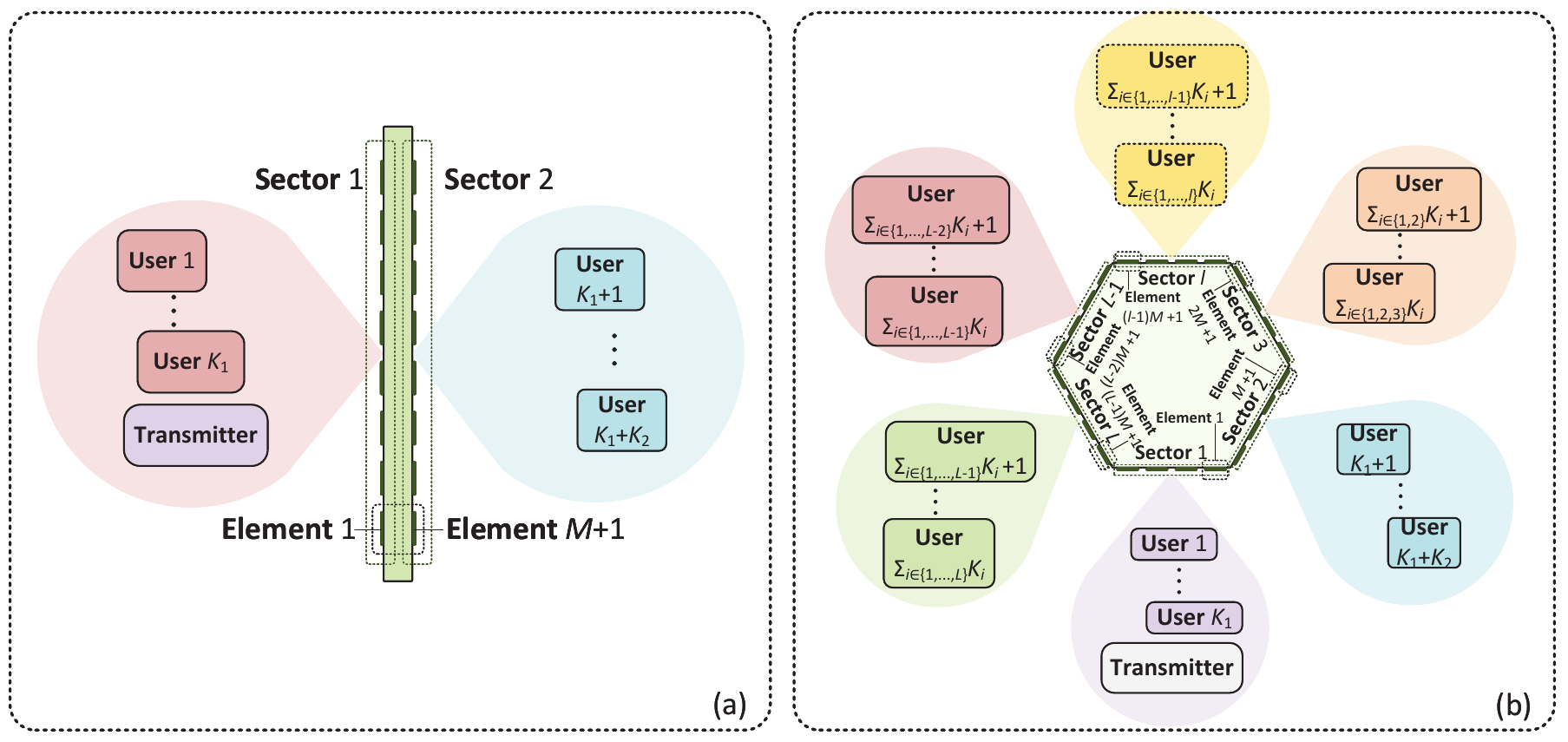}
	\caption{Top views of BD-RIS modes: (a) hybrid transmitting and reflecting mode; (b) multi-sector mode.}
	\label{fig:mode_top}
\end{figure}

\begin{figure}
	\centering 
	\includegraphics[width=0.48\textwidth]{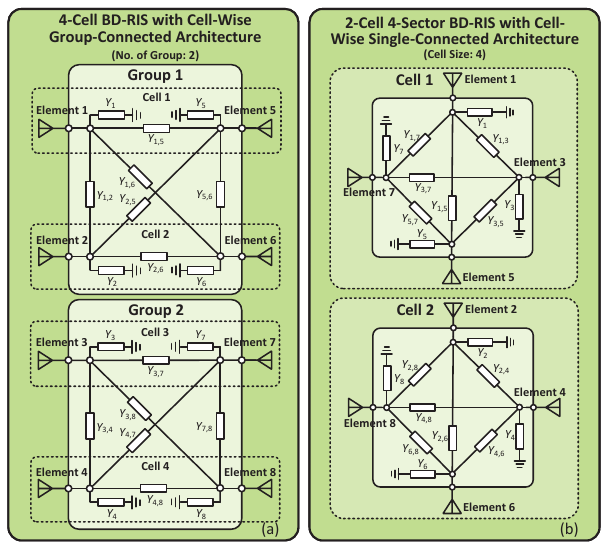}
	\caption{Examples of (a) hybrid mode and (b) multi-sector mode realized by group-connected architectures. $M=8$, $G=2$, and $\bar{M}=4$ for both examples, and $L=4$ for the multi-sector mode.}
	\label{fig:architecture_mode}
\end{figure}

\textit{Example 1: Hybrid Mode Realized by Group-Connected Architecture.} In this case, every two elements are back to back placed and interconnected with each other via a 2-port fully-connected architecture to construct a cell, and every $\frac{\bar{M}}{2}$ cells are interconnected with each other to construct a group, as illustrated in Figs. \ref{fig:mode_top}(a) and \ref{fig:architecture_mode}(a). Therefore, we have the scattering matrix $\mathbf{\Theta}$ blocked as 
\begin{equation}
	\mathbf{\Theta} = \left[\begin{matrix}
		\mathbf{\Theta}_{1,1} & \mathbf{\Theta}_{1,2}\\
		\mathbf{\Theta}_{2,1} & \mathbf{\Theta}_{2,2}
	\end{matrix}\right],
\end{equation} 
where $\mathbf{\Theta}_{i,j}\in\mathbb{C}^{\frac{M}{2}\times \frac{M}{2}}$ describes the power scattering from sector $j$ to sector $i$, $\forall i,j\in\{1,2\}$, and each has a block-diagonal structure:
\begin{equation}
	\mathbf{\Theta}_{i,j} = \mathsf{blkdiag}(\mathbf{\Theta}_{i,j,1},\ldots,\mathbf{\Theta}_{i,j,G}), \forall i,j\in\{1,2\},
\end{equation}
with $\mathbf{\Theta}_{i,j,g}\in\mathbb{C}^{\frac{\bar{M}}{2}\times\frac{\bar{M}}{2}}$, $\forall g\in\mathcal{G}$.
When there is only one source located within the coverage of sector 1 as illustrated in Fig. \ref{fig:mode_top}(a), we introduce more intuitive notations, i.e., $\mathbf{\Theta}_{\mathsf{r}} = \mathbf{\Theta}_{1,1}$, $\mathbf{\Theta}_{\mathsf{t}} = \mathbf{\Theta}_{2,1}$, $\mathbf{\Theta}_{\mathsf{r},g} = \mathbf{\Theta}_{1,1,g}$ and $\mathbf{\Theta}_{\mathsf{t},g} = \mathbf{\Theta}_{2,1,g}$, $\forall g\in\mathcal{G}$ to describe the power reflecting (from sector 1 to sector 1) and transmitting (from sector 1 to sector 2) \cite{li2022beyond}. Then, when the group-connected reconfigurable impedance network is lossless, i.e., $\mathbf{\Theta}^\mathsf{H}\mathbf{\Theta} = \mathbf{I}_{M}$ and $\mathbf{\Theta} = \mathbf{\Theta}^\mathsf{T}$, we have the following constraint:
\begin{equation}
	\mathbf{\Theta}_{\mathsf{r},g}^\mathsf{H}\mathbf{\Theta}_{\mathsf{r},g} + \mathbf{\Theta}_{\mathsf{t},g}^\mathsf{H}\mathbf{\Theta}_{\mathsf{t},g} = \mathbf{I}_{\frac{\bar{M}}{2}}, \mathbf{\Theta}_{\mathsf{r},g} = \mathbf{\Theta}_{\mathsf{r},g}^\mathsf{T}, \label{eq:hybrid}
\end{equation}
which means the sum of the reflected power and transmitted power is conserved without loss.
Specifically when $G = \frac{M}{2}$, both $\mathbf{\Theta}_\mathsf{r}$ and $\mathbf{\Theta}_\mathsf{t}$ are diagonal matrices, which correspond to the STARS, STAR-RIS or IOS \cite{zhang2022intelligent}. 

\textit{Example 2: Multi-Sector Mode Realized by Group-Connected Architecture.} In this case, every $L$ elements are located at the edge of a polygon and interconnected with each other via an $L$-port fully-connected architecture to construct a cell, as illustrated in Figs. \ref{fig:mode_top}(b) and \ref{fig:architecture_mode}(b). Here we focus on the case where cells are isolated from each other for simplicity, while we can also have inter-cell interconnections as in the hybrid mode. Therefore, we have the scattering matrix $\mathbf{\Theta}$ blocked as 
\begin{equation}
	\mathbf{\Theta} = \left[\begin{matrix}
		\mathbf{\Theta}_{1,1} & \mathbf{\Theta}_{1,2} & \cdots & \mathbf{\Theta}_{1,L}\\
		\mathbf{\Theta}_{2,1} & \mathbf{\Theta}_{2,2} & \cdots & \mathbf{\Theta}_{2,L}\\
		\vdots & \vdots & \ddots & \vdots\\
		\mathbf{\Theta}_{L,1} & \mathbf{\Theta}_{L,2} & \cdots & \mathbf{\Theta}_{L,L}
		\end{matrix}\right],
\end{equation}
where $\mathbf{\Theta}_{i,j}\in\mathbb{C}^{\frac{M}{L}\times\frac{M}{L}}$ describes the power scattering from sector $j$ to sector $i$, $\forall i,j\in\mathcal{L}=\{1,\ldots,L\}$, and, given the lack of inter-connections between cells in this example, each has a diagonal structure:
\begin{equation}
	\mathbf{\Theta}_{i,j} = \mathsf{diag}(\varTheta_{i,j,1},\ldots,\varTheta_{i,j,\frac{M}{L}}), \forall i,j\in\mathcal{L},
\end{equation}
with $\varTheta_{i,j,n}\in\mathbb{C}$, $\forall n = 1, \ldots,\frac{M}{L}$. When there is only one source located within the coverage of sector 1, as illustrated in Fig. \ref{fig:mode_top}(b), we focus only on the first block-column of matrix $\mathbf{\Theta}$, i.e., $\mathbf{\Theta}_{l,1}$, $\forall l\in\mathcal{L}$, which describes the power scattering from sector 1 to all other sectors \cite{li2023beyond}.  
Then, when the reconfigurable impedance network is lossless, i.e., $\mathbf{\Theta}^\mathsf{H}\mathbf{\Theta} = \mathbf{I}_{M}$,  we have the constraint:
\begin{equation}
	\sum_{l\in\mathcal{L}}|\varTheta_{l,1,n}|^2 = 1, \forall n = 1, \ldots,\frac{M}{L}.\label{eq:multi_sector}
\end{equation}
A special case with $L=2$ corresponds to the STARS, STAR-RIS, or IOS \cite{zhang2022intelligent}.

Comparing the above two examples, we observe that the same circuit topologies can support various modes by modifying the element arrangements. Alternatively, the same mode can be realized by different architectures as long as elements within each cell are connected. For clarity, we have summarized the architectures, their matrix properties, the supported modes, circuit complexity (the required number of admittance components), and highlights in Table \ref{tab:architecture}.

\begin{table*}[]
	\caption{Matrix Properties, Supported Modes, Circuit Complexity, and Highlights of BD-RIS Architectures}
	\label{tab:architecture}
	\centering
	\begin{tabular}{|c|c|c|c|}
	\hline
	Architectures  & Admittance Matrix Property  & Supported Modes  & Circuit Complexity \\ 
	\hline\hline
	Single-Connected & Diagonal & Reflecting  & $M$ Admittance Components \\ 
	\hline
	Fully-Connected & Full and Symmetric & \multirow{8}{*}{\begin{tabular}[c]{@{}c@{}}Reflecting, \\ Hybrid,\\ Multi-Sector
	\end{tabular}} 
	& $\frac{M(M+1)}{2}$ Admittance Components \\ \cline{1-2} \cline{4-4} 
	Group-Connected & \begin{tabular}[c]{@{}c@{}}Block-Diagonal with \\ Each Block Full and Symmetric
	\end{tabular}   &  & $\frac{M(\bar{M}+1)}{2}$ Admittance Components \\ 
	\cline{1-2} \cline{4-4} 
	Tree-Connected & \begin{tabular}[c]{@{}c@{}}
	Tridiagonal/Arrowhead and Symmetric\end{tabular} &  & $2M-1$ Admittance Components  \\ \cline{1-2} \cline{4-4} 
	Forest-Connected  & \begin{tabular}[c]{@{}c@{}}Block-Diagonal with Each Block\\ Tridiagonal/Arrowhead and Symmetric\end{tabular} &  & $G(2\bar{M}-1)$ Admittance Components \\ \cline{1-2} \cline{4-4} 
	Band/Stem-Connected & \begin{tabular}[c]{@{}c@{}}
		Band/Stem and Symmetric\end{tabular} &  & $(2M-q)\frac{q+1}{2}$ Admittance Components  \\ \cline{1-2} \cline{4-4}
	Dynamically Connected & \begin{tabular}[c]{@{}c@{}}Permuted Block-Diagonal and Symmetric\end{tabular}  &  & \multicolumn{1}{l|}{\begin{tabular}[c]{@{}l@{}}$\frac{M(M+1)}{2}$ Admittances and $\frac{M(M-1)}{2}$ Switches\end{tabular}} \\ \cline{1-2} \cline{4-4} 
	Non-Reciprocal   & Asymmetric    &   & Depends on the Non-Reciprocal Circuit \\ 
	\hline\hline\hline
	\end{tabular}
	\begin{tabular}{|c|l|}
		Architectures  & Highlights \\ 
		\hline\hline
		Single-Connected & The simplest architecture with the least circuit complexity and least beam manipulation flexibility \\ 
		\hline
		Fully-Connected & The most complex architecture with the highest circuit complexity and highest beam manipulation flexibility\\
		\hline
		Group-Connected &  The flexibility-complexity trade-off between single-connected and fully-connected architectures\\ 
		\hline
		Tree-Connected &  The simplest architecture with the least circuit complexity to achieve the optimum in single-user MISO systems\;\;\;\\
		\hline
		Forest-Connected  & The flexibility-complexity trade-off between single-connected and tree-connected architectures \\
		\hline
		Bound/Stem-Connected & The simplest architecture with the least circuit complexity to achieve the optimum in multi-user MIMO systems \\
		\hline
		Dynamically Connected &  The circuit topology can be dynamically modified to adapt to channel state information (CSI)\\ 
		\hline
		Non-Reciprocal   & Can generate asymmetric beams for uplink and downlink to break the wireless channel reciprocity\\ \hline
		\end{tabular}
\end{table*}

\section{Signal Processing for BD-RIS}
\label{sec:signal_process}

In this section, we first focus on a simple SISO scenario to explain the representative tools for BD-RIS optimization. We then summarize the signal processing techniques, including optimization and performance analysis, and channel estimation, for more general BD-RIS-aided wireless communication systems. 

\subsection{Optimizing BD-RIS: A Simple SISO Case}
\label{subsec:opt_siso}

Consider a BD-RIS-aided SISO system, consisting of a single-antenna transmitter, a single-antenna receiver, and an $M$-element BD-RIS. 
Define the channels from transmitter to receiver, from the transmitter to BD-RIS, and from BD-RIS to receiver, respectively, as $h_{RT}\in\mathbb{C}$, $\mathbf{h}_{IT}\in\mathbb{C}^{M\times 1}$, and $\mathbf{h}_{RI}\in\mathbb{C}^{1\times M}$, which are perfectly known at the transmitter. According to \cite{universal2024}, assuming perfect matching, no mutual coupling, no specular reflection, and unilateral approximation at all devices, the overall channel is given by $h(\mathbf{\Theta}) = h_{RT} + \mathbf{h}_{RI}\mathbf{\Theta}\mathbf{h}_{IT}$. 

\subsubsection{Solutions for Unitary $\mathbf{\Theta}$} 
We start by considering the case that BD-RIS is characterized by a unitary matrix, i.e., $\mathbf{\Theta}^\mathsf{H}\mathbf{\Theta}=\mathbf{I}_M$. This corresponds to a lossless and non-reciprocal BD-RIS. Since $\mathbf{\Theta}$ can be easily rotated to align with the direct channel $h_{RT}$ for channel strength maximization, here we assume the direct link is blocked, i.e., $h_{RT} = 0$, for the ease of illustration. Then, we formulated the following problem:
\begin{equation}\label{eq:prob_unitary}
    \begin{aligned}
        \max_{\mathbf{\Theta}} ~ &|\mathbf{h}_{RI}\mathbf{\Theta}\mathbf{h}_{IT}|^2\\
        \mathrm{s.t.} ~&\mathbf{\Theta}^\mathsf{H}\mathbf{\Theta}=\mathbf{I}_M,
    \end{aligned} 
\end{equation}
which can be solved by the following three methods:
\begin{itemize}
    \item \textit{Closed-Form Solution:} 
        Consider the matrix $\mathbf{V}_{RI}\in\mathbb{C}^{M\times M}$ as a unitary matrix containing the right singular vectors of $\mathbf{h}_{RI}$ in its columns.
        In other words, the first column of $\mathbf{V}_{RI}$ is $[\mathbf{V}_{RI}]_{:,1}=\frac{\mathbf{h}_{RI}^\mathsf{H}}{\Vert\mathbf{h}_{RI}\Vert_2}$ and the other columns of $\mathbf{V}_{RI}$ are unit-norm vectors mutually orthogonal and orthogonal to $[\mathbf{V}_{RI}]_{:,1}$.
        Similarly, consider the matrix $\mathbf{U}_{IT}\in\mathbb{C}^{M\times M}$ as a unitary matrix containing the left singular vectors of $\mathbf{h}_{IT}$ in its columns.
        In other words, the first column of $\mathbf{U}_{IT}$ is $[\mathbf{U}_{IT}]_{:,1}=\frac{\mathbf{h}_{IT}}{\Vert\mathbf{h}_{IT}\Vert_2}$ and the other columns of $\mathbf{U}_{IT}$ are unit-norm vectors mutually orthogonal and orthogonal to $[\mathbf{U}_{IT}]_{:,1}$.
        Given $\mathbf{V}_{RI}$ and $\mathbf{U}_{IT}$, it is easy to see that a unitary matrix $\boldsymbol{\Theta}$ globally solving problem (\ref{eq:prob_unitary}) is
        \begin{equation}
        \boldsymbol{\Theta}=\mathbf{V}_{RI}\mathbf{U}_{IT}^\mathsf{H}.
        \end{equation}
        The computational complexity mainly comes from the matrix multiplication and is thus $\mathcal{O}(M^3)$.
    \item \textit{Orthogonal Rotation:} Following the idea in \cite{de2010optimized}, any unitary matrix $\mathbf{\Theta}$ can be expressed using the following parameterization 
    \begin{equation}
        \mathbf{\Theta} = \mathbf{\Theta}_0\prod_{m=1}^{M-1}\prod_{n=m+1}^{M}\mathbf{R}_{m,n},
    \end{equation}
    where $\mathbf{\Theta}_0\in\mathbb{C}^{M\times M}$ is an arbitrary unitary matrix and $\mathbf{R}_{m,n}\in\mathbb{C}^{M\times M}$ denotes a Givens rotation matrix which performs an orthogonal rotation of the $m$-th and $n$-th columns of a unitary matrix with others fixed. A Givens rotation matrix $\mathbf{R}_{m,n}$ is mathematically characterized by two rotation parameters, $\phi_{m,n}$ and $\psi_{m,n}$, and is constructed as 
    \begin{equation}
        [\mathbf{R}_{m,n}]_{i,j} = \begin{cases}
            1, &i = j, i\ne m,n,\\
            \cos\phi_{m,n}, &i=j, i = m,n,\\
            \sin\phi_{m,n}e^{\jmath\psi_{m,n}}, &i=m,j=n,\\
            -\sin\phi_{m,n}e^{-\jmath\psi_{m,n}}, &i=n,j=m,\\
            0, &\text{otherwise}.
        \end{cases}
    \end{equation}
    As such, the original problem is transformed into an unconstrained optimization containing $M(M-1)$ rotation parameters of the corresponding rotation matrices, which can be directly solved by some searching-based methods, e.g., quasi-Newton methods. This solution requires around $\mathcal{O}(I_1M^2(M-1)^2)$ complexity due to the use of quasi-Newton method to solve $M(M-1)$ variables, where $I_1$ denotes the number of iterations.
    \item \textit{Manifold:} It is worth noting that the constraint $\mathbf{\Theta}^\mathsf{H}\mathbf{\Theta}=\mathbf{I}_M$ essentially forms a Manifold \cite{lee2018introduction}. As such, a typical solution is to adopt the Manifold theory to construct an unconstrained optimization on the Manifold space, such that searching-based methods, such as Gradient-Descent, can be used on the Manifold space \cite{zhao2024channel}. This solution requires $\mathcal{O}(I_2M^3)$ complexity, where $I_2$ denotes the number of iterations.
\end{itemize}

\subsubsection{Solutions for Unitary and Symmetric $\mathbf{\Theta}$}
For the case of a reciprocal and lossless BD-RIS with fully-connected architecture, its scattering matrix is symmetric and unitary, that is $\mathbf{\Theta} = \mathbf{\Theta}^\mathsf{T}$ and $\mathbf{\Theta}^\mathsf{H}\mathbf{\Theta} = \mathbf{I}_M$. In this case, the channel strength maximization problem becomes 
\begin{equation}\label{eq:prob_unitary_symm}
    \begin{aligned}
        \max_{\mathbf{\Theta}} ~ &|\mathbf{h}_{RI}\mathbf{\Theta}\mathbf{h}_{IT}|^2\\
        \mathrm{s.t.} ~&\mathbf{\Theta} = \mathbf{\Theta}^\mathsf{T}, \mathbf{\Theta}^\mathsf{H}\mathbf{\Theta}=\mathbf{I}_M.
    \end{aligned} 
\end{equation}
One common idea to simplify the optimization is to decouple the unitary and symmetric constraint of $\mathbf{\Theta}$. This can be done using the following three strategies.
\begin{itemize}
    \item \textit{Matrix Decomposition:} One way to decouple the constraints of $\mathbf{\Theta}$ is to decompose $\mathbf{\Theta}$ as
    $\mathbf{\Theta}=\mathbf{U}\mathbf{D}\mathbf{U}^\mathsf{T}$ \cite{nerini2023closed,sun2024new}, where $\mathbf{U}\in\mathbb{R}^{M\times M}$ is orthonormal, that is $\mathbf{U}^\mathsf{T}\mathbf{U} = \mathbf{I}_M$
    and $\mathbf{D}\in\mathbb{C}^{M\times M}$ is a diagonal matrix whose diagonal entries have unit modulus, that is $\mathbf{D} = \mathsf{diag}(D_1,\ldots,D_M)$, $|D_m| = 1$, $\forall m\in\mathcal{M}$. Then the design of $\mathbf{\Theta}$ is transformed to the design of $\mathbf{D}$ and $\mathbf{U}$:
    \begin{equation}
        \begin{aligned}
            &\mathcal{T}' = \{\{\mathbf{U},\mathbf{D}\}~|~
            \mathbf{U}\in\mathbb{R}^{M\times M}, \mathbf{U}^\mathsf{T}\mathbf{U} = \mathbf{I}_M,\\
            &~~~\mathbf{D} = \mathsf{diag}(D_1,\ldots,D_M), |D_m| = 1, \forall m\},
        \end{aligned}
    \end{equation}
    both having closed-form solutions as detailed in \cite{nerini2023closed,sun2024new}.
    Alternatively, $\mathbf{\Theta}$ can be decomposed as $\mathbf{\Theta} = \mathbf{\Psi}\mathbf{\Psi}^\mathsf{T}$ \cite{santamaria2023snr}. As such, the design of a unitary and symmetric matrix $\mathbf{\Theta}$ is equivalent to that of a unitary matrix $\mathbf{\Psi}\in\mathbb{C}^{M\times M}$, that is
    \begin{equation}
        \mathcal{T}'' = \{\mathbf{\Psi}~|~ \mathbf{\Psi}^\mathsf{H}\mathbf{\Psi} = \mathbf{I}_M\},
    \end{equation}
    which can be obtained in closed form based on the knowledge of channels as detailed in \cite{santamaria2023snr}. The matrix decomposition requires around $\mathcal{O}(M^3)$ due to the use of eigenvalue and Takagi's decomposition.
    \item \textit{Projection:} One can first relax the two constraints of $\mathbf{\Theta}$ and find a solution for a more tractable problem, such as the following problem:
    \begin{equation}\label{eq:prob_relaxed}
        \begin{aligned}
            \max_{\mathbf{\Theta}} ~ &|\mathbf{h}_{RI}\mathbf{\Theta}\mathbf{h}_{IT}|^2\\
            \mathrm{s.t.} ~&\|\mathbf{\Theta}\|_\mathsf{F}^2 \le M,
        \end{aligned} 
    \end{equation}
    and then applies the symmetric unitary projection to satisfy the original constraints \cite{fang2023low}. That is, with $\mathbf{\Theta}_\mathsf{opt}$ as the solution to (\ref{eq:prob_relaxed}) as detailed in \cite{fang2023low}, its symmetric projection is given by 
    \begin{equation}
        \mathbf{\Theta}_\mathsf{sym} = \frac{1}{2}(\mathbf{\Theta}_\mathsf{opt}+\mathbf{\Theta}_\mathsf{opt}^\mathsf{T}).
    \end{equation}
    Then, $\mathbf{\Theta}_\mathsf{sym}$ is further projected to the unitary domain by solving an orthogonal Procrustes problem \cite{manton2002optimization}, i.e., 
    \begin{equation}
        \mathbf{\Theta}_{\mathsf{symuni}} = \arg\min_{\mathbf{\Theta}^\mathsf{H}\mathbf{\Theta}=\mathbf{I}_M}~\|\mathbf{\Theta}-\mathbf{\Theta}_\mathsf{sym}\|_\mathsf{F}^2 = \mathbf{U}_1\mathbf{U}_2^\mathsf{H},
    \end{equation}
    where $\mathbf{U}_1\in\mathbb{C}^{M\times M}$ and $\mathbf{U}_2\in\mathbb{C}^{M\times M}$ are unitary matrices from the singular value decomposition (SVD) of $\mathbf{\Theta}_\mathsf{sym}$, i.e., $\mathbf{\Theta}_\mathsf{sym} = \mathbf{U}_1\mathbf{\Sigma}\mathbf{U}_2^\mathsf{H}$ with $\mathbf{\Sigma}$ being a diagonal matrix containing the singular values of $\mathbf{\Theta}_\mathsf{sym}$ in a decreasing order. The projection requires around $\mathcal{O}(M^3)$ due to the use of SVD.
    \item \textit{Introducing Auxiliary Variables:} Another direct way to decouple the two constraints of $\mathbf{\Theta}$ is to introduce an auxiliary variable $\mathbf{\Phi} = \mathbf{\Theta}$ \cite{zhou2023optimizing}, such that $\mathbf{\Theta}$ is only subject to the unitary constraint and $\mathbf{\Phi}$ is only subject to the symmetric constraint, or vise versa. In this way, the equality constraint $\mathbf{\Phi} = \mathbf{\Theta}$ can be penalized to construct the associated Lagrangian function and problem (\ref{eq:prob_unitary_symm}) can be transformed into a double-variable optimization
    \begin{equation}
        \begin{aligned} \max_{\mathbf{\Theta},\mathbf{\Phi}}~&|\mathbf{h}_{RI}\mathbf{\Theta}\mathbf{h}_{IT}|^2 - \frac{\rho_1}{2}\|\mathbf{\Phi} - \mathbf{\Theta}\|_\mathsf{F}^2\\
        &- \Re\{\mathsf{tr}(\mathbf{\Lambda}^\mathsf{H}(\mathbf{\Phi} - \mathbf{\Theta}))\}\\
        \mathrm{s.t.} ~&\mathbf{\Phi}^\mathsf{H}\mathbf{\Phi} = \mathbf{I}_M, \mathbf{\Theta} = \mathbf{\Theta}^\mathsf{T}, 
        \end{aligned} 
    \end{equation}
    where $\rho_1>0$ is the penalty parameter and $\mathbf{\Lambda}\in\mathbb{C}^{M\times M}$ is the dual variable. Then, the two variables can be alternatively designed. Specifically, the sub-problem for $\mathbf{\Theta}$ with given $\mathbf{\Phi}$ is essentially an unconstrained optimization when focusing only on the diagonal and upper-triangular (or equivalently lower-triangular) entries of $\mathbf{\Theta}$. As such, some well-known searching-based methods for unconstrained optimization, such as Gradient-Descent, can be directly adopted. The sub-problem for $\mathbf{\Phi}$ with given $\mathbf{\Theta}$ is again an orthogonal Procrustes problem \cite{manton2002optimization}, whose closed-form solution can be obtained by performing SVD to $\rho_1\mathbf{\Theta}+\mathbf{\Lambda}$. The overall complexity of such a solution depends on the specific methods used to solve the sub-problem for a symmetric $\mathbf{\Theta}$.
\end{itemize}

\textit{Remark 7:} The matrix decomposition $\mathbf{\Theta}=\mathbf{U}\mathbf{D}\mathbf{U}^\mathsf{T}$ \cite{nerini2023closed,sun2024new} provides interesting insights to understand the relationship between BD-RIS and D-RIS. That is, a BD-RIS can be decoupled as a power divider controlled by $\mathbf{U}$ and a phase shifter network characterized by $\mathbf{D}$. In D-RIS, we have $\mathbf{U}=\mathbf{I}_M$, indicating that signals impinging on one element can only be reflected by the same one. However, in BD-RIS, a flexible power divider is enabled by inter-element connections such that the waves impinging on one element can (partially) flow to other elements and further be reflected.

\subsubsection{Solutions for $\mathbf{Y}_I$}
As discussed in Remark 4, for BD-RIS with tree/forest/stem/band-connected architectures, directly designing the scattering matrix $\mathbf{\Theta}$ cannot fully reflect their mathematical constraints. In this case, we should design the admittance matrix $\mathbf{Y}_I$ based on the following problem:
\begin{equation}\label{eq:prob_adm}
    \begin{aligned} 
        \max_{\mathbf{Y}_I}~~ &|\mathbf{h}_{RI}\mathbf{\Theta}\mathbf{h}_{IT}|^2\\
        \mathrm{s.t.} ~~&\mathbf{\Theta}=(Y_0\mathbf{I}_M+\mathbf{Y}_I)^{-1}(Y_0\mathbf{I}_M -\mathbf{Y}_I),\\
        &\mathbf{Y}_I\in\mathcal{Y},
    \end{aligned}
\end{equation}
where $\mathcal{Y}$ denotes the constraint of $\mathbf{Y}_I$ varying according to architectures illustrated in Section \ref{subsec:architecture}. The main difficulty in optimizing BD-RIS here lies in the matrix inverse due to the nonlinear mapping between $\mathbf{\Theta}$ and $\mathbf{Y}_I$. To tackle this difficulty, there are in general three strategies. 
\begin{itemize}
    \item \textit{Closed-Form Solutions:} As illustrated in Section \ref{sec:example}, the channel strength in (\ref{eq:prob_adm}) is upper-bounded by $|\mathbf{h}_{RI}\mathbf{\Theta}\mathbf{h}_{IT}|^2 \le \|\mathbf{h}_{RI}\|_2^2\|\mathbf{h}_{IT}\|_2^2$,
    where the equality is achieved when 
    \begin{equation}
    \bar{\mathbf{h}}_{RI}^\mathsf{H} = \mathbf{\Theta}\bar{\mathbf{h}}_{IT},\label{eq:opt_cond}
    \end{equation}
    with $\bar{\mathbf{h}}_{RI} = \frac{\mathbf{h}_{RI}}{\|\mathbf{h}_{RI}\|_2}$ and $\bar{\mathbf{h}}_{IT} = \frac{\mathbf{h}_{IT}}{\|\mathbf{h}_{IT}\|_2}$. 
    By (\ref{eq:theta_zy}), (\ref{eq:opt_cond}) can be rewritten as 
    \begin{equation}
        \begin{aligned}
        &(Y_0\mathbf{I}_M + \mathbf{Y}_I)\bar{\mathbf{h}}_{RI}^\mathsf{H} = (Y_0\mathbf{I}_M - \mathbf{Y}_I)\bar{\mathbf{h}}_{IT}\\
        \Rightarrow&\mathbf{Y}_I(\bar{\mathbf{h}}_{RI}^\mathsf{H}+\bar{\mathbf{h}}_{IT}) = Y_0(\bar{\mathbf{h}}_{IT} - \bar{\mathbf{h}}_{RI}^\mathsf{H}).
        \end{aligned}\label{eq:opt_cond1}
    \end{equation}
    It turns out that it is sufficient to solve linear equations in (\ref{eq:opt_cond1}) as the global optimal solution of (\ref{eq:prob_adm}). Note that with a given architecture, one can always focus only on the nonzero entries in $\mathbf{Y}_I$ and solve (\ref{eq:opt_cond1}). Such a closed-form solution for tree/forest-connected architectures has been derived in \cite{nerini2024beyond}, which requires $\mathcal{O}(M^3)$ complexity due to the matrix inversion.
    \item \textit{Searching-Based Methods:} The optimization problem (\ref{eq:prob_adm}) is essentially unconstrained by focusing only on the upper-triangular (or equivalently lower-triangular) and diagonal entries of $\mathbf{Y}_I$. This can be directly solved by some searching-based methods, such as the quasi-Newton method \cite{shen2021}, which requires $\mathcal{O}(I_3\frac{M^2(M+1)^2}{4})$ to solve $\frac{M(M+1)}{2}$ variables, where $I_3$ denotes the number of iterations.
    \item \textit{Introducing Auxiliary Variables:} One can introduce $\mathbf{u} = \mathbf{h}_{RI}\mathbf{\Theta}$, such that the matrix inverse can be eliminated by transferring it to a bilinear constraint \cite{wu2024optimization}:
\begin{equation}
    \begin{aligned}
        &\mathbf{\Theta}=(Y_0\mathbf{I}_M+\mathbf{Y}_I)^{-1}(Y_0\mathbf{I}_M-\mathbf{Y}_I)\\
        \Rightarrow & \mathbf{u}(Y_0\mathbf{I}_M+\mathbf{Y}_I) = \mathbf{h}_{RI}(Y_0\mathbf{I}_M-\mathbf{Y}_I).
    \end{aligned}
\end{equation}
Again, the equality constraint $\mathbf{u}(Y_0\mathbf{I}_M+\mathbf{Y}_I) = \mathbf{h}_{RI}(Y_0\mathbf{I}_M-\mathbf{Y}_I)$ can be penalized to construct the associated Lagrangian function and problem (\ref{eq:prob_adm}) can be transformed into the following form
\begin{equation}
    \begin{aligned}    \max_{\mathbf{u},\mathbf{Y}_I}~&|\mathbf{u}\mathbf{h}_{IT}|^2
        - \frac{\rho_1}{2}\|\mathbf{u}(Y_0\mathbf{I}_M+\mathbf{Y}_I)\\ 
        &- \mathbf{h}_{RI}(Y_0\mathbf{I}_M-\mathbf{Y}_I)\|_2^2
        - \Re\{((\mathbf{u}(Y_0\mathbf{I}_M+\mathbf{Y}_I)\\
        &- \mathbf{h}_{RI}(Y_0\mathbf{I}_M-\mathbf{Y}_I))\bm{\lambda}^\mathsf{H}\}\\
        \mathrm{s.t.} ~&\mathbf{Y}_I\in\mathcal{Y},
        \end{aligned} 
    \end{equation}
    where $\rho_2>0$ denotes the penalty parameter and $\bm{\lambda}\in\mathbb{C}^{1\times M}$ denotes the dual variable. Then the two variables can be alternatively optimized, where the sub-problem for $\mathbf{u}$ is unconstrained and the sub-problem for $\mathbf{Y}_I$ is also unconstrained when focusing on optimizing only the non-zero entries in $\mathbf{Y}_I$. The overall complexity of this solution depends on the specific methods used to solve two unconstrained sub-problems.
\end{itemize}

Based on the above explanations, there are some representative solutions using existing optimization techniques to deal with different BD-RIS constraints.
The mapping between constraints of BD-RIS and their representative solutions is summarized in Table \ref{tab:opt}.  

\begin{table*}[]
	\caption{BD-RIS Architectures and Constraints, Difficulties, and Solutions}
	\label{tab:opt}
	\centering
	\begin{threeparttable}
		\begin{tabular}{|c|c|l|c|c|}
			\hline
			Architectures$\dagger$ and Constraints & Difficulties & Solutions$\ddagger$ & Optimality & Complexity$\natural$\\
			\hline\hline
			\multirow{12}{*}{\begin{tabular}[c]{@{}c@{}} Non-reciprocal architecture\\ with $\mathbf{\Theta}^\mathsf{H}\mathbf{\Theta}= \mathbf{I}_M$\end{tabular}} & \multirow{12}{*}{\begin{tabular}[c]{@{}c@{}} Unitary \\constraint \end{tabular}} & \begin{tabular}[l]{@{}l@{}} \tabitem Construct $\mathbf{\Theta}$ by two unitary matrices based \\ \;\;\;\; on $\mathbf{h}_{RI}$ and $\mathbf{h}_{IT}$, i.e., $\mathbf{\Theta}=\mathbf{V}_{RI}\mathbf{U}_{IT}^\mathsf{H}$  \\ \;\;\;\; with $[\mathbf{V}_{RI}]_{:,1} = \frac{\mathbf{h}_{RI}^\mathsf{H}}{\|\mathbf{h}_{IT}\|_2} = \bar{\mathbf{h}}_{RI}^\mathsf{H}$ and \\ \;\;\;\; $[\mathbf{U}_{IT}]_{:,1} = \frac{\mathbf{h}_{IT}}{\|\mathbf{h}_{IT}\|_2}=\bar{\mathbf{h}}_{IT}$. ($\circledast $) \end{tabular} & Yes & $\mathcal{O}(M^3)$\\
            \cline{3-5}
            & & \begin{tabular}[l]{@{}l@{}} \tabitem Decouple $\mathbf{\Theta} = \mathbf{\Theta}_0\prod_{m=1}^{M-1}\prod_{n=m+1}^{M}\mathbf{R}_{m,n}$, \\ \;\;\;\; transform to an unconstrained optimization \\ \;\;\;\; and use searching-based methods,\\ \;\;\;\; e.g., quasi-Newton method \cite{de2010optimized}. \end{tabular} & No & $\mathcal{O}(I_1M^2(M-1)^2)$\\
            \cline{3-5}
			& & \begin{tabular}[l]{@{}l@{}} \tabitem Adopt the Manifold theory to construct a  \\ \;\;\;\;  Manifold and transform the constrained  \\ \;\;\;\;  optimization to unconstrained optimization \\ \;\;\;\;  on Manifold space, on which the searching\\ \;\;\;\; -based methods, e.g., Gradient-Descent, \\ \;\;\;\; can be used \cite{li2022beyond,zhao2024channel}. \end{tabular} & No & $\mathcal{O}(I_2M^3)$\\
            \hline
			\multirow{10}{*}{\begin{tabular}[c]{@{}c@{}}Fully/group-connected architectures\\ with $\mathbf{\Theta}^\mathsf{H}\mathbf{\Theta}=\mathbf{I}_M$ and $\mathbf{\Theta}=\mathbf{\Theta}^\mathsf{T}$\end{tabular}} & \multirow{10}{*}{\begin{tabular}[l]{@{}l@{}}Coupled unitary\\ and  symmetric \\constraints\end{tabular}} & \begin{tabular}[l]{@{}l@{}} \tabitem Matrix decomposition $\mathbf{\Theta} = \mathbf{U}\mathbf{D}\mathbf{U}^\mathsf{T}$ with \\ \;\;\;\;$\mathcal{T}'$  \cite{nerini2023closed,sun2024new} or $\mathbf{\Theta} = \mathbf{\Psi}\mathbf{\Psi}^\mathsf{T}$ with $\mathcal{T}''$  \cite{santamaria2023snr};\end{tabular} & Yes & $\mathcal{O}(M^3)$\\
            \cline{3-5}
			& & \begin{tabular}[l]{@{}l@{}} \tabitem  Symmetric and unitary projections by \\ \;\;\;\; $\mathbf{\Theta}_{\mathsf{sym}}=\frac{1}{2}(\mathbf{\Theta}_{\mathsf{opt}}+\mathbf{\Theta}_{\mathsf{opt}}^\mathsf{T})$ and solving\\
			\;\;\;\; the orthogonal Procrustes problem \\ \;\;\;\; $\mathbf{\Theta} = \arg\min_{\mathbf{\Theta}^\mathsf{H}\mathbf{\Theta}=\mathbf{I}_M}~\|\mathbf{\Theta}-\mathbf{\Theta}_\mathsf{sym}\|_\mathsf{F}^2$ \cite{fang2023low};\end{tabular} & No & $\mathcal{O}(M^3)$\\
            \cline{3-5}
			& & \begin{tabular}[l]{@{}l@{}} \tabitem Introduce auxiliary variables $\mathbf{\Phi} = \mathbf{\Theta}$, \\ \;\;\;\; with $\mathbf{\Phi} =\mathbf{\Phi}^\mathsf{T}$ and $\mathbf{\Theta}^\mathsf{H}\mathbf{\Theta} = \mathbf{I}_M$,\\
				\;\;\;\; or vise versa \cite{zhou2023optimizing}. \end{tabular} & No & $\circleddash$\\
			\hline
			\multirow{6}{*}{\begin{tabular}[c]{@{}c@{}}Arbitrary reciprocal architectures\\ with $\mathbf{\Theta}=(\mathbf{Y}_0\mathbf{I}_M + \mathbf{Y}_I)^{-1}$\\ $\times(\mathbf{Y}_0\mathbf{I}_M - \mathbf{Y}_I)$ and $\mathbf{Y}_I\in\mathcal{Y}$\end{tabular}} & \multirow{6}{*}{\begin{tabular}[c]{@{}c@{}}Matrix inverse \\from $\mathbf{\Theta}$ to $\mathbf{Y}_I$\end{tabular}} & \begin{tabular}[l]{@{}l@{}} \tabitem Extract non-zero entries in $\mathbf{Y}_I$ and globally  \\ \;\;\;\; solve linear equations\\ \;\;\;\; $\mathbf{Y}_I(\bar{\mathbf{h}}_{RI}^\mathsf{H}+\bar{\mathbf{h}}_{IT}) = Y_0(\bar{\mathbf{h}}_{IT} - \bar{\mathbf{h}}_{RI}^\mathsf{H})$. ($\circledast $)\end{tabular} & Yes & $\mathcal{O}(M^3)$\\
            \cline{3-5}
            & & \begin{tabular}[l]{@{}l@{}} \tabitem Searching-based methods, e.g., quasi-Newton \\ \;\;\;\; method \cite{shen2021} not sensitive to matrix structures. \end{tabular} & No & $\mathcal{O}\big(I_3\frac{M^2(M+1)^2}{4}\big)$ \\
            \cline{3-5}
			& & \begin{tabular}[l]{@{}l@{}} \tabitem Introduce auxiliary variables, e.g., $\mathbf{u}=\mathbf{h}_{RI}\mathbf{\Theta}$, \\ \;\;\;\;  to eliminate the matrix inverse by \\ \;\;\;\; $\mathbf{\Theta}=(Y_0\mathbf{I}_M+\mathbf{Y}_I)^{-1}(Y_0\mathbf{I}_M-\mathbf{Y}_I)
            \Rightarrow$\\ \;\;\;\; $\mathbf{u}(Y_0\mathbf{I}_M+\mathbf{Y}_I) = \mathbf{h}_{RI}(Y_0\mathbf{I}_M-\mathbf{Y}_I)$ \cite{wu2024optimization}. \end{tabular} & No & $\circleddash$\\
			\hline
		\end{tabular}
		\begin{tablenotes}
			\footnotesize
			\item[$\dagger$] The architectures are all reciprocal unless otherwise stated.
			\item[$\ddagger$] The solutions that end up with ($\circledast$) can only be used in SISO systems, while others are applicable to multi-antenna/user scenarios.
			\item[$\natural$] The notation $I_i,\forall i\in\{1,2,3\}$ denotes the number of iterations to guarantee the convergence of different methods.
			\item[$\natural$] The complexity of solutions that marked as $\circleddash$ is determined by specific methods used to solve unconstrained optimization problems. 
		\end{tablenotes}
	\end{threeparttable}
\end{table*}

\subsection{Survey on BD-RIS Optimization and Performance Analysis}
\label{subsec:beamforming_r}

The representative solutions for BD-RIS optimization in SISO systems can be flexibly used to solve some extracted sub-problems in more general scenarios, such as MIMO and multi-user systems. In this subsection, we will provide a comprehensive survey on BD-RIS optimization and performance analysis in various scenarios. 

Consider a general BD-RIS-aided multi-user MIMO system consisting of an $N$-antenna transmitter, an $M$-element BD-RIS, and $K$ multi-antenna users, each of which has $N_k$ antennas, $\forall k\in\mathcal{K}=\{1,\ldots,K\}$. Define the wireless channels from the transmitter to user $k$ as $\mathbf{H}_{{RT},k}\in\mathbb{C}^{N_k\times N}$, from the transmitter to BD-RIS as $\mathbf{H}_{{IT}}\in\mathbb{C}^{M\times N}$, and from the BD-RIS to user $k$ as $\mathbf{H}_{{RI},k}\in\mathbb{C}^{N_k\times M}$. According to \cite{universal2024}, assuming perfect matching, no mutual coupling, no specular reflection, and unilateral approximation at all devices, the overall channel $\mathbf{H}_k(\mathbf{\Theta})\in\mathbb{C}^{N_k\times N}$ from the transmitter to user $k$, which is a function of the scattering matrix $\mathbf{\Theta}$ of BD-RIS, writes as 
\begin{equation}
	\mathbf{H}_k(\mathbf{\Theta}) = \mathbf{H}_{{RT},k} + \mathbf{H}_{{RI},k}\mathbf{\Theta}\mathbf{H}_{{IT}}, \forall k\in\mathcal{K}.
\end{equation}
Define a precoder matrix $\mathbf{W} = [\mathbf{W}_1,\mathbf{W}_2,\ldots,\mathbf{W}_K]\in\mathbb{C}^{N\times (\sum_{k\in\mathcal{K}}N_{\mathsf{s},k})}$, where $\mathbf{W}_k\in\mathbb{C}^{N\times N_{\mathsf{s},k}}$ denotes the precoder matrix for user $k$ and $N_{\mathsf{s},k}$ denotes the number of data streams to be transmitted to user $k$. 
Assuming perfect CSI is known at the transmitter, the joint transmit precoder and BD-RIS design problem can be formulated as the following admittance matrix based form
\begin{equation}\label{prob:admittance}
	\begin{aligned} 
		\max_{\mathbf{W},\mathbf{\Theta},\mathbf{Y}_I} &F(\mathbf{W},\{\mathbf{H}_k(\mathbf{\Theta})\}_{\forall k})\\
		\mathrm{s.t.} ~~&\mathbf{\Theta}=(Y_0\mathbf{I}_M+\mathbf{Y}_I)^{-1}(Y_0\mathbf{I}_M -\mathbf{Y}_I),\\
		&\mathbf{Y}_I\in\mathcal{Y},\\
		&\mathbf{W}\in\mathcal{W},
	\end{aligned}
\end{equation}
where $F(\mathbf{W},\{\mathbf{H}_k(\mathbf{\Theta})\}_{\forall k})$ denotes a general utility function varying according to performance metrics, $\mathcal{Y}$ denotes the constraint of $\mathbf{Y}_I$ varying according to architectures, and $\mathcal{W}$ denotes the constraint of the precoder matrix varying according to budget requirements. Note that (\ref{prob:admittance}) can be used to optimize all reciprocal architectures in Section \ref{subsec:architecture} where $\mathbf{Y}_I$ is explicitly defined and constrained.
Alternatively, the optimization problem can be formulated as the following scattering matrix based form
\begin{equation}\label{prob:scattering}
	\begin{aligned} 
		\max_{\mathbf{W},\mathbf{\Theta}}~~ &F(\mathbf{W},\{\mathbf{H}_k(\mathbf{\Theta})\}_{\forall k})\\
		\mathrm{s.t.} ~~&\mathbf{\Theta}\in\mathcal{T},\\
		&\mathbf{W}\in\mathcal{W},
	\end{aligned}
\end{equation}
where $\mathcal{T}$ denotes the constraints of $\mathbf{\Theta}$ and varying according to architectures. Note that (\ref{prob:scattering}) can be used to optimize single-, group-, fully-connected architectures in Section \ref{subsec:architecture} and non-reciprocal architectures in Section \ref{subsec:architecture_nr}.

The expressions of utility functions and the constraints of variables adapting to various wireless scenarios and corresponding beamforming solutions to tackle the above difficulties are specified below.

\subsubsection{Received Power Maximization for Multiple Input Single Output (MISO) and Single Input Multiple Output (SIMO)}
In MISO systems, we have $N_k=K=N_{\mathsf{s},k}=1$, and the utility function becomes 
\begin{equation}
	F_\mathsf{MISO}(\mathbf{w},\mathbf{h}(\mathbf{\Theta})) = |\mathbf{h}(\mathbf{\Theta})\mathbf{w}|^2,
\end{equation}
where $\mathbf{h}(\mathbf{\Theta})=\mathbf{h}_{RT} + \mathbf{h}_{RI}\mathbf{\Theta}\mathbf{H}_{IT}$ with $\mathbf{h}_{RT}\in\mathbb{C}^{1\times N}$ and $\mathbf{w}\in\mathbb{C}^{N\times 1}$ denotes the precoder vector constrained by the set
\begin{equation}
	\mathcal{W}_\mathsf{MISO} = \{\mathbf{w}~|~\|\mathbf{w}\|_2^2\le P\},
\end{equation}
where $P$ denotes the power budget at the transmitter. In this case, the optimization problem has multiple variables and the typical solution is to iteratively design the BD-RIS and the transmit precoder until the convergence of the objective function is guaranteed. Specifically, when the BD-RIS is given, the optimal transmit precoder takes the form of the maximum ratio transmission scheme. When the transmit precoder is given, the design of BD-RIS boils down to the SISO case such that the solutions for various fixed reciprocal and non-reciprocal architectures summarized in Section \ref{subsec:opt_siso} can be directly used. 
In the sequel, we elaborate the solutions for some dynamic BD-RIS architectures. 

For reciprocal BD-RIS, with the form of (\ref{prob:scattering}) and the constraint of dynamically connected architectures, \cite{nerini2024static} has proposed an offline grouping strategy design adapting to the static CSI. 
For non-reciprocal BD-RIS, the following beamforming design and performance analysis studies have been conducted.
With the form of (\ref{prob:scattering}) and a non-reciprocal BD-RIS architecture having non-diagonal scattering matrices being constrained by the set 
\begin{equation}\label{eq:constraint_nd}
    \begin{aligned}
    &\mathcal{T}_{\mathsf{non-diag}} = \{\mathbf{\Theta}~|~\mathbf{\Theta} = \mathbf{\Gamma}_\mathsf{r}\bar{\mathbf{\Theta}}\mathbf{\Gamma}_\mathsf{t}, \\
    &~~~~\bar{\mathbf{\Theta}} = \mathsf{diag}(e^{\jmath\theta_1},\ldots,e^{\jmath\theta_M}), \theta_m\in[0,2\pi), \forall m\},
    \end{aligned}
\end{equation}
closed-form solutions can be derived. Focusing primarily on SISO systems with the utility function $F_{\mathsf{SISO}}(h(\mathbf{\Theta}))$, \cite{li2022reconfigurable} has derived the performance upper-bound and the closed-form solution for BD-RIS. The beamforming design is later extended to MISO systems with the utility function $F_\mathsf{MISO}(\mathbf{w},\mathbf{h}(\mathbf{\Theta}))$. 
With the form of (\ref{prob:scattering}), \cite{li2024coordinated} has further proposed a coordinated non-reciprocal group-connected architecture with scattering matrices having more than $M$ non-zero entries, and derived the optimal grouping strategy which could maximize $F_\mathsf{MISO}(\mathbf{w},\mathbf{h}(\mathbf{\Theta}))$. This architecture is further used to support the multi-sector mode and maximize $F_\mathsf{MISO}(\mathbf{w},\mathbf{h}(\mathbf{\Theta}))$ related to specific users \cite{dong2024reconfigurable}.

From the problem formulation and optimization perspectives, the single input multiple output (SIMO) system is equivalent to the MISO system. Therefore, the aforementioned solutions for SISO systems in Section \ref{subsec:opt_siso} and MISO systems are readily applicable for SIMO systems.

\subsubsection{Capacity Maximization for MIMO} In this case, we have $K = 1$ and $N_k = N_\mathsf{r}$. The utility function becomes 
\begin{equation}
	F_\mathsf{MIMO}(\mathbf{Q},\mathbf{H}(\mathbf{\Theta})) = \log_2\det\left(\mathbf{I}_{N_\mathsf{r}} + \frac{1}{\sigma^2}\mathbf{H}(\mathbf{\Theta})\mathbf{Q}\mathbf{H}^\mathsf{H}(\mathbf{\Theta})\right),
\end{equation} 
where $\mathbf{H}(\mathbf{\Theta}) = \mathbf{H}_{RT} + \mathbf{H}_{RI}\mathbf{\Theta}\mathbf{H}_{IT}$ with $\mathbf{H}_{RT}\in\mathbb{C}^{N_\mathsf{r}\times N}$ and $\mathbf{H}_{RI}\in\mathbb{C}^{N_\mathsf{r}\times M}$, $\sigma^2$ denotes the noise power. We also replace the precoder with the transmitting covariance matrix $\mathbf{Q}\in\mathbb{C}^{N\times N}$, which satisfies
\begin{equation}
	\mathbf{Q}\in\mathcal{Q} = \{\mathbf{Q}~|~\mathsf{tr}(\mathbf{Q})\le P, \mathbf{Q}\succeq \mathbf{0}_{N\times N}\}.
\end{equation} 

With the form of (\ref{prob:scattering}), the following beamforming design studies have been conducted. 
Focusing on reciprocal and fully-connected architectures constrained by $\mathcal{T}_{\mathsf{group-conn}}$ ($G=1$)
    \begin{equation}
        \begin{aligned}
            &\mathcal{T}_\mathsf{group-conn} = \{\mathbf{\Theta} = \mathsf{blkdiag}(\mathbf{\Theta}_{1},\ldots,\mathbf{\Theta}_{G})~|\\
            &~~~~~~\mathbf{\Theta}_g = \mathbf{\Theta}_g^\mathsf{T},\mathbf{\Theta}_g^\mathsf{H}\mathbf{\Theta}_g = \mathbf{I}_{\bar{M}}, \forall g\in\mathcal{G}\},
        \end{aligned}
    \end{equation}
    a common strategy is to iteratively design BD-RIS and the covariance matrix \cite{santamaria2024mimo}. Specifically, the BD-RIS design can generally follow the idea of decoupling $\mathbf{\Theta}$ into $\mathbf{\Theta} = \mathbf{\Psi}^\mathsf{H}\mathbf{\Psi}$ with $\mathbf{\Psi}$ being unitary. For more specific LoS MIMO channels (both $\mathbf{H}_{RI}$ and $\mathbf{H}_{IT}$ have rank 1), the closed-form solution of $\mathbf{\Theta}$ can be obtained by exploiting the rank-1 property of channels $\mathbf{H}_{RI}$ and $\mathbf{H}_{IT}$ \cite{santamaria2025rate}. Focusing on non-reciprocal BD-RIS architecture constrained by the set
	\begin{equation}
		\mathcal{T}_{\mathsf{non-recip}} = \{\mathbf{\Theta}~|~\mathbf{\Theta}^\mathsf{H}\mathbf{\Theta} = \mathbf{I}_M\},\label{eq:constraint_nr}
	\end{equation}
	the problem has been solved by the Manifold method \cite{zhao2024channel}. Matrix decomposition based solutions have also been proposed in closed form considering either near-field \cite{bartoli2023spatial} or far-field \cite{bjornson2024capacity} scenarios, which are proved to be globally optimal.

\subsubsection{Sum-Rate Maximization for Multi-User MISO}
In multi-user MISO systems, we have $N_k = N_{\mathsf{s},k}=1$, $\forall k\in\mathcal{K}$. 
The utility function for sum-rate maximization writes as
\begin{equation}
	\begin{aligned}
		&F_\mathsf{MU-MISO}^\mathsf{sum-rate}(\mathbf{W},\{\mathbf{h}_k(\mathbf{\Theta})\}_{\forall k})\\
		 &~~~~~~~= \sum_{k\in\mathcal{K}}\log_2\left(1+\gamma_k(\mathbf{W},\mathbf{h}_k(\mathbf{\Theta}))\right),\\
		&\gamma_k(\mathbf{W},\mathbf{h}_k(\mathbf{\Theta})) = \frac{|\mathbf{h}_k(\mathbf{\Theta})\mathbf{w}_k|^2}{\sum_{i\ne k}|\mathbf{h}_k(\mathbf{\Theta})\mathbf{w}_i|^2 + \sigma^2}, \forall k\in\mathcal{K},
	\end{aligned}
\end{equation}
where $\mathbf{h}_k(\mathbf{\Theta})=\mathbf{h}_{RT,k} + \mathbf{h}_{RI,k}\mathbf{\Theta}\mathbf{H}_{IT}$ with $\mathbf{h}_{RT,k}\in\mathbb{C}^{1\times N}$ and $\mathbf{h}_{RI,k}\in\mathbb{C}^{1\times M}$, and $\mathbf{W}=[\mathbf{w}_1,\ldots,\mathbf{w}_K]\in\mathbb{C}^{N\times K}$ is constrained by the set
\begin{equation}
	\mathcal{W}_\mathsf{MU-MISO}^\mathsf{sum-rate} = \{\mathbf{W}~|~\|\mathbf{W}\|_\mathsf{F}^2\le P\}.
\end{equation}
This is again a multi-variable optimization problem, which can be solved either by separately designing BD-RIS and transmit precoder, or by block coordinate descent (BCD) methods \cite{boyd2004convex}. 

For BD-RIS having reciprocal architectures, the following beamforming design and performance analysis studies have been conducted.
With the form of (\ref{prob:scattering}) and group/fully-connected architectures constrained by $\mathcal{T}_\mathsf{group-conn}$, a two-stage BD-RIS and precoder design has been proposed in \cite{fang2023low} with low computational complexity. Specifically, the optimization of BD-RIS matrix is based on the symmetric unitary projection as detailed in Section \ref{subsec:opt_siso}. Following the idea of symmetric unitary projection, a joint precoder and BD-RIS design has been further proposed in \cite{zhou2025joint}. 
With the form of (\ref{prob:admittance}) and group/fully-connected architectures constrained by $\mathcal{Y}_\mathsf{group-conn}$, a quasi-Newton based solution together with a heuristic user scheduling scheme has been proposed in \cite{kim2023scattering}.
With the form of (\ref{prob:scattering}) and multi-sector mode BD-RIS constrained by the set
\begin{equation}
    \begin{aligned} 
        \mathcal{T}_\mathsf{multi-sec} = \Big\{\mathbf{\Theta}~|~\sum_{l\in\mathcal{L}}|\varTheta_{l,1,n}|^2=1,\forall n=1,\ldots,\frac{M}{L}\Big\},
    \end{aligned}
\end{equation}
where $\varTheta_{l,1,n}$, $\forall l\in\mathcal{L}$, $\forall n = 1,\ldots,\frac{M}{L}$ are extracted from $\mathbf{\Theta}$ according to Example 2 of Section \ref{subsec:unified}, an iteratively closed-form solution can be obtained by deriving the Karush-Kuhn-Tucker conditions \cite{li2023beyond}. In addition, \cite{samy2024enhancing} has derived the closed-form expression of the achievable sum-rate aided by multi-sector BD-RIS, with a further extension to other performance metrics, such as energy efficiency, error probability, outage probability \cite{samy2024beyond}. 

For BD-RIS having non-reciprocal architectures, the following beamforming design studies have been conducted. 
For BD-RIS constrained by the set $\mathcal{T}_{\mathsf{non-recip}}$, \cite{sobhi2024joint} and \cite{loli2024meta} have proposed learning-based methods which support larger-dimensional optimizations. 
For BD-RIS with hybrid mode and proper non-reciprocal architectures, the constraint $\mathbf{\Theta}_\mathsf{r}=\mathbf{\Theta}_\mathsf{r}^\mathsf{T}$ from Example 1 of Section \ref{subsec:unified} can be dropped such that $\mathbf{\Theta}$ is constrained by the set 
\begin{equation}
    \begin{aligned} 
        \mathcal{T}_\mathsf{hyb} = \{\mathbf{\Theta}~|~\mathbf{\Theta}_{\mathsf{r},g}^\mathsf{H}\mathbf{\Theta}_{\mathsf{r},g}+\mathbf{\Theta}_{\mathsf{t},g}^\mathsf{H}\mathbf{\Theta}_{\mathsf{t},g} = \mathbf{I}_{\frac{\bar{M}}{2}\times\frac{\bar{M}}{2}},\forall g\in\mathcal{G}\},
    \end{aligned}
\end{equation}
where $\mathbf{\Theta}_{\mathsf{r},g}$ and $\mathbf{\Theta}_{\mathsf{t},g}$, $\forall g\in\mathcal{G}$ are extracted from $\mathbf{\Theta}$ according to Example 1 of Section \ref{subsec:unified}. It should be noted here that the constraint of hybrid BD-RIS essentially constructs a Stiefel Manifold \cite{lee2018introduction}. Therefore, an iterative searching-based solution can be obtained by using the Manifold theory \cite{li2022beyond}. 
For BD-RIS with hybrid mode and dynamically connected architectures adapting to the instantaneous CSI, \cite{li2023dynamic} has proposed a heuristic grouping strategy design.

\subsubsection{Energy Efficiency Maximization and Power Minimization for Multi-User MISO}
In multi-user MISO systems, the utility functions for energy-efficiency maximization and power minimization are respectively given by 
\begin{equation}
	F_\mathsf{MU-MISO}^\mathsf{energy-effi}(\mathbf{W},\{\mathbf{h}_k(\mathbf{\Theta})\}_{\forall k}) = \frac{F_\mathsf{MU-MISO}^\mathsf{sum-rate}(\mathbf{W},\{\mathbf{h}_k(\mathbf{\Theta})\}_{\forall k})}{\eta\|\mathbf{W}\|_\mathsf{F}^2 + P_\mathsf{d}},
\end{equation}
\begin{equation}
	F_\mathsf{MU-MISO}^\mathsf{power-min}(\mathbf{W},\{\mathbf{h}_k(\mathbf{\Theta})\}_{\forall k}) = -\|\mathbf{W}\|_\mathsf{F}^2 - \sum_{k\in\mathcal{K}} \mathbb{I}_{\mathcal{S}_k} (\mathbf{W},\mathbf{\Theta}),
\end{equation}
where $\eta$ denotes the power amplifier efficiency, $P_\mathsf{d}$ denotes the power used for device operating, and $\mathbb{I}_{\mathcal{S}_k}(\mathbf{W},\mathbf{\Theta})$ denotes the indicator function of set $\mathcal{S}_k$ 
\begin{equation}
	\mathcal{S}_k = \{(\mathbf{W},\mathbf{\Theta})~|~\gamma_k(\mathbf{W},\mathbf{h}_k(\mathbf{\Theta}))\ge \bar{\gamma}_k\}, \forall k\in\mathcal{K},
\end{equation}
with $\bar{\gamma}_k$ being the signal-to-interference-plus-noise (SINR) threshold for user $k$.

For BD-RIS having reciprocal architectures, the following beamforming design and performance analysis studies have been conducted.
With the form of (\ref{prob:scattering}) and group/fully-connected architectures constrained by $\mathcal{T}_\mathsf{group-conn}$, \cite{zhou2023optimizing} has proposed an iterative framework by decoupling the unitary and symmetric constraints of $\mathbf{\Theta}$ with a newly introduced auxiliary variable $\mathbf{\Phi}=\mathbf{\Theta}$, such that $\mathbf{\Theta}$ is only subject to the unitary constraint and $\mathbf{\Phi}$ is only subject to the symmetric constraint, or vise versa. The proposed framework is applicable to sum-rate maximization, energy-efficiency maximization, and power minimization problems. 
With the form of (\ref{prob:scattering}) and STARS, which is essentially a special case of multi-sector BD-RIS constrained by $\mathcal{T}_\mathsf{multi-sec}$ when $L=2$ and also a special case of hybrid BD-RIS constrained by $\mathcal{T}_\mathsf{hyb}$ when $\bar{M}=2$, \cite{farhadi2025meta} has proposed a meta-learning approach for joint optimization of BD-RIS, resource allocation, and antenna selection.
With the form of (\ref{prob:admittance}), \cite{wu2024optimization} has recently proposed a universal framework applicable to arbitrary constraints of BD-RIS admittance matrix, the aforementioned three utility functions, and sum-rate maximization for more general multi-user MIMO scenarios. The main idea is to introduce auxiliary variables $\mathbf{u}_k = \mathbf{h}_{RI,k}\mathbf{\Theta}$, $\forall k\in\mathcal{K}$, such that the matrix inverse in (\ref{prob:admittance}) can be eliminated by transferring it to bilinear constrains to facilitate the optimization. The applicable BD-RIS constraints include but are not limited to those for group/fully-, forest/tree-, and band/stem-connected architectures in Section \ref{subsec:architecture}.

The above literature about optimization and performance analysis for BD-RIS is summarized in Table \ref{tab:beamforming}. 

\begin{table*}[]
	\caption{BD-RIS Literature on Optimization and Performance Analysis}
	\label{tab:beamforming}
	\centering
	\begin{threeparttable}
		\begin{tabular}{|c|c|c|c|c|l|}
			\hline
			Ref. & Architecture$^\dagger$ & Mode & Scenario & Metric & Highlights \\
			\hline\hline
			\cite{shen2021} &  \multirow{2}{*}{Fully/Group-Connected}   & \multirow{8}{*}{Reflecting}   &  \multirow{3}{*}{SISO} & \multirow{8}{*}{\begin{tabular}[c]{@{}c@{}}Received\\ Power Max.\end{tabular}}  & \begin{tabular}[l]{@{}l@{}}Derive the performance upper-bound and \\scaling law w.r.t the number of elements\end{tabular}\\
			\cline{1-1}\cline{6-6}
			\cite{nerini2023closed,santamaria2023snr} &  &  &  &  & \multirow{2}{*}{Propose closed-form global optimal solutions}     \\
			\cline{1-2}
			\cite{sun2024new} & Fully-Connected & & & & \\
			\cline{1-2}\cline{4-4}\cline{6-6}
			\cite{nerini2024beyond} & Tree/Forest-Connected &  & \multirow{5}{*}{\begin{tabular}[c]{@{}c@{}}MISO,\\SIMO\end{tabular}} & & Propose a closed-form solution to achieve MISO optimum\\
			\cline{1-2}\cline{6-6}
			\cite{nerini2024static} & Dynamically Connected & & & & \begin{tabular}[l]{@{}l@{}}Propose an offline grouping strategy\\ design adapting to static CSI\end{tabular}\\
			\cline{1-2}\cline{6-6}
			\cite{li2023reconfigurable,li2024coordinated} & \multirow{2}{*}{Non-Reciprocal} & & & & Derive the optimal permuting strategy\\ 
			\cline{1-1}\cline{3-3}\cline{6-6}
			\cite{dong2024reconfigurable} & & Multi-Sector & & & Use non-reciprocal architectures to support multi-sector mode\\
			\hline
			\cite{santamaria2024mimo} & Fully-Connected & \multirow{7}{*}{Reflecting} & \multirow{7}{*}{MIMO}  & \multirow{7}{*}{\begin{tabular}[c]{@{}c@{}}Capacity\\ Max.\end{tabular}} & Jointly design BD-RIS and transmitter covariance matrix \\
			\cline{1-2}\cline{6-6}
            \cite{santamaria2025rate} & Fully/Group-Connected & & & & \begin{tabular}[l]{@{}l@{}}Derive the closed-form solution of BD-RIS\\ for rank-1 channels\end{tabular}\\
            \cline{1-2}\cline{6-6}
            \cite{zhao2024channel} &\multirow{5}{*}{Non-Reciprocal} & & & & Propose a Manifold-based solution with fast convergence\\
            \cline{1-1}\cline{6-6}
			\cite{bartoli2023spatial} &  & & & & \begin{tabular}[l]{@{}l@{}}Derive nearly-optimal solution of BD-RIS\\ for near-field channels\end{tabular}\\
			\cline{1-1}\cline{6-6}
			\cite{bjornson2024capacity} & & & & & \begin{tabular}[l]{@{}l@{}}Derive the closed-form solution of BD-RIS\\ for far-field channels\end{tabular}\\
			\hline
			\cite{fang2023low} & Fully/Group-Connected & \multirow{4}{*}{Reflecting} & \multirow{15}{*}{\begin{tabular}[c]{@{}c@{}}Multi-User\\ MISO\end{tabular}} & \multirow{7}{*}{\begin{tabular}[c]{@{}c@{}}Sum-Rate\\ Max.\end{tabular}} &Propose a two-stage design with low complexity\\  
            \cline{1-2}\cline{6-6}
            \cite{zhou2025joint} & Fully-Connected & & & & Propose a joint design based on unitary symmetric projection\\
			\cline{1-2}\cline{6-6}
			\cite{kim2023scattering} & Fully/Group-Connected & & & & Propose a heuristic user scheduling scheme\\
			\cline{1-2}\cline{6-6}
			\cite{sobhi2024joint,loli2024meta} & \multirow{4}{*}{Non-Reciprocal} & & & & Propose learning-based beamforming design algorithms\\
			\cline{1-1}\cline{3-3}\cline{6-6}
			\cite{li2022beyond} & & \multirow{2}{*}{Hybrid} & & & Propose an iterative algorithm using Manifold theory\\
			\cline{1-1}\cline{6-6}
			\cite{li2023dynamic} & & & & &\begin{tabular}[l]{@{}l@{}} Propose a heuristic grouping strategy\\ adapting to instantaneous CSI\end{tabular}\\
			\cline{1-3}\cline{6-6}
			\cite{li2023beyond} &\multirow{4}{*}{Group-Connected} & \multirow{4}{*}{Multi-Sector} & & & Propose an iteratively closed-form solution\\
			\cline{1-1}\cline{5-6}
			\cite{samy2024enhancing,samy2024beyond} & & & & \begin{tabular}[c]{@{}c@{}}Sum-Rate\\ Max., Energy \\Effi. Max. \end{tabular} & \begin{tabular}[l]{@{}l@{}} Derive the closed-form expression of achievable sum-rate\\ and energy efficiency aided by multi-sector BD-RIS \end{tabular} \\
			\cline{1-3}\cline{5-6}
            \cite{farhadi2025meta} & Group-Connected & \begin{tabular}[c]{@{}c@{}}Hybrid\\ (STAR)\end{tabular} & & \begin{tabular}[c]{@{}c@{}}Energy Effi.\\ Max. \end{tabular} & Propose a meta-learning approach\\
            \cline{1-3}\cline{5-6}
			\cite{zhou2023optimizing} & Fully/Group-Connected & \multirow{3}{*}{Reflecting} & & \multirow{2}{*}{\begin{tabular}[c]{@{}c@{}}Sum-Rate\\ Max., Energy \\Effi. Max.,\\Power Min. \end{tabular}} & \begin{tabular}[l]{@{}l@{}}Propose a unified optimization framework for sum-rate\\ maximization, energy efficiency  maximization, and \\power minimization \end{tabular}\\
			\cline{1-2}\cline{4-4}\cline{6-6}
			\cite{wu2024optimization,wu2025beyond} & Arbitrary & &\begin{tabular}[c]{@{}c@{}}Multi-User\\ MIMO \end{tabular}& & \begin{tabular}[l]{@{}l@{}}Propose a unified optimization framework for commonly\\ used metrics and arbitrary architectures \end{tabular}\\
			\hline
		\end{tabular}
		\begin{tablenotes}
			\footnotesize
			\item[$\dagger$] The architectures are all reciprocal unless otherwise stated.
		\end{tablenotes}
	\end{threeparttable}\vspace{-0.2 cm}
\end{table*}

\subsection{Channel Estimation}
\label{subsec:estimation}

Channel estimation is crucial for BD-RIS-aided wireless systems since the performance benefits of BD-RIS are supported by proper optimization, beamforming design methods and performance analysis, all of which rely highly on accurate CSI.
In D-RIS systems, there are generally two approaches to acquire instantaneous CSI, namely semi-passive channel estimation where RIS is mounted with RF chains for sensing signals, and passive channel estimation where RIS does not have the ability to sense signals \cite{zheng2022survey}. 
The semi-passive channel estimation approach results in separate base station-RIS and RIS-user channels with relatively low training overhead, both of which are independent of RIS architectures. Therefore, this approach used in D-RIS systems \cite{hu2021semi,alexandropoulos2020hardware,taha2019deep} is readily applicable to BD-RIS systems. However, the drawback of this approach is that the channel estimation performance relates tightly to the number of RF chains. Intuitively, the more the RF chains, the better the estimation performance. This indicates that a satisfactory estimation performance is achieved at the cost of significant power consumption. 
The passive channel estimation approach is an alternative approach which eases the requirement for RF chains \cite{swindlehurst2022channel}. 
Take the time-division duplex (TDD) system as an example, the passive channel estimation usually happens in uplink. Specifically, on the uplink, the user consecutively transmits pilot signals to the base station through BD-RIS and the BD-RIS k.pdf varying its response based on a pre-defined pattern.  
In this sense, it is in general not easy to obtain exactly the separate base station-RIS and RIS-user channels. Instead, the cascaded channel, which is the combination of base station-RIS and RIS-user channels, can be estimated. The passive channel estimation approach has been widely studied in D-RIS literature, starting by simple orthogonal RIS pattern design \cite{yang2020intelligent,zheng2019intelligent,you2020channel} with the aim to demonstrate the feasibility of this approach, followed by advanced pattern and protocol design with reduced training overhead \cite{wang2020channel,guan2021anchor,zhou2022channel}.
Nevertheless, it is worth noting that the above passive channel estimation studies for D-RIS do not work for BD-RIS due to the following reasons. 
\begin{itemize}
	\item The structure of the cascaded channel is mathematically determined by the structure of BD-RIS scattering matrix $\mathbf{\Theta}$, such that D-RIS with a diagonal scattering matrix leads to a cascaded channel different from BD-RIS with beyond-diagonal scattering matrix. 
	\item The pattern design of D-RIS is subject to constraints different from BD-RIS architectures. 
\end{itemize}
These two points will be explained in more details, based on a specific least-squares estimation.

\subsubsection{Least-Squares Estimation}

Consider a narrowband BD-RIS-aided MISO system consisting of an $N$-antenna base station, an $M$-antenna BD-RIS working on reflecting mode, and a single-antenna user.
In the uplink, denote the channels from the user to BD-RIS as $\mathbf{h}_{IR}\in\mathbb{C}^{M\times 1}$ and from the BD-RIS to the base station as $\mathbf{H}_{TI}\in\mathbb{C}^{N\times M}$. We assume the direct link from the user to the base station is blocked\footnote{In the case where a direct link exists, one can first turn OFF the BD-RIS by letting $\mathbf{\Theta}$ be a zero matrix and estimate the direct channel using well-established methods in conventional MIMO systems, and then estimate the cascaded channel related to
BD-RIS by removing the contribution of the direct channel from the received data and using the proposed pattern design.}, $\mathbf{h}_{TR}=\mathbf{0}_{N\times 1}$, and focus only on the estimation of BD-RIS-aided channels. The uplink channel is thus 
\begin{equation}
    \label{eq:uplink_channel}
	\mathbf{h}_\mathsf{up} = \mathbf{H}_{TI}\mathbf{\Theta}\mathbf{h}_{IR} = \underbrace{\mathbf{h}^\mathsf{T}_{IR}\otimes\mathbf{H}_{TI}}_{=\mathbf{H}_\mathsf{cas}^\mathsf{BD}\in\mathbb{C}^{N\times M^2}}\mathsf{vec}(\mathbf{\Theta}),
\end{equation}
where $\mathbf{H}_\mathsf{cas}^\mathsf{BD}$ is the cascaded channel to be estimated. Note that the expression in (\ref{eq:uplink_channel}) holds for $\mathbf{\Theta}$ with any structures, such as being diagonal, block-diagonal, or full. Specifically for D-RIS with $\mathbf{\Theta} = \mathsf{diag}(\varTheta_1,\ldots,\varTheta_M)$, the uplink channel can be expressed differently as
\begin{equation}
	\mathbf{h}_\mathsf{up} = \mathbf{H}_{TI}\mathbf{\Theta}\mathbf{h}_{IR} = \sum_{m=1}^M\underbrace{[\mathbf{h}_{IR}]_{m}[\mathbf{H}_{TI}]_{:,m}}_{=\mathbf{h}_{\mathsf{cas},m}^\mathsf{D}\in\mathbb{C}^{N\times 1}}\varTheta_m = \mathbf{H}_\mathsf{cas}^\mathsf{D}\bm{\theta},
\end{equation}
where $\bm{\theta}=[\varTheta_1,\ldots,\varTheta_M]^\mathsf{T}$ and the cascaded channel to be estimated is given by
\begin{equation}
	\mathbf{H}_\mathsf{cas}^\mathsf{D} = [\mathbf{h}_{\mathsf{cas},1}^\mathsf{D},\ldots,\mathbf{h}_{\mathsf{cas},M}^\mathsf{D}]\in\mathbb{C}^{N\times M}.
\end{equation}
Comparing $\mathbf{H}_\mathsf{cas}^\mathsf{BD}$ and $\mathbf{H}_\mathsf{cas}^\mathsf{D}$, we could observe that the two cascaded channels have different dimensions and mathematical structures, such that the passive channel estimation designed for D-RIS does not work for BD-RIS. In the following, we will elaborate more on the difference in pattern design between BD-RIS and D-RIS. To do so, we first briefly revisit the channel estimation process \cite{li2023channel}, which generally works for any BD-RIS architectures and D-RIS.

Assuming the user sends pilot symbol vector $x_j$, $|x_j|=1$ at time slot $j$, $\forall j\in\mathcal{J} = \{1,\ldots,J\}$, the signal received at the base station is 
\begin{equation}
	\mathbf{y}_j = \sqrt{P_\mathsf{u}}\mathbf{H}_\mathsf{cas}^\mathsf{BD}\mathsf{vec}(\ddot{\mathbf{\Theta}}_j)x_j + \mathbf{n}_j, 
\end{equation}
where $P_\mathsf{u}$ denotes the power transmitted at the user, $\ddot{\mathbf{\Theta}}_j$ denotes the BD-RIS scattering matrix at time slot $j$, $\mathbf{n}_j\sim\mathcal{CN}(\mathbf{0}_{N\times 1},\sigma^2\mathbf{I}_N)$ denotes the noise with power $\sigma^2$.
Assuming $x_j=1$, $\forall j\in\mathcal{J}$ without loss of channel estimation performance, the overall received signal stacking the received signal from all times slots is 
\begin{equation}
	\mathbf{Y}_{\mathsf{all}} = \sqrt{P_\mathsf{u}}\mathbf{H}_\mathsf{cas}^\mathsf{BD}\mathbf{\Theta}_\mathsf{all}^\mathsf{BD} + \mathbf{N},
\end{equation} 
where we define $\mathbf{Y}_\mathsf{all} = [\mathbf{y}_1,\ldots,\mathbf{y}_J]$, $\mathbf{\Theta}_\mathsf{all}^\mathsf{BD}=[\mathsf{vec}(\ddot{\mathbf{\Theta}}_1),\ldots,\mathsf{vec}(\ddot{\mathbf{\Theta}}_J)]$, and $\mathbf{N}=[\mathbf{n}_1,\ldots,\mathbf{n}_J]$. 
The simplest way to estimate $\mathbf{H}_\mathsf{cas}^\mathsf{BD}$ is the least-squares method, resulting in the estimator
\begin{equation}
	\widehat{\mathbf{H}}_\mathsf{cas}^\mathsf{BD} = \sqrt{P_\mathsf{u}^{-1}}\mathbf{Y}_\mathsf{all}(\mathbf{\Theta}_\mathsf{all}^\mathsf{BD})^\mathsf{H}(\mathbf{\Theta}_\mathsf{all}^\mathsf{BD}(\mathbf{\Theta}_\mathsf{all}^\mathsf{BD})^\mathsf{H})^{-1},
\end{equation}
with $J\ge M^2$ to guarantee the successful estimation. This leads to the mean square error (MSE) as $\mathrm{err}_{\widehat{\mathbf{H}}_\mathsf{cas}^\mathsf{BD}}=\mathbb{E}\{\|\mathbf{H}_\mathsf{cas}^\mathsf{BD}-\widehat{\mathbf{H}}_\mathsf{cas}^\mathsf{BD}\|_\mathsf{F}^2\} = \frac{N\sigma^2}{P_\mathsf{u}}\mathsf{tr}((\mathbf{\Theta}_\mathsf{all}^\mathsf{BD}(\mathbf{\Theta}_\mathsf{all}^\mathsf{BD})^\mathsf{H})^{-1})$, which indicates that the channel estimation performance depends purely on the design of $\mathbf{\Theta}_\mathsf{all}^\mathsf{BD}$. 
We assume the BD-RIS has a non-reciprocal architecture, where the scattering matrices $\ddot{\mathbf{\Theta}}_j$, $\forall j\in\mathcal{J}$ are constrained by $\mathcal{T}_\mathsf{non-recip}$ in (\ref{eq:constraint_nr}), for the ease of illustration and comparison.
This motivates the following problem 
\begin{equation}\label{prob:ce}
	\begin{aligned}
		\min_{\mathbf{\Theta}_\mathsf{all}^\mathsf{BD}}~  &\mathsf{tr}((\mathbf{\Theta}_\mathsf{all}^\mathsf{BD}(\mathbf{\Theta}_\mathsf{all}^\mathsf{BD})^\mathsf{H})^{-1})\\
		\mathrm{s.t.}~&\mathbf{\Theta}_\mathsf{all}^\mathsf{BD}=[\mathsf{vec}(\ddot{\mathbf{\Theta}}_1),\ldots,\mathsf{vec}(\ddot{\mathbf{\Theta}}_J)],\\
		&\ddot{\mathbf{\Theta}}_j^\mathsf{H}\ddot{\mathbf{\Theta}}_j = \mathbf{I}_M, \forall j\in\mathcal{J},\\
		&\mathsf{rank}(\mathbf{\Theta}_\mathsf{all}^\mathsf{BD}) = M^2,
	\end{aligned}
\end{equation}
where the training overhead is minimized as $J_\mathsf{LS}^\mathsf{BD} = M^2$ without introducing estimation ambiguities.  
Similarly, for the case of D-RIS, we define the RIS pattern $\mathbf{\Theta}_\mathsf{all}^\mathsf{D} = [\bm{\theta}_1,\ldots,\bm{\theta}_J]\in\mathbb{C}^{M\times J}$, where $\bm{\theta}_j=[\ddot{\varTheta}_{j,1},\ldots,\ddot{\varTheta}_{j,M}]^\mathsf{T}$ collects the diagonal entries in $\ddot{\mathbf{\Theta}}_j = \mathsf{diag}(\ddot{\varTheta}_{j,1},\ldots,\ddot{\varTheta}_{j,M})$ of D-RIS at time slot $j$, with $|\ddot{\varTheta}_{j,m}|=1$, $\forall j\in\mathcal{J}$ when the RIS is lossless. To minimize the channel estimation error, we have
\begin{equation}\label{prob:ce_conv}
	\begin{aligned}
		\min_{\mathbf{\Theta}_\mathsf{all}^\mathsf{D}}~  &\mathsf{tr}((\mathbf{\Theta}_\mathsf{all}^\mathsf{D}(\mathbf{\Theta}_\mathsf{all}^\mathsf{D})^\mathsf{H})^{-1})\\
		\mathrm{s.t.}~&\mathbf{\Theta}_\mathsf{all}^\mathsf{D}=[\bm{\theta}_1,\ldots,\bm{\theta}_J],\\
		&|\ddot{\varTheta}_{j,m}| = 1, \forall j\in\mathcal{J},\forall m\in\mathcal{M},\\
		&\mathsf{rank}(\mathbf{\Theta}_\mathsf{all}^\mathsf{D}) = M,
	\end{aligned}
\end{equation}
where the training overhead is minimized as $J_\mathsf{LS}^\mathsf{D} = M$ here for the same reason as in problem (\ref{prob:ce}).
Problems (\ref{prob:ce}) and (\ref{prob:ce_conv}) clearly show the difference between BD-RIS and D-RIS in pattern design. Specifically, for (\ref{prob:ce_conv}), one can directly use orthogonal matrices, such as the discrete Fourier matrix or the Hadamard matrix, to construct $\mathbf{\Theta}_\mathsf{all}^\mathsf{D}$ \cite{swindlehurst2022channel}. However, the solutions for (\ref{prob:ce_conv}) are not feasible for (\ref{prob:ce}) due to the unique constraints of BD-RIS.

\begin{table*}[]
	\caption{BD-RIS Literature on Channel Estimation}
	\label{tab:ce}
	\centering
	\begin{threeparttable}
		\begin{tabular}{|c|c|c|l|}
			\hline
			Ref.$\ddagger$ & Estimate Type & Overhead$\dagger$ & Highlights \\
			\hline\hline
			\cite{li2023channel,li2024channel} & \multirow{4}{*}{Cascaded} & $M^2$ & \begin{tabular}[l]{@{}l@{}} The first least-squares estimation scheme based on a closed-form BD-RIS pattern \end{tabular}\\
			\cline{1-1}\cline{3-4}
			\cite{wang2024low,wang2025low} & & $2M$ & A two-phase estimation scheme by exploring the cascaded channel characteristics\\
            \cline{1-1}\cline{3-4}
            \cite{samy2025low} & & $3M$ & A three-stage estimation scheme by exploring the property of non-diagonal scattering matrices\\
			\hline
			\cite{sokal2024decoupled} & \multirow{4}{*}{Separate} & $M^2$ & A decoupled estimation method based on Khatri-Rao Factorization\\
			\cline{1-1}\cline{3-4}
			\cite{de2024channel,ginige2024efficient} &  & Much smaller than $M^2$ & A decoupled estimation method based on Tucker decomposition and alternating least squares\\
            \cline{1-1}\cline{3-4}
            \cite{de2024semi} & & / & A semi-blind estimation method that avoids the pilot-assisted training stage\\
			\hline
		\end{tabular}
		\begin{tablenotes}
			\footnotesize
			\item[$\dagger$] The overhead is calculated for a MISO system, which will scale with the number of users and/or antennas for more complex multi-user MIMO systems.
                \item[$\ddagger$] \cite{li2023channel,li2024channel,wang2024low,wang2025low,sokal2024decoupled,de2024channel,ginige2024efficient,de2024semi} focus on non-reciprocal BD-RIS constrained by $\mathcal{T}_\mathsf{non-recip}$ and \cite{samy2025low} focuses on non-reciprocal BD-RIS constrained by $\mathcal{T}_\mathsf{non-diag}$.
		\end{tablenotes}
	\end{threeparttable}
\end{table*}

\subsubsection{Related Works and Discussions}

To perfectly capture the constraints of BD-RIS architectures and facilitate the least-squares estimation, \cite{li2023channel} has proposed a global optimal closed-form solution to (\ref{prob:ce}), followed by a further extension to BD-RIS with hybrid/multi-sector modes in more general multi-user MIMO systems \cite{li2024channel}. Based on the proposed pattern design, \cite{li2023channel,li2024channel} for the first time provide a comprehensive study for BD-RIS-aided systems from channel estimation to beamforming design and data transmission. However, the drawbacks of the method in \cite{li2023channel,li2024channel} are 1) the high training overhead, e.g., $J_\mathsf{LS}^\mathsf{BD} = M^2$ for a non-reciprocal BD-RIS constrained by $\mathcal{T}_\mathsf{non-recip}$, 2) the limited performance without exploring the structural characteristics of the cascaded channel, and 3) the channel estimation error for non-reciprocal BD-RIS that theoretically scales with the group size $\bar{M}$ \cite{li2024channel}:
\begin{equation}
\mathrm{err}^\mathsf{BD} = \frac{\sigma^2}{P_\mathsf{u}}N\bar{M},
\end{equation}
which indicates that the estimation error increases with the circuit complexity of BD-RIS (i.e., a larger $\bar{M}$ leads to higher estimation error).
To further reduce the channel estimation error, tensor decomposition has been applied in \cite{sokal2024decoupled} based on the designed pattern in \cite{li2023channel}, which leads to the possibility of separate estimation of base station-RIS and RIS-user channels.
Beyond that, two tensor decomposition based algorithms have been proposed in \cite{de2024channel} with a milder requirement of the training overhead and enhanced estimation performance. 
\cite{ginige2024efficient} again explores the tensor decomposition for BD-RIS-aided channel estimation, which further facilitates the channel prediction with channel aging.
\cite{de2024semi} proposes a semi-blind channel and symbol estimation method that avoids the pilot-assisted estimation by exploring also the tensor decomposition for BD-RIS-aided channels.  
In addition, \cite{wang2024low} has focused on the protocol design for a single-user MISO system and proposed a novel two-phase estimation scheme, where the cascaded channel related to the first element of BD-RIS is 
estimated, followed by fast estimation of the other channel coefficients. The two-phase estimation scheme explores the dependence between entries of the cascaded channel $\mathbf{H}_\mathsf{cas}^\mathsf{BD}$ and thus helps to significantly reduce the training overhead from $J_\mathsf{LS}^\mathsf{BD} = M^2$ to $J_\mathsf{2-phase}^\mathsf{BD} = 2M$.
This channel estimation scheme has been further extended to more general multi-user MIMO systems, with the training overhead in the same order as in D-RIS-aided system \cite{wang2025low}.
Besides, \cite{samy2025low} has proposed a three-stage protocol for a non-reciprocal BD-RIS aided MISO system where the BD-RIS is characterized by a non-diagonal scattering matrix constrained by $\mathcal{T}_\mathsf{non-diag}$ in (\ref{eq:constraint_nd}). Using the special signal flowing property in the non-diagonal scattering matrix, i.e., the signal impinging on one element is purely reflected by another one, the training overhead can be as low as $J_\mathsf{3-stage}^{\mathsf{ND}} = 3M$.
For clarity, the aforementioned works have been summarized in Table \ref{tab:ce}.

The channel estimation study for BD-RIS-aided wireless systems is still at the very early stage. It is worth noting that, the passive channel estimation approach relies on the BD-RIS pattern design and results in a cascaded channel (or separate channels based on tensor decomposition whose process relies on BD-RIS constraints), both of which are related to BD-RIS architectures. In this sense, the existing literature \cite{li2023channel,li2024channel,sokal2024decoupled,de2024channel,wang2024low,wang2025low,ginige2024efficient,samy2025low} based on non-reciprocal architectures cannot be readily extended to other BD-RIS architectures. Moreover, in practice, BD-RIS inevitably has hardware impairments that will induce new mathematical constraints and communications models, as will be detailed in Section \ref{sec:hardware}. Therefore, those works based on BD-RIS with perfect hardware may not be applicable to BD-RIS with practical hardware impairments.
Another critical issue for passive channel estimation is the training overhead that grows with the number of elements and the circuit complexity. Therefore, it remains unexplored, but it is important to develop BD-RIS pattern designs automatically adapting to architectures and hardware impairments, and to study better channel estimation schemes with affordable training overhead.

\section{Benefits of BD-RIS}
\label{sec:benefits}

Thanks to the flexible interconnections between elements to support various architectures and modes, BD-RIS has multiple benefits, such as enhancing channel gain, increasing transmission quality, enlarging coverage, etc. In this section, we summarize the key benefits of BD-RIS, each of which is supported by numerical results. 

\subsection{Boosting Received Power and Rates}

\begin{figure}
    \centering
    \includegraphics[width=0.45\textwidth]{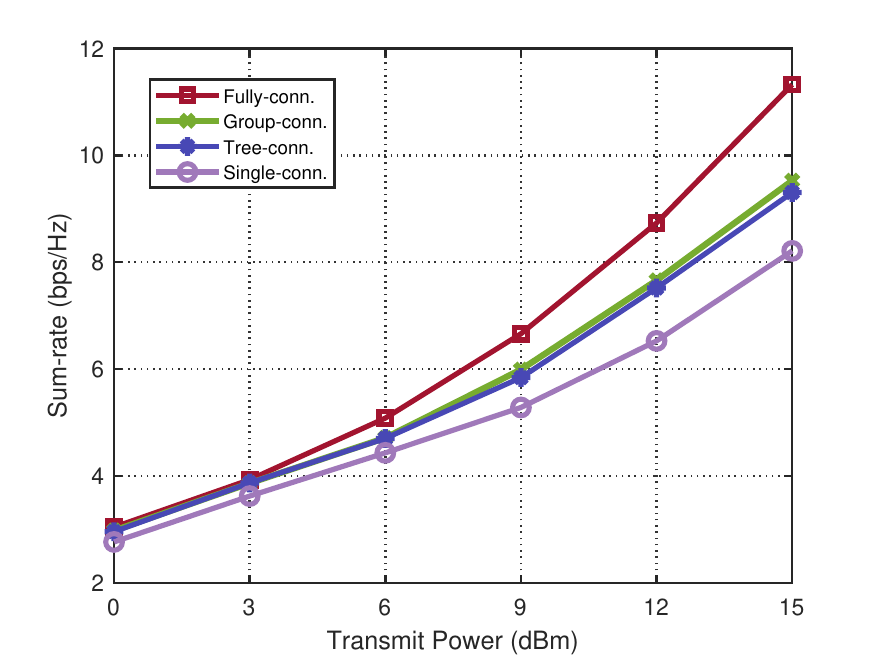}
    \caption{ Sum-rate of a 4-user system with the aid of a 32-element BD-RIS versus transmit power for different BD-RIS architectures. Channels through BD-RIS follow Rician fading with a Rician factor 2 dB. The direct transmitter-user links are assumed to be blocked. The transmitter-RIS distance is set as 50 m and the RIS-user distance is set as 2.5 m \cite{wu2024optimization}.}
    \label{fig:sum_rate_opt}
\end{figure}

Compared to (lossless) D-RIS which can only manipulate the phase shift of the diagonal entries in the scattering matrix, BD-RIS has higher flexibility in manipulating both amplitude and phase shift of diagonal and off-diagonal entries in its scattering matrix. 
This flexibility boosts the performance of various systems, such as increasing the received power by up to 62\% for SISO as shown in Fig. \ref{fig:gain} and increasing the sum-rate for MU-MISO by up to 43\% as shown in Fig. \ref{fig:sum_rate_opt}.


\subsection{Enabling Low-Complexity Architectures with High Performance}

The flexibility provided by the interconnections between BD-RIS elements enables various circuit topology designs to support different architectures as illustrated in Sections \ref{subsec:architecture} and \ref{subsec:architecture_nr}. More importantly, this flexibility enables architectures with least hardware complexity (i.e., number of tunable components) to achieve optimal performance in various systems.

\begin{figure}
    \centering 
    \includegraphics[width=0.45\textwidth]{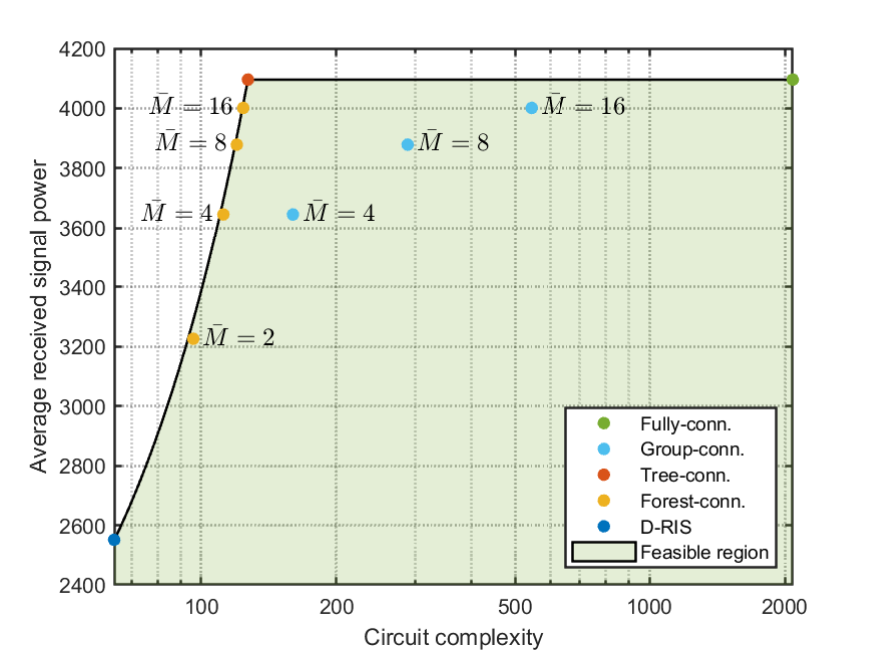}
    \caption{Pareto frontier between performance and circuit complexity including single-, group-, fully-, forest-, and tree-connected BD-RIS, with different group sizes $\bar{M}$ labeled on two sides of the curve \cite{nerini2023pareto}. The direct transmitter-receiver channel is assumed to be blocked and channels through BD-RIS follow i.i.d. Rayleigh fading.}
    \label{fig:pareto_siso}\vspace{-0.2 cm}
\end{figure}

For the MISO system, \cite{nerini2024beyond} has theoretically proven that the least circuit complexity for MISO optimal BD-RIS is 
\begin{equation}
C_\mathsf{MISO}^\mathsf{opt} = 2M-1.
\end{equation}
This condition essentially forms a tree-connected architecture, and two representative examples are arrowhead BD-RIS and tridiagonal BD-RIS. 
Based on this conclusion, \cite{nerini2023pareto} has analyzed the Pareto frontier between performance and circuit complexity for a SISO system. Fig. \ref{fig:pareto_siso} shows that tree-connected architecture can reach the performance achieved by fully-connected BD-RIS with the lowest possible complexity, thus achieving the best performance-complexity trade-off. 

\begin{figure}
    \centering
    \includegraphics[width=0.45\textwidth]{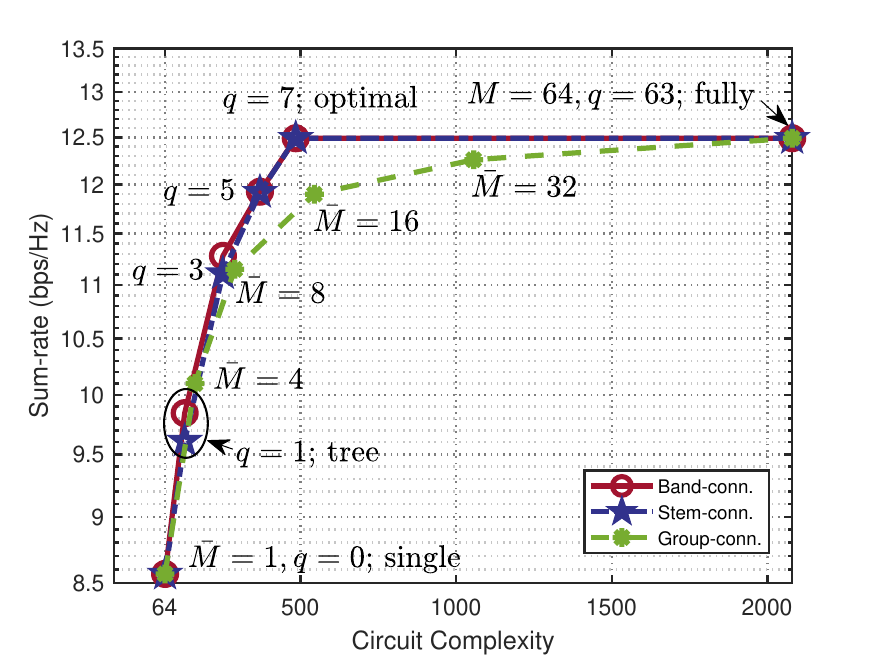}
    \caption{Pareto frontier between performance and circuit complexity including single-, group-, fully-, stem-, and band-connected BD-RIS, with different group sizes $\bar{M}$ and stem/band width $q$ labeled on two sides of the curves. The direct transmitter-user channels are assumed to be blocked and channels through BD-RIS follow Rician fading with a Rician factor 2 dB. The transmitter-RIS distance is set as 50 m and the RIS-user distance is set as 2.5 m. ($N=K=4$, $M=64$, $P=10$ dBm) \cite{wu2025beyond}.}\vspace{-0.2 cm}
    \label{fig:perato_sumrate}
\end{figure}


For more general multi-user MIMO systems, \cite{wu2025beyond} has theoretically proved that the least circuit complexity for multi-user MIMO optimal BD-RIS is 
\begin{equation}
    C_\mathsf{MU-MIMO}^\mathsf{opt}=\min\left\{D,\frac{M}{2}\right\}\left(2M - 2\min\left\{D,\frac{M}{2}\right\} + 1\right),
\end{equation}
where $D=\min\{\sum_{k\in\mathcal{K}}N_k,N\}$ is the DoF of the multi-user MIMO channel. Two representative examples which can reach such condition are band- and stem-connected BD-RIS with band and stem width $q = 2\min\{D,\frac{M}{2}\} - 1$. Fig. \ref{fig:perato_sumrate} reports the Pareto frontier between performance and circuit complexity achieved by BD-RIS-aided multi-user MISO systems, indicating that band/stem-connected BD-RIS with proper band/stem width $q$ can reach the performance achieved by fully-connected BD-RIS, with much less circuit complexity.

\begin{figure}
    \centering
    \includegraphics[width=0.47\textwidth]{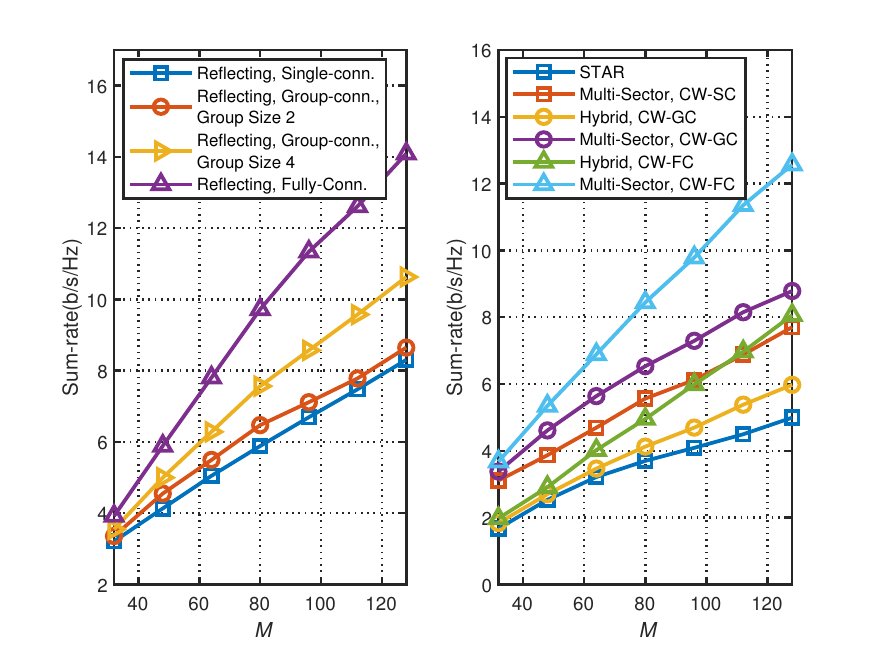}
    \caption{Sum-rate for a BD-RIS-aided multi-user MISO system versus the number of elements $M$. ``CW-FC'', ``CW-GC'', and ``CW-SC'', respectively, refer to cell-wise fully-, group-, and single-connected. Channels through BD-RIS follow Rician fading with a Rician factor 0 dB. The transmitter-user channels are fully blocked. The transmitter-RIS distance is set as 100 m and the RIS-user distance is set as 10 m. The group size for hybrid and multi-sector modes with CW-GC architectures are respectively set as 4 and 8. The number of sectors for the multi-sector mode is set as 4. $N=K=4$ with four users distributed evenly across the sectors covered by BD-RIS, that is, all users on the same side for reflecting mode, every two users on one side for hybrid mode, and one user within each sector for multi-sector mode.}
    \label{fig:hybrid_mode_results}\vspace{-0.2 cm}
\end{figure}


\vspace{-0.2 cm}

\subsection{Enabling Flexible Modes with Highly-Directional Wireless Coverage}

The inter-element connections in BD-RIS not only boost the channel strength and system performance, but also provide possibility to flexibly arrange the locations and orientations of elements. This leads to different modes with enlarged wireless coverage, such as hybrid mode proposed in \cite{li2022beyond} and multi-sector modes proposed in \cite{li2023beyond}. 
Results in Fig. \ref{fig:hybrid_mode_results} demonstrate that a reflecting BD-RIS can achieve up to 75\% of sum-rate improvement over D-RIS (single-conn.) when $M = 128$. More importantly, a 4-sector BD-RIS with cell-wise fully-connected architecture can increase the sum-rate by 150\% over STAR-RIS with $M = 128$, thanks to the higher antenna gains provided by more directional antennas.



\subsection{Providing Orders of Magnitude Gains in Distributed Deployments}

\begin{figure}
    \centering
    \includegraphics[width=0.45\textwidth]{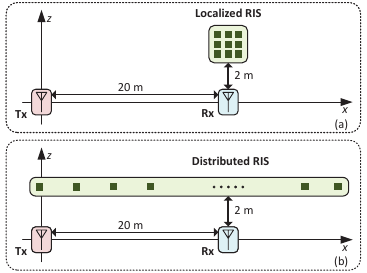}
    \caption{Illustration of (a) localized and (b) distributed RIS-aided wireless communication systems.}
    \label{fig:distributed_RIS}\vspace{-0.2 cm}
\end{figure}

In most existing literature, an RIS has been regarded as an antenna array whose inter-element spacing is comparable with (typically smaller than) the wavelength. This setting results in a localized RIS whose elements are localized in a specific site. The localized RIS has been widely studied in far-field scenarios, where the wireless channels related to each RIS element share approximately the same large-scale fading, as shown in Fig. \ref{fig:distributed_RIS}(a). In this sense, the gain of BD-RIS over D-RIS mainly comes from the better exploration of small-scale fading effects. 
To understand if the joint effects of small- and large-scale fading in wireless channels can provide additional flexibility to be captured by BD-RIS architectures, \cite{nerini2025localized} has proposed the concept of distributed RIS. In the distributed RIS, elements are distributed over a wide region with inter-element spacing much larger than the wavelength, such that the large-scale fading between transmitter/receiver and each RIS element can be quite different, as shown in Fig. \ref{fig:distributed_RIS}(b). 

\begin{figure}
    \centering
    \subfigure[Gain of distributed over localized BD-RIS $G^\mathsf{dis}$]{\includegraphics[width=0.43\textwidth]{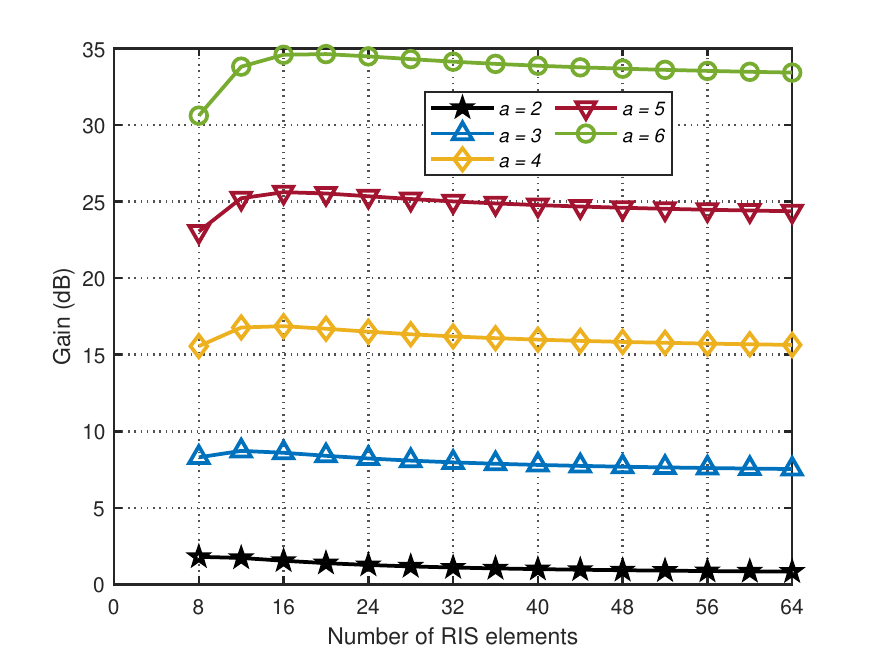}}
    \subfigure[Gain of distributed over localized BD-RIS $G^\mathsf{dis}$]{\includegraphics[width=0.43\textwidth]{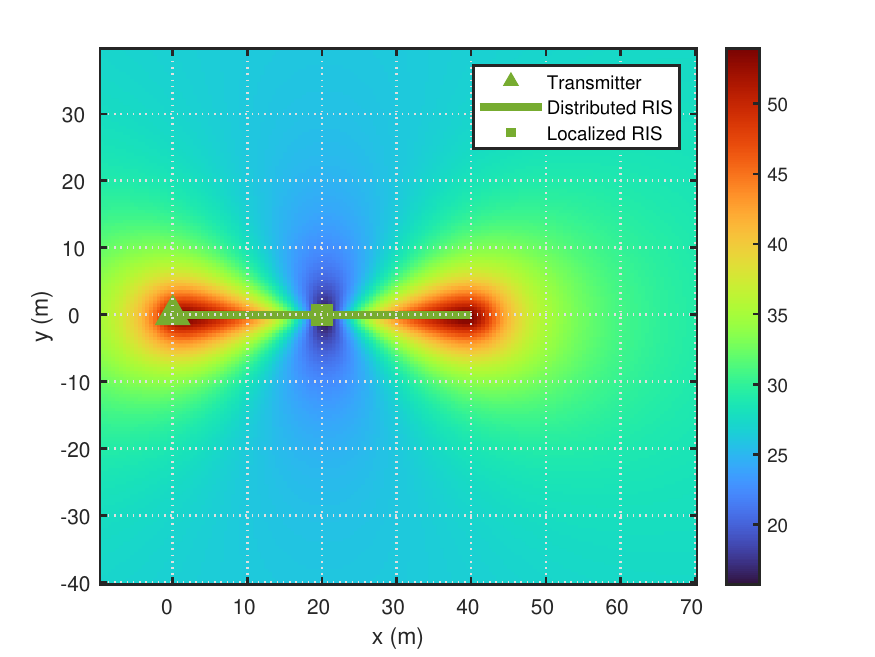}}
    \caption{$G^\mathsf{dis}$ (in dB) for different values of path-loss exponent, number of RIS elements, and locations of receiver \cite{nerini2025localized}. }
    \label{fig:results_distributed}\vspace{-0.4 cm}
\end{figure}

To quantify the benefit of distributed RIS, \cite{nerini2025localized} has derived the gain of distributed over localized fully-connected BD-RIS in SISO systems under Rayleigh fading channels, which is bounded by
\begin{subequations}\label{eq:gain_distributed}
    \begin{align}
        G^\mathsf{dis} &> \left(\frac{d_Rd_T}{\sqrt[a]{M^2}\min\{\mathbf{d}_R\}\min\{\mathbf{d}_T\}}\right)^a,\\
        G^\mathsf{dis} &< \left(\frac{d_Rd_T}{\min\{\mathbf{d}_R\}\min\{\mathbf{d}_T\}}\right)^a,
    \end{align}
\end{subequations}
where $d_R$ and $d_T$, respectively, denote the distance between the localized RIS and the receiver/transmitter; $\mathbf{d}_R\in\mathbb{R}^{M\times 1}$ and $\mathbf{d}_T\in\mathbb{R}^{M\times 1}$, respectively, denote the distance between all $M$ elements of the distributed RIS and the receiver/transmitter; $a$ denotes the path-loss exponent. The bounds in (\ref{eq:gain_distributed}) indicate that to guarantee $G^\mathsf{dis} >1$, a sufficient condition is to have $d_Rd_T > \sqrt[a]{M^2}\min\{\mathbf{d}_R\}\min\{\mathbf{d}_T\}$ and a necessary condition is to have $d_Rd_T > \min\{\mathbf{d}_R\}\min\{\mathbf{d}_T\}$, both of which are mild and can be easily achieved. 
This is also numerically evaluated in Fig. \ref{fig:results_distributed}(a), showing that $G^\mathsf{dis}$ grows exponentially with $a$ and reaches as high as several orders of magnitude. The impact of locations of the receiver is reported in Fig. \ref{fig:results_distributed}(b), showing that the distributed arrangement has the most significant performance benefit when the receiver is far from localized BD-RIS while being close to distributed BD-RIS. 

\subsection{Enabling Simultaneously Optimal Transmissions for Uplink and Downlink with Non-Reciprocal Architectures}

One key property of non-reciprocal BD-RIS is that its admittance/impedance/scattering matrices are asymmetric, which naturally break the uplink-downlink reciprocity of the wireless channels. For example, given an uplink BD-RIS-aided MISO channel $\mathbf{h}_\mathsf{up} = \mathbf{H}_{TI}\mathbf{\Theta}\mathbf{h}_{IR}$ as shown in (\ref{eq:uplink_channel}), 
the downlink channel $\mathbf{h}_\mathsf{down} = \mathbf{h}_{RI}\mathbf{\Theta}\mathbf{H}_{IT} = \mathbf{h}_{IR}^\mathsf{T}\mathbf{\Theta}\mathbf{H}_{TI}^\mathsf{T}$ is not equal to the transpose of the uplink channel, i.e., $\mathbf{h}_\mathsf{down}\ne\mathbf{h}_\mathsf{up}^\mathsf{T}$ since $\mathbf{\Theta} \ne \mathbf{\Theta}^\mathsf{T}$. 
This property provides unique benefits to scenarios where uplink and downlink transmissions behave differently, such as full-duplex systems \cite{zhang2016full}. 

\begin{figure}
    \centering
    \includegraphics[width=0.485\textwidth]{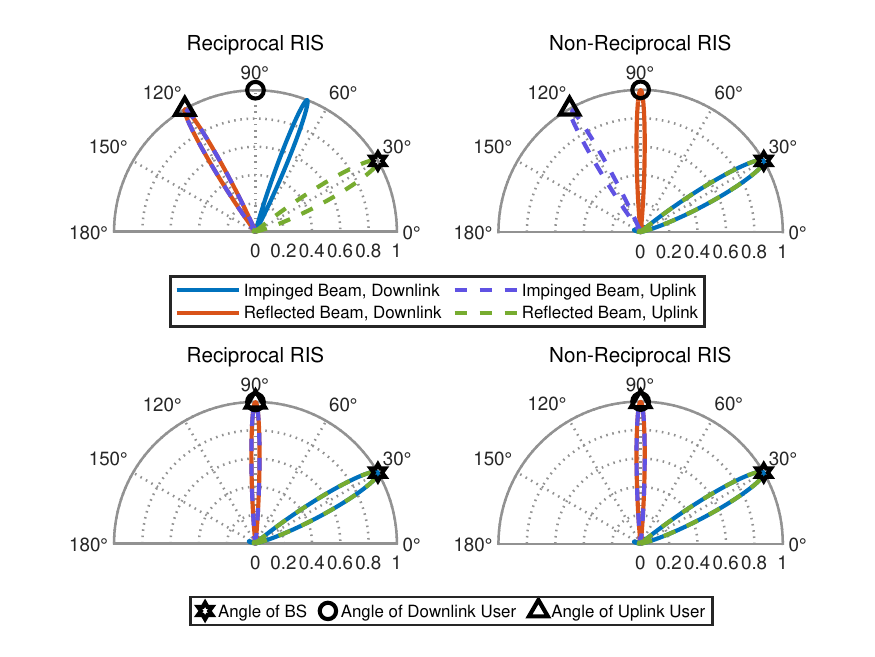}
    \caption{The impinging and reflected beam patterns of reciprocal and non-reciprocal BD-RISs. The channels between devices are assumed to be pure LoS. The base station (marked as ``BS'') is located at $\frac{\pi}{6}$; the downlink user is located at $\frac{\pi}{2}$; and the uplink user is located at $\frac{2\pi}{3}$ in the top figures and at $\frac{\pi}{2}$ in the bottom figures \cite{li2024non}.}
    \label{fig:results_full_duplex}\vspace{-0.2 cm}
\end{figure}

To visualize the benefit of applying non-reciprocal BD-RIS in full-duplex systems, \cite{li2024non} has derived the general RIS-aided full-duplex system model, and the theoretical conditions for non-reciprocal BD-RIS to simultaneously maximize the received powers of the signal of interest in the uplink and downlink. Results in the top two figures of Fig. \ref{fig:results_full_duplex} show that, when the uplink and downlink users are not aligned, both the impinging and reflected beams of non-reciprocal BD-RIS can exactly point to the directions of the signal of interest with the maximum power, while those of reciprocal BD-RIS fail to point to the downlink user. Meanwhile, the bottom two figures in Fig. \ref{fig:results_full_duplex} show that, only when the uplink and downlink users are aligned, both the impinging and reflected beams of reciprocal and non-reciprocal BD-RIS can exactly point to the directions of signal of interest. Due to the unique property of non-reciprocal BD-RIS, if a wave hits the non-reciprocal BD-RIS from one direction, the surface behaves differently than if it hits from the opposite direction. This finally enables an uplink user and a downlink user at different locations to optimally communicate with the same full-duplex base station. While \cite{li2024non} primarily focuses on a single-antenna scenario, the benefit of non-reciprocal BD-RIS in full-duplex systems has been recently shown in more general multi-user multi-antenna scenarios \cite{liu2024non}.

\begin{figure}
    \centering
    \includegraphics[width=0.45\textwidth]{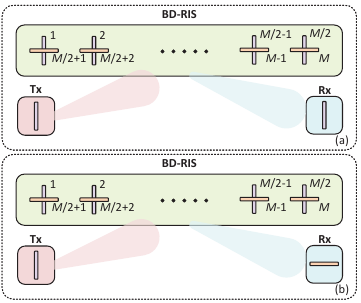}
    \caption{Dual-polarized BD-RIS-aided system where transmitter and receiver (a) have the same polarization and (b) have the opposite polarization.}
    \label{fig:dual_polarized}\vspace{-0.2 cm}
\end{figure}

\subsection{Providing Enhanced Gains in Dual-Polarized Systems}
Existing BD-RIS literature has focused on uni-polarized systems for the ease of analysis. However, modern MIMO systems use dual-polarized antenna arrays to have more antennas within limited space to provide more beamforming gains \cite{kim2010limited}. 
Dual-polarized D-RIS-aided systems have been studied in \cite{han2021dual,zheng2024ris}, while it remains unknown if BD-RIS still has performance gains over D-RIS in dual-polarized systems.
Taking into account this practical consideration, \cite{nerini2025dual} has modeled and analyzed a dual-polorized BD-RIS-aided SISO system, where the BD-RIS has half vertically polarized elements and half horizontally polarized elements, and the receiving and transmitting antennas can have the same or opposite polarizations, as illustrated in Fig. \ref{fig:dual_polarized}. 

\begin{figure}
    \centering
    \includegraphics[width=0.45\textwidth]{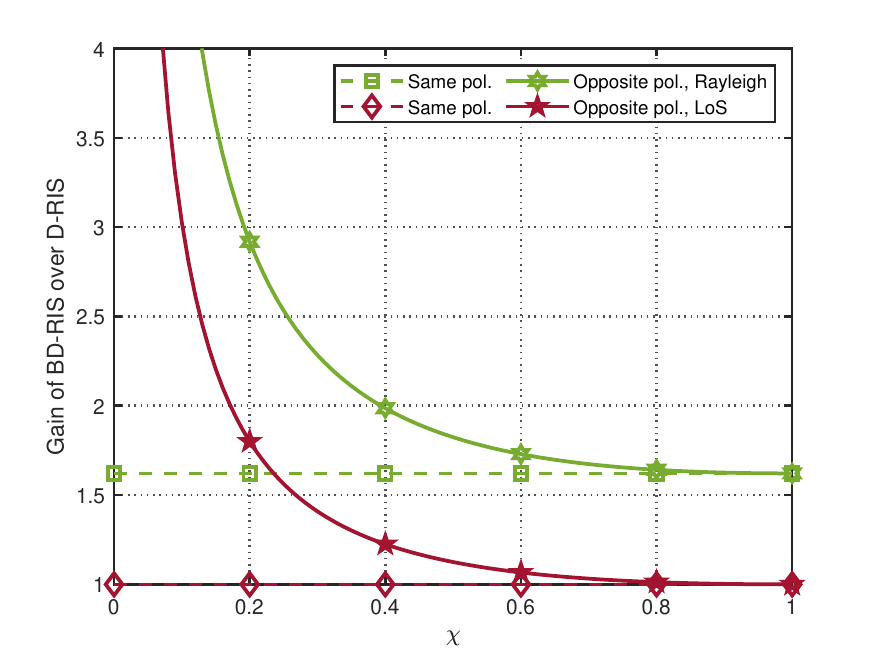}
    \caption{Gain of BD-RIS over D-RIS $\lim_{M\rightarrow\infty}G^\mathsf{dual-polarized}$ as a function of $\chi$ when the transmitter and receiver have the same/opposite polarization and with Rayleigh/LoS channels \cite{nerini2025dual}.}
    \label{fig:results_dualpol}\vspace{-0.2 cm}
\end{figure}

When the transmitter and receiver have the same polarizations, as illustrated in Fig. \ref{fig:dual_polarized}(a), the limit of the power gain of dual-polarized fully-connected BD-RIS over D-RIS remains the same as the uni-polarized case \cite{shen2021}. That is, dual-polarized BD-RIS can increase the received power in SISO by up to 62\% compared to dual-polarized D-RIS. 
When the transmitter and receiver have opposite polarizations, the limit of the power gain of dual-polarized fully-connected BD-RIS over D-RIS varies with channel fadings, and is expressed as \cite{nerini2025dual}
\begin{equation}
    \lim_{M\rightarrow\infty}G^\mathsf{dual-polarized} = \begin{cases}
        \frac{4(1 + \chi)^2}{\pi^2\chi}, &\text{Rayleigh},\\
        \frac{(1+\chi)^2}{4\chi}, &\text{LoS},
    \end{cases}
\end{equation}
where $0<\chi<1$ denotes the inverse of the cross-polar discrimination. This analytical result indicates that BD-RIS can offer a gain $G^\mathsf{dual-polarized} >1$ for both Rayleigh fading and LoS channels. To visualize the impact of $\chi$ on performance gain, Fig. \ref{fig:results_dualpol} illustrates the gain of BD-RIS over D-RIS as a function of $\chi$. These analytical results show that BD-RIS has more significant benefits over D-RIS with a smaller $\chi$, as small values of $\chi$ increase disparity between the channel entries.

\section{BD-RIS with Hardware Impairments}
\label{sec:hardware}

The discussions on BD-RIS so far have mainly focused on idealized and simplified hardware, such as lossless reconfigurable impedance networks with continuously tunable admittance components, with no frequency dependence, and with no mutual coupling between elements. These idealized assumptions help to understand the fundamentals and study the performance limit of BD-RIS. However, they cannot be perfectly achieved in real-world implementations. In this section, we thus briefly revisit some important hardware impairments of BD-RIS and their impacts on system performance. Similar to Section \ref{subsec:estimation} where we discussed channel estimation errors alone assuming perfect hardware at BD-RIS, below we will also show the impact of each hardware impairment assuming perfect other factors to guide readers from scratch. Analyzing real-world BD-RIS-aided scenarios by jointly considering multiple practical factors can be a meaningful future direction.

\subsection{Discrete-Value Impedance and Admittance}

\begin{figure}
    \centering
    \includegraphics[width=0.45\textwidth]{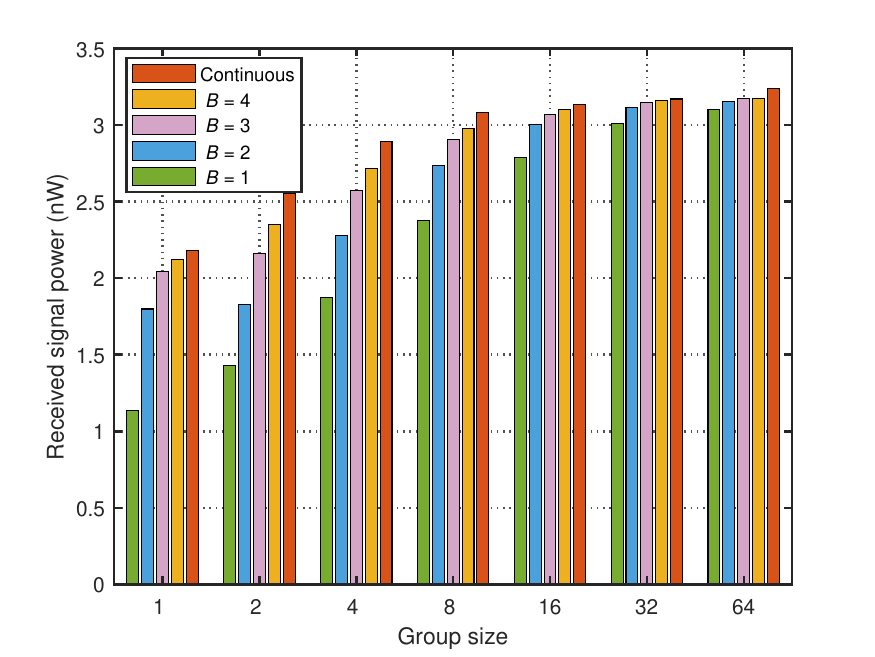}
    \caption{Average received signal power versus the group size. The transmitter, RIS, and receiver are located at a 3D coordinate system with respectively (5,-250,25), (0,0,5), and (5,5,1.5) in meters (m). The channels from the transmitter to receiver and through RIS have i.i.d. Rayleigh fading. ($N = 4$, $N_\mathsf{r}=2$, $M = 64$) \cite{nerini2023discrete}.}
    \label{fig:results_discrete}
\end{figure}

In Sections \ref{sec:model} and \ref{sec:signal_process}, the reconfigurable impedance network of BD-RIS is assumed to have continuous-value admittance and scattering matrices for the ease of modeling, optimization and performance analysis, as well as channel estimation. Nevertheless, admittance components tunable with finer resolution require a much more complex circuit design, especially for BD-RIS architectures with sophisticated interconnections. In D-RIS, the modeling of discrete-value scattering matrices can be very simple by uniformly quantifying the phase of each reflection coefficient. Specifically, for D-RIS with a diagonal scattering matrix $\mathbf{\Theta} = \mathsf{diag}(\varTheta_1,\ldots,\varTheta_M)$, $|\varTheta_m|=1$, its discrete-value set is expressed as a codebook
\begin{equation}
	\varTheta_m \in\Big\{e^{\jmath\frac{2\pi}{2^B}b}~|~b\in\mathcal{B}=\{0,1,2^B-1\}\Big\}, \forall m\in\mathcal{M}, \label{eq:conv_dis}
\end{equation}
where $B$ denotes the number of resolution bits. In D-RIS literature, typical beamforming and channel estimation solutions are to quantize continuous-value results to (\ref{eq:conv_dis}), or to directly design each scattering coefficient by selecting from (\ref{eq:conv_dis}). 
This, however, does not work for BD-RIS with scattering matrices whose entries are dependent on each other. To model the discrete-value BD-RIS, \cite{nerini2023discrete} has focused on the lossless fully/group-connected architectures, and, for the first time, proposed to model the discrete-value reactance matrix $\mathbf{X}_I = \Im\{\mathbf{Z}_I\} = \mathsf{blkdiag}(\mathbf{X}_{I,1},\ldots,\mathbf{X}_{I,g})$, where $\mathbf{X}_{I,g}\in\mathbb{R}^{\bar{M}\times\bar{M}}$, $\mathbf{X}_{I,g} = \mathbf{X}_{I,g}^\mathsf{T}$ denotes the reactance matrix for group $g$ of the group-connected architecture. Then, each entry of $\mathbf{X}_{I,g}$ is constrained by\footnote{In D-RIS, the reactance matrix $\mathbf{X}_I$ boils down to a diagonal matrix and each diagonal entry $[\mathbf{X}_I]_{m,m}$ is linked to $\varTheta_m$ in $\mathbf{\Theta}$ as $\varTheta_m = \frac{\jmath[\mathbf{X}_I]_{m,m} + Z_0}{\jmath[\mathbf{X}_I]_{m,m} - Z_0}$, such that one can directly obtain the discrete-value phase shift $\angle\varTheta_m$ with a discrete-value reactance $[\mathbf{X}_I]_{m,m}$. }
\begin{equation}
	[\mathbf{X}_{I,g}]_{m,m'}\in\{\pm X_{I,b}~|~b \in\mathcal{B}\},\forall g\in\mathcal{G},\forall m,m'\in\bar{\mathcal{M}},\label{eq:dis}
\end{equation}
where $X_{I,b}>0$, $\forall b\in\mathcal{B}$.
Comparing (\ref{eq:conv_dis}) and (\ref{eq:dis}), we observe that the codebook for D-RIS is known based on a finite interval, while that for BD-RIS contains unknown candidates $\pm X_{I,b}$. Therefore, the design of discrete-value BD-RIS includes an additional step, that is to determine the codebook first. To effectively solve this problem, \cite{nerini2023discrete} has proposed an offline learning method to obtain the codebook, which relies on a training set including sufficient channel realizations. With the determined codebook, the online deployment based on instantaneous CSI is conducted by iteratively designing individual reactance $[\mathbf{X}_{I,g}]_{m,m'}$ through a one-dimensional search.
Based on the above design, results in Fig. \ref{fig:results_discrete} show that one resolution bit is sufficient in fully-connected BD-RIS to achieve satisfactory performance close to the continuous-value case. Such a mild requirement of the resolution bits is beneficial for the practical implementation of BD-RIS. 

The design in \cite{nerini2023discrete} is also applicable to other reciprocal architectures. To have a unified framework of the discrete-value BD-RIS design, it is suggested to discretize the nonzero entries of admittance matrices as explained in Remarks 4 and 5. That is, given a lossless and reciprocal reconfigurable impedance network with admittance matrix $\mathbf{Y}_I$, each nonzero entry of the corresponding susceptance matrix $\mathbf{B}_I = \Im\{\mathbf{Y}_I\}\in\mathbb{R}^{M\times M}$ is constrained by 
\begin{equation}
	\left[\mathbf{B}_{I}\right]_{m,m'}\in\mathcal{C}=\{\pm B_{I,b}~|~b\in\mathcal{B}\},
\end{equation} 
for $[\mathbf{B}_I]_{m,m'}\ne 0, m,m'\in\mathcal{M}$, where $B_{I,b}>0$, $\forall b\in\mathcal{B}$.
The offline codebook design problem aiming at maximizing the average strength of a MIMO channel is formulated as 
\begin{equation}\label{prob:codebook}
	\begin{aligned}
		\max_{\mathcal{C}}~ &\mathbb{E}\{\|\mathbf{H}(\mathbf{\Theta})\|_\mathsf{F}^2\}\\
		\mathrm{s.t.}~~&\mathbf{\Theta} = (Y_0\mathbf{I}_M + \mathbf{Y}_I)^{-1}(Y_0\mathbf{I}_M - \mathbf{Y}_I),\\
		&\mathbf{Y}_I\in\mathcal{Y}, \mathbf{B}_I = \Im\{\mathbf{Y}_I\}, \\
		&\left[\mathbf{B}_{I}\right]_{m,m'}\in\mathcal{C},~
		\forall~[\mathbf{B}_I]_{m,m'}\ne 0, m,m'\in\mathcal{M},\\
		&B_{I,b}>0, \forall b\in\mathcal{B},
	\end{aligned}
\end{equation}
where $\mathbf{H}(\mathbf{\Theta}) = \mathbf{H}_{RT} + \mathbf{H}_{RI}\mathbf{\Theta}\mathbf{H}_{IT}$ according to Section \ref{subsec:beamforming_r}. The proposed offline learning method in \cite{nerini2023discrete} is readily adapted to BD-RIS architectures with different constraints $\mathcal{Y}$ as illustrated in Section \ref{subsec:architecture}.
Once the codebook design is obtained, the discrete-value BD-RIS can be designed based on iteratively exhaustive search as in \cite{nerini2023discrete}, or by directly quantizing the continuous-value solutions from Section \ref{subsec:beamforming_r}.

The discrete-value expression and design for non-reciprocal BD-RIS still remain unexplored. For non-reciprocal architectures, at the current stage there is no clear mapping between the circuit topology and its admittance/impedance matrices. Alternatively, it is possible to directly discretize its scattering matrix $\mathbf{\Theta}$. Specifically, given a non-reciprocal BD-RIS architecture constrained by $\mathcal{T}_\mathsf{non-recip}$, the discrete-value expression could be 
\begin{equation}
	\mathbf{\Theta}\in\mathcal{C}_\mathsf{non-recip} = \{\mathbf{\Phi}_b~|~\forall b\in\mathcal{B}\},
\end{equation}
where $\mathbf{\Phi}_b^\mathsf{H}\mathbf{\Phi}_b = \mathbf{I}_M$, $\forall b\in\mathcal{B}$. The simplest way is to randomly generate unitary matrices to construct $\mathcal{C}_\mathsf{non-recip}$, which may suffer significant performance loss. Alternatively, we can follow the idea for reciprocal architectures, and formulate the following offline codebook design problem
\begin{equation}\label{prob:codebook_nr}
	\begin{aligned}
		\max_{\mathcal{C}_\mathsf{non-recip}}~ &\mathbb{E}\{\|\mathbf{H}(\mathbf{\Theta})\|_\mathsf{F}^2\}\\
		\mathrm{s.t.}~~&\mathbf{\Theta} \in\mathcal{C}_\mathsf{non-recip},\\
		&\mathbf{\Phi}_b^\mathsf{H}\mathbf{\Phi}_b = \mathbf{I}_M, \forall b\in\mathcal{B}.
	\end{aligned}
\end{equation}
There has not been existing literature solving exactly the problem (\ref{prob:codebook_nr}), while it can be viewed as a vector quantization problem and potentially solved by the Lloyd algorithm \cite{sabin1986global}. The detailed design will be left as an interesting future work.

\subsection{Lossy Interconnections and Admittance Components}

Power loss is a very important factor in hardware devices. Modeling the hardware without capturing accurately the loss could cause non-negligible performance degradation. In D-RIS, the power loss comes from the circuit of each reconfigurable component \cite{abeywickrama2020intelligent}. However, in BD-RIS architectures, the power loss comes from two aspects: 1) the interconnection loss between ports of the reconfigurable impedance network and 2) the loss from the circuit of reconfigurable admittance components. 
These two kinds of losses have been recently modeled and analyzed, respectively in \cite{nerini2025localized} and \cite{peng2025lossy}, both focusing on reciprocal BD-RIS architectures. Below we will briefly revisit how to capture these two kinds of losses in BD-RIS architectures. 

\subsubsection{Lossy Interconnections}

\begin{figure}
	\centering 
	\includegraphics[width=0.45\textwidth]{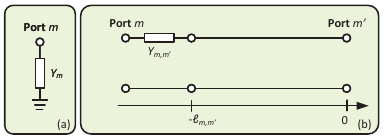}
	\caption{Illustration of lossy interconnections. (a) Port $m$ connected to ground through $Y_m$ and (b) port $m$ connected to port $m'$ through $Y_{m,m'}$ and a transmission line of length $\ell_{m,m'}$.}
	\label{fig:lossy_interconn}
\end{figure}

\begin{figure}[t]
    \centering
    \includegraphics[width=0.45\textwidth]{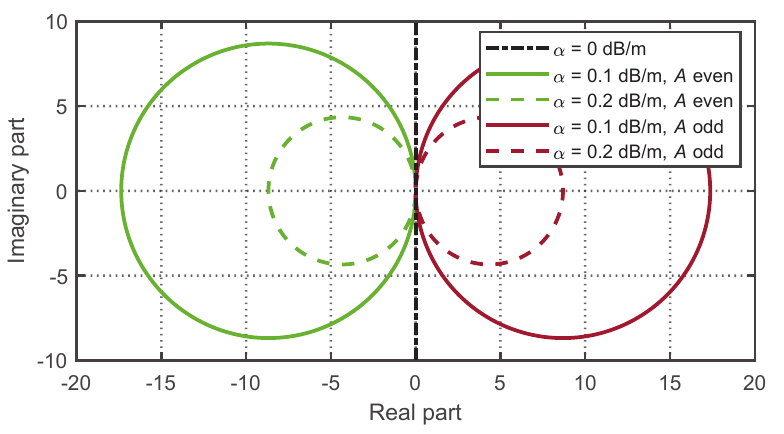}
    \caption{Values of $[\mathbf{Y}_I]_{m,m'}$ modeled as (\ref{eq:lossy_interconn_simplified}) with $Z_0 = 50\Omega$ and $\ell_{m,m'}=0.1$ m \cite{nerini2025localized}.}\vspace{-0.3 cm}
    \label{fig:results_lossy_ynm}
\end{figure}

\begin{figure}[t]
    \centering
    \includegraphics[width=0.45\textwidth]{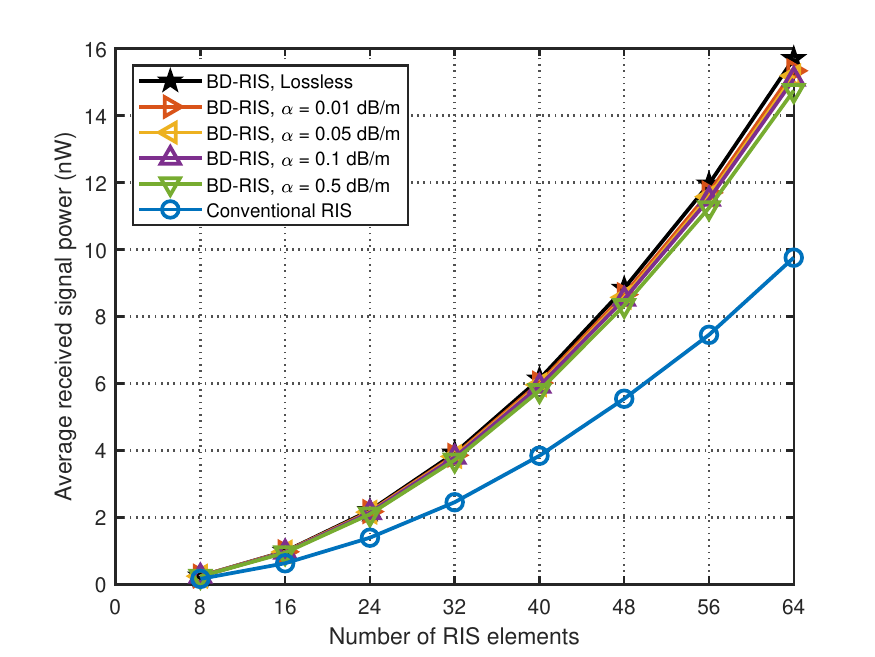}
    \caption{Average received signal power versus the number of RIS elements with different values of $\alpha$. The transmitter, RIS, and, receiver are located at a 3D coordinate system with respectively (0,0,0), (20,0,2), and (20,0,0) in meters (m). The inter-element spacing in BD-RIS is set as 0.05 m. The direct transmitter-receiver channel is assumed to be fully obstructed. Channels through RIS follow i.i.d. Rayleigh fading ($N = N_\mathsf{r}=1$) \cite{nerini2025localized}. }\vspace{-0.3 cm}
    \label{fig:results_lossy_pow}
\end{figure}

The reciprocal BD-RIS with lossy interconnections is modeled as an $M$-port reconfigurable impedance network, characterized by its admittance matrix $\mathbf{Y}_I\in\mathbb{C}^{M\times M}$. Similar to the model in Section \ref{subsec:model}, each port $m$ is connected to a reconfigurable admittance component $Y_m$ to ground, as illustrated in Fig. \ref{fig:lossy_interconn}(a). The main difference is port $m$ and port $m'$, $m'\ne m$, if interconnected, are interconnected via a reconfigurable admittance component $Y_{m,m'}$ in series with a transmission line with length $\ell_{m,m'}$, $\forall m,m'\in\mathcal{M}$, as illustrated in Fig. \ref{fig:lossy_interconn}(b). According to the derivation in \cite{nerini2025localized}, the entries in $\mathbf{Y}_I$ have expressions
\begin{equation}
	\label{eq:lossy_interconn}
	[\mathbf{Y}_I]_{m,m'} = \begin{cases}
		\frac{-2}{Y_{m,m'}^{-1}\zeta_{m,m'}^+ + Z_0\zeta _{m,m'}^-}, & m \ne m',\\
		Y_m - \sum_{n\ne m}\frac{\zeta_{m,m'}^+}{2}[\mathbf{Y}_I]_{m,n}, & m = m',
	\end{cases}
\end{equation}
where $\zeta_{m,m'}^+ = e^{\zeta\ell_{m,m'}} + e^{-\zeta\ell_{m,m'}}$ and $\beta_{m,m'}^- = e^{\zeta\ell_{m,m'}} - e^{-\zeta\ell_{m,m'}}$ with $\zeta\in\mathbb{C}$ being the propagation constant of the transmission line. Specifically, in $\zeta = \alpha + \jmath\beta$, the real part $\alpha$ denotes the attenuation constant\footnote{The value of $\alpha$ depends on the transmission line parameters, including series resistance and inductance, and shunt conductance and capacitance per unit length of the lumped-element equivalent circuit \cite{pozar2021microwave}.} and the imaginary part $\beta$ denotes the phase constant. 
It is difficult to gain insights from (\ref{eq:lossy_interconn}) given its complex expressions. To understand how the lossy interconnections in BD-RIS impact the feasible range of entries in $\mathbf{Y}_I$, \cite{nerini2025localized} has simplified (\ref{eq:lossy_interconn}) by assuming a special case where $\ell_{m,m'}$ is a multiple of half of wavelength $\lambda$, $\lambda = \frac{2\pi}{\beta}$. That is, $\ell_{m,m'} = \frac{\pi}{\beta}A_{m,m'}$, where $A_{m,m'}\in\mathbb{Z}$. In addition, to focus purely on the impact of lossy interconnections, each reconfigurable admittance component is assumed to be lossless, i.e., $\Re\{Y_m\} = 0$, $\Re\{Y_{m,m'}\} = 0$, $\forall m,m'\in\mathcal{M}$. Then, the constraint of $[\mathbf{Y}_I]_{m,m'}$ with $m\ne m'$ can be simplified to 
\begin{equation}
	[\mathbf{Y}_I]_{m,m'} = \frac{-(-1)^{A_{m,m'}}}{\underbrace{Y_{m,m'}^{-1}\cosh(\alpha\ell_{m,m'})}_{\text{purely imaginary}}+\underbrace{Z_0\sinh(\alpha\ell_{m,m'})}_{\text{purely real}}},\label{eq:lossy_interconn_simplified}
\end{equation}
indicating that $[\mathbf{Y}_I]_{m,m'}$, if is not zero, must have a non-zero real part resulting in power loss. More interestingly, the existence of lossy interconnections makes the value of $[\mathbf{Y}_I]_{m,m'}$ being constrained on a circle in the complex plane. An extreme case ($\alpha = 0$) that the real parts of $[\mathbf{Y}_I]_{m,m'}$ are zeros corresponds to BD-RIS with lossless interconnections. The constraint (\ref{eq:lossy_interconn_simplified}) referring to BD-RIS with lossy interconnections and lossless BD-RIS can be visualized in Fig. \ref{fig:results_lossy_ynm}. From Fig. \ref{fig:results_lossy_ynm} we observe that the real and imaginary parts of $[\mathbf{Y}_I]_{m,m'}$ are dependent on each other in lossy cases and the possible value of $[\mathbf{Y}_I]_{m,m'}$ depends on the value of $\alpha$, while the imaginary part of $[\mathbf{Y}_I]_{m,m'}$ for lossless cases can take arbitrary values, consistent with existing BD-RIS literature \cite{shen2021,nerini2024beyond,wu2024optimization}. The impact of lossy interconnections in BD-RIS to system performance has also been shown in Fig. \ref{fig:results_lossy_pow}, from which we observe that the impact of lossy interconnections in BD-RIS is negligible due to the small inter-element spacing. 

\begin{figure}[t]
    \centering
    \includegraphics[width=0.45\textwidth]{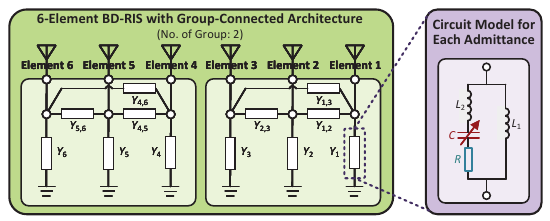}
    \caption{An example of a 6-element BD-RIS with group-connected architectures and the equivalent lossy circuit for each admittance component.}
    \label{fig:admittance_loss}\vspace{-0.3 cm}
\end{figure}

\subsubsection{Lossy Admittance Components}
\begin{figure}
    \centering
    \includegraphics[width=0.45\textwidth]{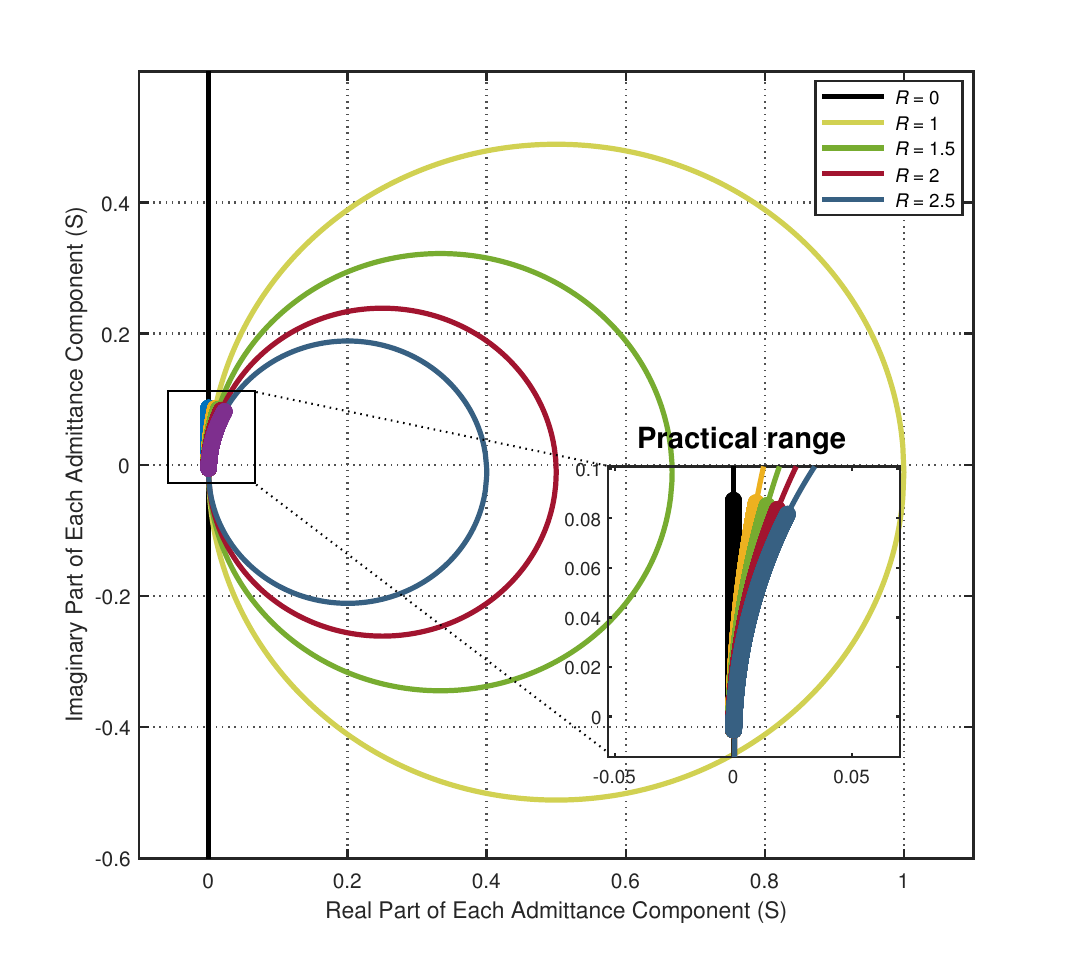}
    \caption{Values of $Y_{m,m'}$ modeled as (\ref{eq:loss}). The frequency is set as $f_\mathsf{c} = 2.4$ GHz. $L_1 = 6$ nH, $L_2 = 0.7$ nH. Circles represent all possible values of $Y_{m,m'}$ for varing $R$, while the practical range is constrained by tha value of capacitance $C\in [0.35,3.2]$ pF \cite{peng2025lossy}.}
    \label{fig:circle}
\end{figure}

\begin{figure}
    \centering
    \includegraphics[width=0.45\textwidth]{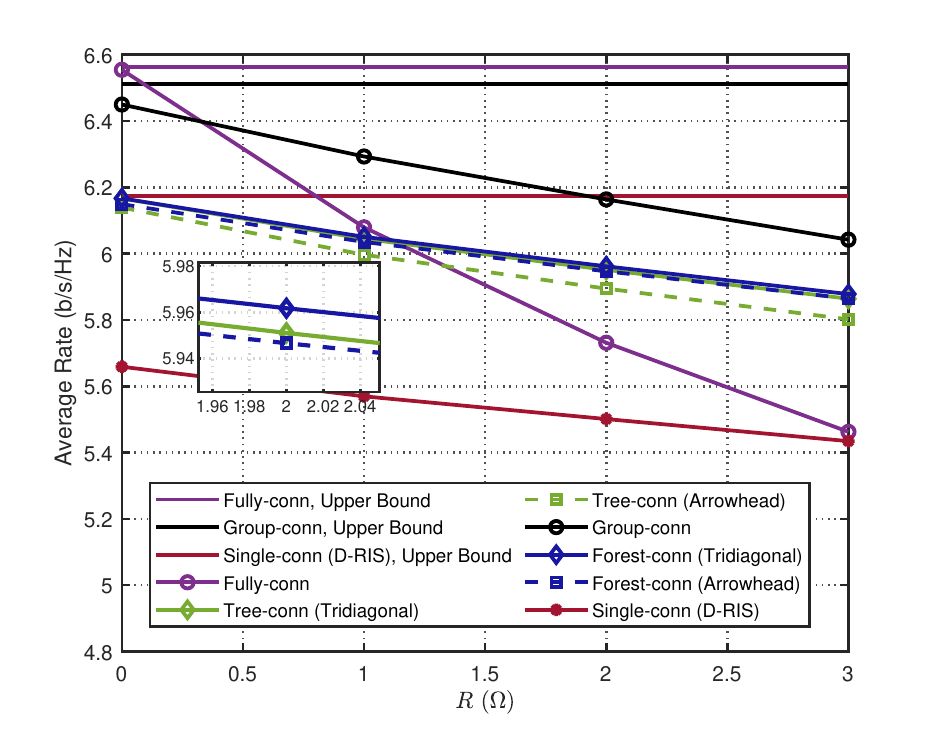}
    \caption{Average rate versus $R$ for a lossy BD-RIS-aided SISO system. The channels through BD-RIS follow Rician fading with a Rician factor 2 dB, and the direct channel from the transmitter to the receiver follows Rayleigh fading. The transmitter-receiver, transmitter-RIS, and RIS-receiver distances are respectively set as 52 m, 50 m, and 2.5 m ($M=30$, $\bar{M} = 6$ for the group/forest-connected architecture, $P=20$ dBm) \cite{peng2025lossy}.}
    \label{fig:results_lossy_BD_RIS}
\end{figure}

In D-RIS, the power loss of each reconfigurable component can be individually characterized in the amplitude of its reflection coefficients \cite{abeywickrama2020intelligent}. That is, for a lossy D-RIS, each entry $\varTheta_m$ in the scattering matrix has $|\varTheta_m| < 1$. The smaller the value of $|\varTheta_m|$, the larger the loss in D-RIS. 
However, it is not possible to model individually the entries in the scattering matrix $\mathbf{\Theta}$ of BD-RIS since they are coupled with each other due to interconnections. To tackle this difficulty, \cite{peng2025lossy} has recently proposed to directly analyze the loss of each reconfigurable admittance component, and the scattering matrix for lossy BD-RIS can be naturally obtained by (\ref{eq:theta_zy}) and (\ref{eq:mapping_y}). This is done by modeling each reconfigurable admittance component
as a lumped circuit consisting of inductors $L_1$, $L_2$, a tunable capacitance $C$, and an equivalent resistor $R$ that accounts for the parasitic resistance of the varactor, as illustrated in Fig. \ref{fig:admittance_loss}.
The admittance (if is not zero), with a given frequency of signals $\omega_\mathsf{c} = 2\pi f_\mathsf{c}$, is a function of the tunable capacitance $C$ given by 
\begin{equation}
    \begin{aligned}
        Y_{m,m'}(C) &= \frac{1}{\jmath\omega_\mathsf{c} L_1} + \frac{1}{\jmath\omega_\mathsf{c} L_2 + \frac{1}{\jmath\omega_\mathsf{c} C} + R}, \\
        &= \jmath\left(-\frac{1}{\omega_\mathsf{c} L_1} + \frac{-\omega_\mathsf{c}L_2 + \frac{1}{\omega_\mathsf{c} C}}{R^2 + \big(\omega_\mathsf{c}L_2 - \frac{1}{\omega_\mathsf{c}C}\big)^2}\right)\\
        & ~~ + \frac{R}{R^2+\big(\omega_\mathsf{c}L_2 - \frac{1}{\omega_\mathsf{c}C}\big)^2}, \forall m,m'\in\mathcal{M},
    \end{aligned}\label{eq:loss}
\end{equation}
where $Y_{m,m} = Y_m$. 
Specifically, the value of $R$ reflects the amount of power loss in each reconfigurable component and $R \ne 0$ in practical devices. 
Interestingly, the existence of losses in each reconfigurable admittance component makes the real part $\Re\{Y_{m,m'}\}$ and the imaginary part $\Im\{Y_{m,m'}\}$ being constrained by a circle in the complex plane, given by 
\begin{equation}
\Big(\Re\{Y_{m,m'}\} - \frac{1}{2R}\Big)^2 + \Big(\Im\{Y_{m,m'}\} - \Big(-\frac{1}{\omega_\mathsf{c} L_1}\Big)\Big)^2 = \Big(\frac{1}{2R}\Big)^2.
\end{equation}
An extreme case ($R=0$) that the real parts of $Y_{m,m'}$ are zeros corresponds to BD-RIS with lossless reconfigurable admittance components, implying that the feasible range of $Y_{m,m'}$ lies in the y-axis of the complex plane. This phenomenon can be visualized in Fig. \ref{fig:circle} based on a practical varactor SMV2020-079. The impact of lossy admittance components in BD-RIS to system performance is also shown in Fig. \ref{fig:results_lossy_BD_RIS}, from which we observe that the optimum of fully/tree-connected architectures in SISO systems can be destroyed by significant power losses from a large number of lossy reconfigurable admittance components. 

The above modeling of lossy interconnections and admittance components is applicable to all reciprocal BD-RIS architectures as illustrated in Section \ref{subsec:architecture}. Given a specific circuit topology, one can simply set $Y_{m,m'}=0$ to indicate that there are no direct interconnections between port $m$ and $m'$, and adopt (\ref{eq:lossy_interconn}) or (\ref{eq:lossy_interconn_simplified}) to construct the admittance matrix with lossy interconnections, and to adopt (\ref{eq:loss}) and (\ref{eq:mapping_y}) to construct the admittance matrix with lossy reconfigurable admittance components. However, the joint consideration of the impact of two kinds of losses to system performance remains unexplored and is a meaningful future research direction.

\subsection{Wideband Effect}

Frequency dependence is an intrinsic phenomenon in circuits, where the response of inductance and capacitance varies with the frequency of an applied signal. Therefore, the response of RIS, i.e., its impedance/admittance/scattering matrices of the reconfigurable impedance network, is naturally frequency dependent \cite{hu2023wideband}. In narrowband communications, since the bandwidth of signals is much less than the central frequency, it is reasonable to approximately assume the response of RIS to be frequency independent. However, when it comes to wideband communications, the frequency dependence in RIS is not negligible. To show how significant is the frequency dependence in RIS-aided wideband communication systems, there have been a few representative works on the wideband modeling, optimization, and implementation of D-RIS \cite{li2021intelligent,zhang2021joint,wang2024wideband}.
Nevertheless, these works model the wideband effect of D-RIS by characterizing the frequency dependence of the individual entries of its scattering matrix. This does not work for BD-RIS with interconnected elements, and thus coupled entries in its scattering matrix, making it difficult to accurately yet efficiently capture the frequency dependence in BD-RIS. 
In addition, there have been a few works on applying BD-RIS in wideband systems \cite{soleymani2024maximizing,demir2024wideband}, while they primarily assume a frequency-independent BD-RIS model, which is essentially not consistent with the behavior of BD-RIS in wideband scenarios. 
To establish a physically consistent BD-RIS model suitable for wideband systems, \cite{li2025beyond} and \cite{de2024beyond} have proposed to model the frequency dependence of the individual reconfigurable admittance component based on specific circuit designs. Below, we will briefly revisit how to model the frequency dependence in BD-RIS. 


\begin{figure}
    \centering
    \subfigure[Susceptance versus frequency]{\includegraphics[width=0.24\textwidth]{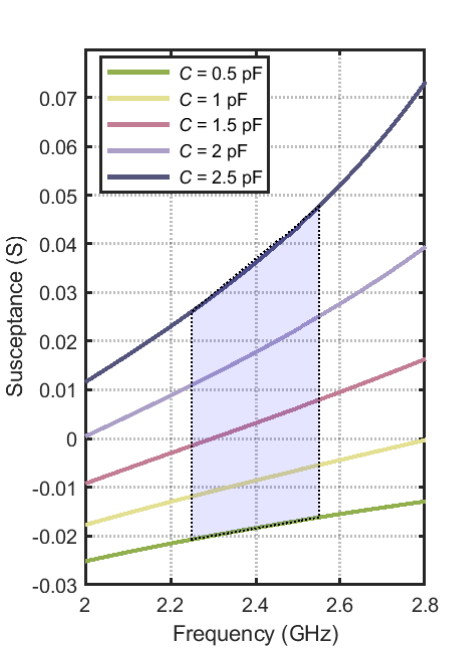}}
    \subfigure[Susceptance variation]{\includegraphics[width=0.24\textwidth]{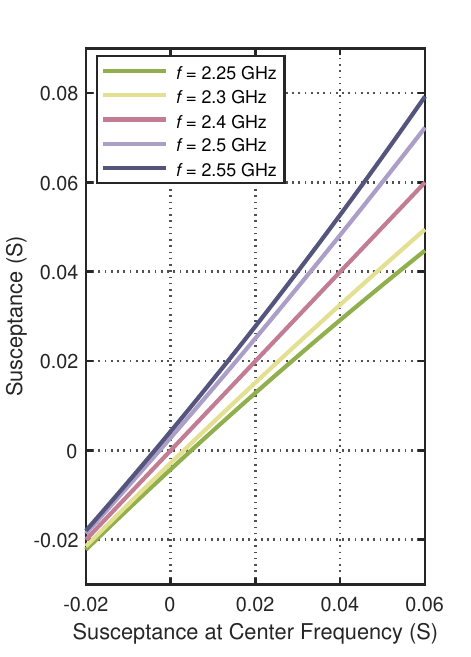}}
    \caption{The susceptance $\Im\{Y_{m,m'}\}$ as a function of (a) frequency and (b) the value of susceptance at the central frequency $f_\mathsf{c}=2.4$ GHz with $L_1 = 2.5$ nH, $L_2 = 0.7$ nH, and $C\in[0.2,3]$ pF for a practical varactor diode \cite{li2025beyond}.}
    \label{fig:susceptance_frequency}
\end{figure}

\begin{figure}[t]
    \centering 
    \subfigure[Group-connected BD-RIS with and without wideband modeling]{\includegraphics[width=0.43\textwidth]{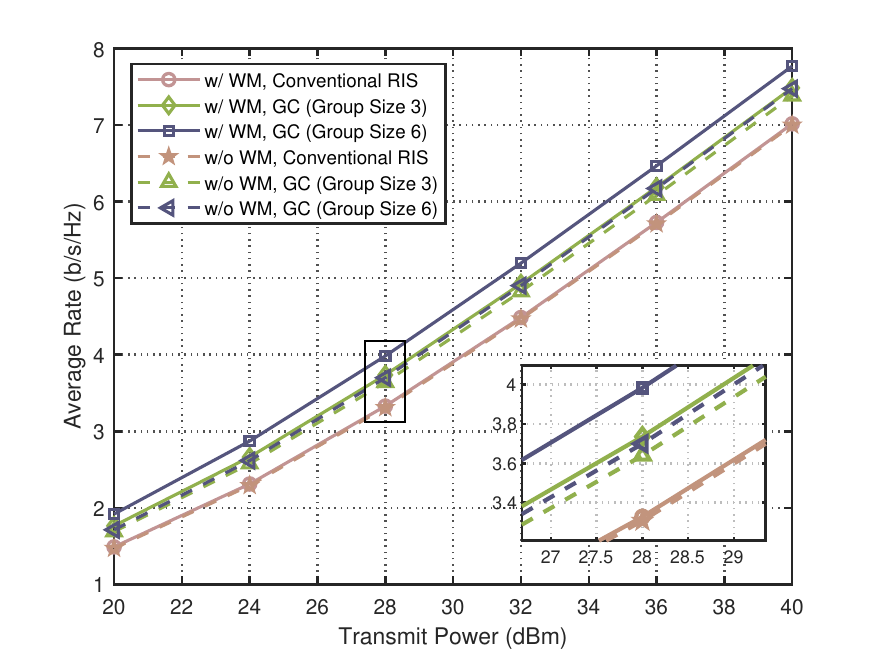}}
    \subfigure[Forest-connected BD-RIS with and without wideband modeling]{\includegraphics[width=0.43\textwidth]{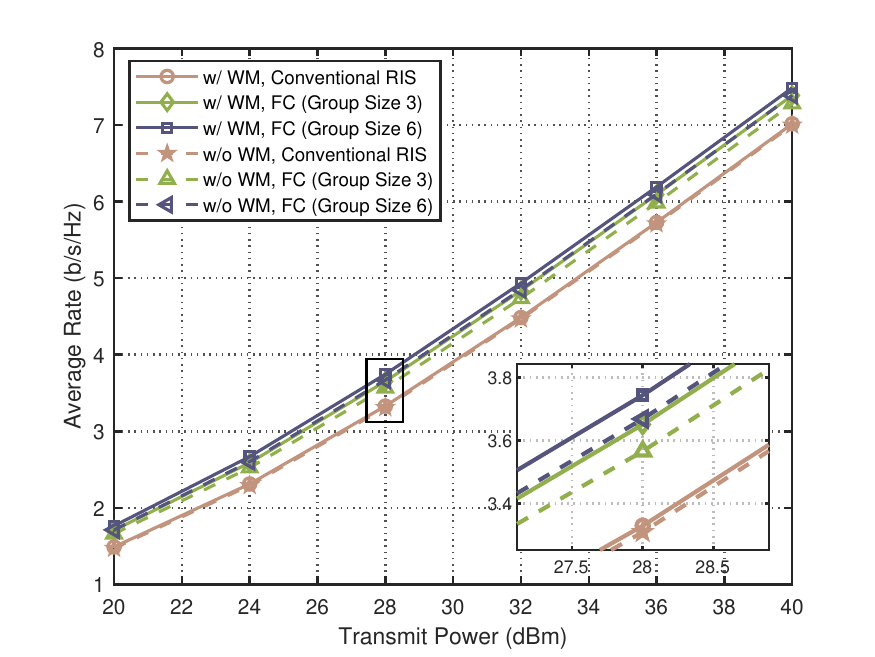}}
    \caption{Average rate versus transmit power $P$ with BD-RIS having different reciprocal architectures ($M=48$, $\bar{M}\in\{1,3,6\}$). The OFDM channels from the transmitter to the reciever and through BD-RIS are modeled as multi-tap finite duration impulse response sequences with i.i.d. complex Gaussian random variables. The legend ``WM'' is short for wideband modeling; ``GC'' is short for group-connected; ``FC'' is short for forest-connected. For both ``GC'' and ``FC'' architectures, the case of $\bar{M}=1$ refers to the D-RIS \cite{li2025beyond}.}
    \label{fig:average_rate}
\end{figure}

In reciprocal BD-RIS architectures, each reconfigurable admittance component can be modeled as a lumped circuit consisting of inductors $L_1$, $L_2$, and one tunable capacitor $C$\footnote{Here we consider a lossless model with $R=0$ in Fig. \ref{fig:admittance_loss} to focus purely on the modeling and analysis of frequency dependence in BD-RIS.}. Accordingly, the admittance of this circuit is a function of $C$ and the angular frequency $\omega=2\pi f$ for the incident signals:
\begin{equation}
Y_{m,m'}(C,\omega) = \frac{1}{\jmath\omega L_1} + \frac{1}{\jmath\omega L_2 + \frac{1}{\jmath\omega C}}, \forall m,m'\in\mathcal{M}, \label{eq:frequency}
\end{equation}
where $Y_{m,m}$ boils down to the admittance component $Y_m$ connected to ground.
We observe from (\ref{eq:frequency}) that, the values of $Y_{m,m'}$ at different frequencies are linked to each other by a common capacitor $C$. The relationship can be numerically illustrated in Fig. \ref{fig:susceptance_frequency}, based on a wideband system with central frequency $f_\mathsf{c}=2.4$ GHz and a practical varactor diode SMV1231-079 \cite{li2025beyond}. Results in Fig. \ref{fig:susceptance_frequency}(a) show that within some practical bandwidth for wideband signals, the susceptance $\Im\{Y_{m,m'}\}$ can be regarded as a linear function of frequency. This further implies that $\Im\{Y_{m,m'}\}$ varies approximately linearly with the susceptance at central frequency, as illustrated in Fig. \ref{fig:susceptance_frequency}(b). This motivates possible simplifications to (\ref{eq:frequency}) to benefit BD-RIS optimization. 
The impact of having frequency-dependent BD-RIS matrices on the performance of a SISO orthogonal frequency division multiplexing (OFDM) system is shown in Fig. \ref{fig:average_rate}. Results show that, with increasing circuit complexity (more admittance components) in BD-RIS architectures, an increasing performance gap appears between taking into account the wideband modeling when designing BD-RIS or not. This is attributed to the more significant variation of BD-RIS matrices between subcarriers, so ignoring this variation will cause non-negligible performance loss.

\begin{table*}[]
	\caption{BD-RIS Literature on Wideband Modeling and Optimization}
	\label{tab:wideband}
	\centering
	\begin{threeparttable}
		\begin{tabular}{|c|c|c|l|}
			\hline
			Ref. & Architecture$\dagger$ & Frequency Dependence & Highlights \\
			\hline\hline
			\cite{soleymani2024maximizing} & \begin{tabular}[c]{@{}c@{}}Group-Connected\\ (Multi-Sector Mode)\end{tabular} & \multirow{2}{*}{No} &  \begin{tabular}[l]{@{}l@{}} Optimize the spectral and energy efficiency of a multi-user MIMO\\ OFDM system aided by multi-sector BD-RIS \end{tabular}\\
			\cline{1-2}\cline{4-4}
			\cite{demir2024wideband} & Group-Connected &  & \begin{tabular}[l]{@{}l@{}} Derive the wideband system model aided by BD-RIS from time \\ domain and maximize the capacity of a MIMO-OFDM system\end{tabular} \\
			\hline
			\cite{li2025beyond,de2024beyond} & \begin{tabular}[c]{@{}c@{}}Proposed for Group- and Forest\\ -Connected but also applicable to\\ other Reciprocal Architectures\end{tabular} & \multirow{2}{*}{Yes} & \begin{tabular}[l]{@{}l@{}}Propose the wideband modeling of the individual reconfigurable \\ admittance component in BD-RIS based on lumped circuit models \end{tabular}\\
			\cline{1-2}\cline{4-4}
			\cite{katsanos2024multi} & Non-Diagonal & &\begin{tabular}[l]{@{}l@{}}Propose the wideband modeling of the individual phase shifter in \\non-diagonal RIS using scattering parameter analysis\end{tabular} \\
			\hline
		\end{tabular}
		\begin{tablenotes}
			\footnotesize
			\item[$\dagger$] The BD-RIS works on the reflecting mode unless otherwise stated.
		\end{tablenotes}
	\end{threeparttable}
\end{table*}

The above modeling is again applicable to all reciprocal BD-RIS architectures as illustrated in Section \ref{subsec:architecture}. Intuitively, the higher the circuit complexity (the larger the required number of admittance components) in BD-RIS architectures, the more significant the impact of wideband modeling at BD-RIS. In addition, \cite{katsanos2024multi} has proposed to model and analyze the frequency dependence of non-reciprocal BD-RIS with constraint $\mathcal{T}_{\mathsf{non-diag}}$, by individually modeling the frequency-dependent scattering parameter of each reconfigurable element. Nevertheless, due to the lack of a unified circuit topology illustration as in reciprocal architectures, the wideband modeling and optimization for more general non-reciprocal BD-RIS architectures still remain unexplored.  
For clarity, the BD-RIS literature on wideband modeling and optimization has been summarized in Table \ref{tab:wideband}. 

\vspace{-0.3 cm}

\subsection{Mutual Coupling Effect}

Mutual coupling, which refers to the electromagnetic interaction between antenna elements in an array, is an important factor in array design. Mutual coupling is in general inversely proportional to the spacing between elements: the smaller the inter-element spacing, the stronger the mutual coupling. In most existing RIS literature, the mutual coupling is ignored by assuming sufficiently large inter-element spacing. However, this strong assumption  cannot be easily achieved in practice, given that RIS usually consists of numerous densely spaced elements within a limited aperture to provide sufficient beamforming gain \cite{di2020smart,di2022communication}. There have been existing works on the modeling and analysis of the mutual coupling between D-RIS elements \cite{gradoni2021end,qian2021mutual,akrout2023physically}, which have been recently extended to BD-RIS-aided channels \cite{li2024beyond,semmler2024decoupling,nerini2024global}. Below, we will briefly revisit the modeling of mutual coupling aware BD-RIS-aided channels, and how the existence of mutual coupling impacts the system performance.

Consider a BD-RIS-aided MIMO system consisting of an $N$-antenna transmitter, an $M$-element BD-RIS, and an $N_\mathsf{r}$ receiver, as also illustrated in Section \ref{subsec:beamforming_r}. 
To capture explicitly the mutual coupling at BD-RIS, \cite{universal2024,li2024beyond,gradoni2021end,akrout2023physically} have made physics-consistent assumptions and derived the following three equivalent wireless channel models\footnote{Note that the scattering matrices $\mathbf{S}_{RT}$, $\mathbf{S}_{RI}$, and $\mathbf{S}_{IT}$ in (\ref{eq:channel_s}) are not one-to-one mapped to the channels $\mathbf{H}_{RT}$, $\mathbf{H}_{RI}$, and $\mathbf{H}_{IT}$ in the widely adopted model $\mathbf{H}(\mathbf{\Theta}) = \mathbf{H}_{RT} + \mathbf{H}_{RI}\mathbf{\Theta}\mathbf{H}_{IT}$. The detailed analysis can be found in \cite{universal2024,nossek2024physically,nerini2024physics}.}
\begin{subequations}\label{eq:channels}
    \begin{align}
        \mathbf{H}_\mathsf{S}(\mathbf{\Theta}) &= \mathbf{S}_{RT} + \mathbf{S}_{RI}(\mathbf{I}_M - \mathbf{\Theta}\mathbf{S}_{II})^{-1}\mathbf{\Theta}\mathbf{S}_{IT},
        \label{eq:channel_s}\\
        \mathbf{H}_\mathsf{Z}(\mathbf{Z}_I) &= \frac{1}{2Z_0}(\mathbf{Z}_{RT} - \mathbf{Z}_{RI}(\mathbf{Z}_{II} + \mathbf{Z}_I)^{-1}\mathbf{Z}_{IT}),\label{eq:channel_z}\\
        \mathbf{H}_\mathsf{Y}(\mathbf{Y}_I) & = \frac{1}{2Y_0}(-\mathbf{Y}_{RT} + \mathbf{Y}_{RI}(\mathbf{Y}_{II} + \mathbf{Y}_{I})^{-1}\mathbf{Y}_{IT}),
    \end{align}
\end{subequations}
where $\mathbf{S}_{RT}\in\mathbb{C}^{N_\mathsf{r}\times N}$, $\mathbf{Z}_{RT}\in\mathbb{C}^{N_\mathsf{r}\times N}$, and $\mathbf{Y}_{RT}\in\mathbb{C}^{N_\mathsf{r}\times N}$, respectively, denote the transmission scattering, impedance, and admittance matrices from the transmitter to receiver; $\mathbf{S}_{RI}\in\mathbb{C}^{N_\mathsf{r}\times M}$, $\mathbf{Z}_{RI}\in\mathbb{C}^{N_\mathsf{r}\times M}$, and $\mathbf{Y}_{RI}\in\mathbb{C}^{N_\mathsf{r}\times M}$, respectively, denote the transmission scattering, impedance, and admittance matrices from RIS to the receiver;
$\mathbf{S}_{IT}\in\mathbb{C}^{M \times N}$, $\mathbf{Z}_{IT}\in\mathbb{C}^{M\times N}$, and $\mathbf{Y}_{IT}\in\mathbb{C}^{M\times N}$, respectively, denote the transmission scattering, impedance, and admittance matrices from the transmitter to RIS; $\mathbf{S}_{II}\in\mathbb{C}^{M\times M}$, $\mathbf{Z}_{II}\in\mathbb{C}^{M\times M}$, and $\mathbf{Y}_{II}\in\mathbb{C}^{M\times M}$, respectively, denote the mutual coupling scattering, impedance, and admittance matrices at RIS. Based on the derivations in \cite{universal2024} and \cite{li2024beyond}, these terms are related to one another by
$\mathbf{S}_{RT} = \frac{\mathbf{Z}_{RT}}{2Z_0} - \frac{\mathbf{Z}_{RI}}{2Z_0}(\mathbf{Z}_{II} + Z_0\mathbf{I}_M)^{-1}\mathbf{Z}_{IT}$, $\mathbf{S}_{RI} = \mathbf{Z}_{RI}(\mathbf{Z}_{II} + Z_0\mathbf{I}_M)^{-1}$, $\mathbf{S}_{IT} = (\mathbf{Z}_{II} + Z_0\mathbf{I}_M)^{-1}\mathbf{Z}_{IT}$, $\mathbf{S}_{II} = (\mathbf{Z}_{II} + Z_0\mathbf{I}_M)^{-1}(\mathbf{Z}_{II}-Z_0\mathbf{I}_M)$,
$\mathbf{Y}_{RT} = \frac{1}{Z_0^2}(-\mathbf{Z}_{RT} + \mathbf{Z}_{RI}\mathbf{Z}_{II}^{-1}\mathbf{Z}_{IT})$,
$\mathbf{Y}_{RI} = -\frac{\mathbf{Z}_{RI}\mathbf{Z}_{II}^{-1}}{Z_0}$,
$\mathbf{Y}_{IT} = -\frac{\mathbf{Z}_{II}^{-1}\mathbf{Z}_{IT}}{Z_0}$, and $\mathbf{Y}_{II}=\mathbf{Z}_{II}^{-1}$.

\textit{Remark 8:}
The sum of $\mathbf{Z}_I + \mathbf{Z}_{II}$ in (\ref{eq:channel_z}) really highlights the essence of BD-RIS, i.e., $\mathbf{Z}_I + \mathbf{Z}_{II}$ acts as a general impedance matrix that captures the joint effect of artificial coupling (induced by inter-element interconnections) and the physical mutual coupling (induced by the closely spaced elements). In the presence of non-zero off-diagonal elements in $\mathbf{Z}_{II}$, the non-zero off-diagonal elements of $\mathbf{Z}_I$ can be engineered and reconfigured, through inter-element inter-connections, to either compensate or exploit the physical mutual coupling. These non-zero off-diagonal elements highlight the underpinning new DoF offered by inter-connecting ports/elements in the surfaces/multi-port impedance network and hence artificially engineering and reconfiguring the coupling across elements of the surface.

\textit{Remark 9:} 
We would like to clarify again based on (\ref{eq:channel_s}) that the mutual coupling here refers to the physical coupling between antenna elements and is characterized by the matrix $\mathbf{S}_{II}$, which is determined by the design of the antenna array and is independent of the scattering matrix $\mathbf{\Theta}$ of the reconfigurable impedance network. 
Readers may argue that by rewriting the channel model $\mathbf{H}_\mathsf{S}(\mathbf{\Theta})$ in (\ref{eq:channel_s}) as
\begin{equation}
    \mathbf{H}_\mathsf{S}(\mathbf{\Theta}) = \mathbf{S}_{RT} + \mathbf{S}_{RI}\mathbf{\Theta}_\mathsf{eff}\mathbf{S}_{IT},\label{eq:channel_s1}
\end{equation}
where $\mathbf{\Theta}_\mathsf{eff} = (\mathbf{I}_M-\mathbf{\Theta}\mathbf{S}_{II})^{-1}\mathbf{\Theta}$, one can also obtain a beyond-diagonal matrix $\mathbf{\Theta}_\mathsf{eff}$ for D-RIS.
However, the BD-RIS terminology is used to describe the scattering matrix $\mathbf{\Theta}$.
In this sense, within the BD-RIS, a kind of tunable coupling is artificially introduced between ports of the reconfigurable impedance network, thereby generating $\mathbf{\Theta}$ itself that is not limited to be diagonal.

\textit{Remark 10:} Another interpretation of BD-RIS has been recently presented in \cite{del2025physics}, whose main idea is that an $M$-element BD-RIS-aided channel can be represented as a channel involving an effective D-RIS matrix $\tilde{\mathbf{\Theta}}\in\mathbb{C}^{\tilde{M}\times \tilde{M}}$, where $\tilde{M}$ denotes the number of reconfigurable admittance components in BD-RIS. Specifically, the channel model can be described as 
\begin{equation}
    \mathbf{H}_\mathsf{S}(\tilde{\mathbf{\Theta}}) = \tilde{\mathbf{S}}_{RT} + \tilde{\mathbf{S}}_{RI}(\mathbf{I}_{\tilde{M}} - \tilde{\mathbf{\Theta}}\tilde{\mathbf{S}}_{II})^{-1}\tilde{\mathbf{\Theta}}\tilde{\mathbf{S}}_{IT},
\end{equation}
where $\tilde{\mathbf{S}}_{RT}\in\mathbb{C}^{N_\mathsf{r}\times N}$, $\tilde{\mathbf{S}}_{RI}\in\mathbb{C}^{N_\mathsf{r}\times\tilde{M}}$, and $\tilde{\mathbf{S}}_{IT}\in\mathbb{C}^{\tilde{M}\times N}$ can be regarded as some transformations of $\mathbf{S}_{RT}$, $\mathbf{S}_{RI}$, and $\mathbf{S}_{IT}$. Meanwhile, $\tilde{\mathbf{S}}_{II}\in\mathbb{C}^{\tilde{M}\times\tilde{M}}$ can be regarded as an equivalent ``mutual coupling'' matrix. In this way, a BD-RIS-aided wireless channel can be transformed to an effective D-RIS-aided wireless channel with some ``mutual coupling''.

The impedance matrix $\mathbf{Z}_{II}$ characterizes the mismatching and mutual coupling at the RIS elements. Specifically, the diagonal entries of $\mathbf{Z}_{II}$ refer to the self impedance and the off-diagonal entries refer to the mutual coupling depending on the inter-element spacing. Most existing literature assumes perfect matching and no mutual coupling at RIS elements, which is mathematically described as $\mathbf{Z}_{II} = Z_0\mathbf{I}_M$. This assumption makes $\mathbf{S}_{II} = \mathbf{0}_{M\times M}$ and the wireless channel $\mathbf{H}_\mathsf{S}(\mathbf{\Theta})$ in (\ref{eq:channels}) a linear function of $\mathbf{\Theta}$, as widely assumed in most existing RIS literature. However, as discussed above, it is difficult to achieve such a strong assumption in practice. That is, the off-diagonal entries of $\mathbf{Z}_{II}$ are generally nonzero, and can be modeled as functions of inter-element spacing \cite{akrout2023physically,gradoni2021end}.

\begin{figure}
    \centering
    \includegraphics[width=0.48\textwidth]{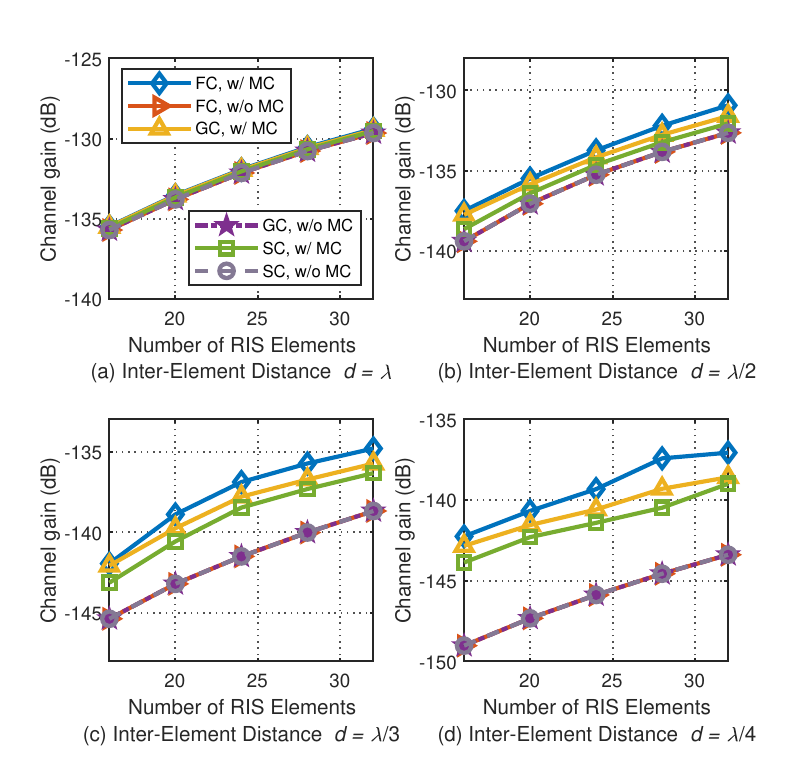}
    \caption{Channel gain versus the number of BD-RIS elements ($M = 16$, $\bar{M}=4$ for group-connected architecture). The direct transmitter-receiver link are assumed to be fully blocked. The markers ``FC'', ``GC'', ``SC'' respectively refer to fully-connected, group-connected, and single-connected (D-RIS); ``w/ MC'' refers to the case accounting for mutual coupling at BD-RIS and ``w/o MC'' refers to the case where the off-diagonal entries of $\mathbf{Z}_{II}$ are forced to zero \cite{li2024beyond}.}
    \label{fig:gain_mc}
\end{figure}

To evaluate how the existence of mutual coupling impacts the system performance, \cite{li2024beyond} has primarily focused on a group/fully-connected BD-RIS-aided SISO system and proposed an iterative algorithm to design the BD-RIS impedance matrix. Simulation results in Fig. \ref{fig:gain_mc} demonstrate that, under purely LoS channels, the performance gap between BD-RIS and D-RIS increases with decreasing inter-element distance since the stronger mutual coupling is better exploited by BD-RIS architectures. 
This indicates that BD-RIS is suitable for compact deployment with small inter-element spacing. 
In addition, \cite{semmler2024decoupling} has proposed decoupling networks as a solution to compensate for the mutual coupling at BD-RIS. The decoupling network can transform the system into a structure equivalent to a system without mutual coupling. As such, the beamforming design algorithms primarily proposed for systems without mutual coupling can be directly used for the case where the mutual coupling exists. Following the idea of using decoupling networks, \cite{nerini2024global} has proposed global optimal closed-form solutions for fully-connected and tree-connected BD-RIS-aided SISO systems, and provided the channel gain scaling law in the presence of mutual coupling. 
Specifically, for a SISO system with Rayleigh fading channels, \cite{nerini2024global} has derived the average power gain in the presence of mutual coupling at fully/tree-connected BD-RIS over that without mutual coupling, that is 
\begin{equation}
    G^\mathsf{MC} = Z_{II}^2\frac{\mathsf{tr}(\mathbf{R}^2) + \mathsf{tr}^2(\mathbf{R})+\sqrt{\pi\mathsf{tr}(\mathbf{R}^2)}\mathsf{tr}(\mathbf{R})}{M + M^2 + \sqrt{\pi M}M},
\end{equation}
where $\mathbf{R} = \Re\{\mathbf{Z}_{II}\}^{-1}$ and $[\mathbf{Z}_{II}]_{m,m} = Z_{II}$, $\forall m\in\mathcal{M}$, and theoretically proved that $G^\mathsf{MC}\ge 1$ using the fact that $\mathsf{tr}(\mathbf{R})\ge\frac{M}{Z_{II}}$ and that $\mathsf{tr}(\mathbf{R}^2)\ge\frac{M}{Z_{II}^2}$. This highlights that mutual coupling in fully/tree-connected BD-RIS can increase the average channel gain under Rayleigh fading channels. 
Not limited to SISO systems, \cite{wijekoon2024physically} has considered more general multi-user multi-antenna systems, derived the mutual coupling aware channel model, and optimized BD-RIS under different modes to maximize the system sum-rate. Results show that not accurately capturing the mutual coupling effect at BD-RIS will cause a non-negligible sum-rate performance loss.
More recently, \cite{wu2025beyondnew} has proven that band-connected architecture can achieve the same channel shaping capability as fully-connected architecture in multi-user MIMO systems with practical channel models capturing mutual coupling and other electromagnetic factors, which further generalizes the results in \cite{nerini2024beyond,wu2025beyond,nerini2024global}.
While the aforementioned works \cite{li2024beyond,semmler2024decoupling,nerini2024global,wijekoon2024physically,wu2025beyondnew} assume idealized BD-RIS hardware to better understand the impact of mutual coupling, \cite{zhou2025beyond} has focused on more general situations to discover optimal BD-RIS architectures that have the best performance-complexity tradeoff with mutual coupling, losses, and discrete values.


\section{Applications of BD-RIS}
\label{sec:applications}

The appealing benefits of BD-RIS have spawned an explosion of application-oriented works together with other promising 5G/6G techniques, such as advanced multiple access techniques, sensing, and WPT. 
In this section, we summarize the emerging applications of BD-RIS, as detailed below and summarized in Table \ref{tab:app}.

\subsection{BD-RIS for Communications}

\subsubsection{BD-RIS for Channel Shaping}
Most existing BD-RIS literature focuses on specific metrics, such as channel gain, sum-rate, transmit power consumption, to evaluate the performance benefit of BD-RIS, while a fundamental question which has not been answered is, to what extent a BD-RIS can shape the wireless channels (in terms of their singular values). To answer this question, \cite{zhao2024channel} has theoretically proven that BD-RIS is able to reach a larger dynamic range of channel singular values than D-RIS.
Results in \cite{zhao2024channel} have shown that this capability of BD-RIS can help to increase power gain and achievable rates. 
Further, \cite{yahya2024beyond,santamaria2025interference} have recently shown that BD-RIS can achieve better interference nulling/minimization to MIMO interference channels, which can be explained by the potential of BD-RIS to reduce the DoF.  

\subsubsection{BD-RIS Aiding Channel Reciprocity Attack and Physical Layer Security}
According to microwave engineering, conventional wireless channels are naturally reciprocal, in the sense that the uplink channel is the transpose of the downlink channel. 
However, as explained in the previous subsection, introducing non-reciprocal BD-RIS in the wireless propagation environment can break the uplink-downlink reciprocity of wireless channels. From the perspective of physical layer security, this property of non-reciprocal BD-RIS can be beneficial to support simultaneously optimal uplink and downlink transmission when downlink and uplink legitimate users are not aligned, as illustrated in Fig. \ref{fig:results_full_duplex}. 
Non-reciprocal BD-RIS can also be helpful to enhance the secrecy rate and secrecy outage probability \cite{agarwal2023enhanced} and enable a wireless circulator for one-way secure communications \cite{liu2025secure} in the presence of eavesdroppers. 
From an attacker's perspective, this property can still be useful since a non-reciprocal BD-RIS can be potentially deployed to maliciously degrade the downlink network performance \cite{wang2024channel,wang2024beyond}. 

\begin{table*}[]
	\caption{Literature on Applications of BD-RIS}
	\label{tab:app}
	\centering
		\begin{tabular}{|c|c|c|c|c|}
			\hline
			Ref. & Application & Architecture & Mode & Metric \\
			\hline\hline
			\cite{zhao2024channel} & \multirow{3}{*}{Channel Shaping} & Non-Reciprocal & \multirow{6}{*}{Reflecting} & Channel Gain \& Sum-Rate Max.\\
            \cline{1-1}\cline{3-3}\cline{5-5}
            \cite{yahya2024beyond} & & \multirow{2}{*}{Fully/Group-Connected} & & Sum-Rate Max. \& Interference Nulling\\
            \cline{1-1}\cline{5-5}
            \cite{santamaria2025interference} & & & & Interference Min.\\
            \cline{1-3}\cline{5-5}
            \cite{agarwal2023enhanced} & \multirow{2}{*}{Physical Layer Security} & \multirow{5}{*}{Non-Reciprocal} & & Secrecy Rate Max. \& Secrecy Outage Prob. Min. \\
            \cline{1-1}\cline{5-5}
            \cite{liu2025secure} & & & & Sum-rate Max.\\
            \cline{1-2}\cline{5-5}
            \cite{wang2024channel,wang2024beyond} & Reciprocity Attack & & & Sum-Rate Min. \\
            \cline{1-2}\cline{4-5}
            \cite{mahmood2024enhancing} & THz Commun. & & Hybrid & Sum-Rate Max. \\
            \cline{1-2}\cline{4-5}
            \cite{mishra2023transmitter} & \multirow{3}{*}{Massive MIMO} & & Reflecting & Max. Min. Rate\\
            \cline{1-1}\cline{3-5}
            \cite{raeisi2025modern} & &Fully/Group-Connected & \multirow{2}{*}{Hybrid} &  Bit Error Rate Min. \& Rate. Max.\\
            \cline{1-1}\cline{3-3}\cline{5-5}
            \cite{li2025beamforming} & & Non-Reciprocal & & Sum-Rate Max. \\
            \hline
                     \cite{zhang2024full} & Wireless Sensing & Group-Connected & Multi-Sector & Mean Square Error Min.\\
            \hline
            \cite{raeisi2024efficient} & Localization & Fully-Connected & Reflecting &  Cram$\acute{\text{e}}$r-Rao Lower Bound Analysis\\
            \hline
            \cite{wang2024dual} & \multirow{6}{*}{ISAC} & Non-Reciprocal & Hybrid & Max. Min. Signal-to-Clutter-Plus-Noise Ratio\\
            \cline{1-1}\cline{3-5}
            \cite{esmaeilbeig2024beyond} & & \multirow{7}{*}{Fully-Connected} & \multirow{7}{*}{Reflecting} & Weighted Sum of Radar and Commmun. SNRs Max. \\
            \cline{1-1}\cline{5-5}
            \cite{guang2024power} & & & & Power Min. \\
            \cline{1-1}\cline{5-5}
            \cite{liu2024enhancing} & & & & Sum-Rate (Throughput) Max. \\
            \cline{1-1}\cline{5-5}
            \cite{chen2024transmitter} & & & & Joint Sum-Rate Max. \& Cram$\acute{\text{e}}$r-Rao Bound Min. \\
            \cline{1-1}\cline{5-5}
            \cite{nguyen2025beyond} & & & & Outage Probability Analysis\\ 
            \cline{1-2} \cline{5-5} \cite{azarbahram2025beyond,azarbahram2025beamforming} & WPT & & & Direct Current Harvested Power Max. \\
            \cline{1-2}\cline{5-5}
            \cite{hua2024cell} & SWIPT & & & Sum-Rate Max. \& Harvested Power Max.\\
            \hline
            \cite{nerini2024physically} & SIM & Fully/Group-Connected & Hybrid & Channel Gain Max. \\
            \hline
            \cite{zhang2024beyond} & \multirow{2}{*}{NOMA} & Fully/Group-Connected & \multirow{4}{*}{Reflecting} & Sum-Rate Max.\\
            \cline{1-1}\cline{3-3}\cline{5-5}
            \cite{agarwal2025fairness} & & Non-Reciprocal & & Max. Min. Rate\\
            \cline{1-3}\cline{5-5}
            \cite{fang2022fully,kim2023group,khisa2024gradient} & \multirow{3}{*}{RSMA} & \multirow{2}{*}{Fully/Group-Connected} & & Sum-Rate Max.\\
            \cline{1-1}\cline{5-5}
            \cite{soleymani2023optimization} & & & & Max. Min. Rate\\
            \cline{1-1}\cline{3-5}
            \cite{li2024synergizing} & & Group-Connected & Multi-Sector & Ergodic Sum-Rate Max. (Imperfect CSI)\\
            \hline
            \cite{huroon2023optimized} & \multirow{3}{*}{UAV} &\multirow{2}{*}{Non-Reciprocal} & \multirow{5}{*}{Reflecting} & Sum-Rate Max. \\
            \cline{1-1}\cline{5-5}
             \cite{mahmood2023joint} & & & & Min. Max. Computational Time\\
             \cline{1-1}\cline{3-3}\cline{5-5}
            \cite{lin2024securing} & & Fully/Group-Connected & & Zero Secrecy Rate Prob. Min. \\
            \cline{1-3}\cline{5-5}
            \cite{khan2025enhancing} & \multirow{2}{*}{NTN} &\multirow{2}{*}{Non-Reciprocal} & & Secrecy Rate Max. \\
            \cline{1-1}\cline{5-5}
            \cite{khan2025beyondnew} & & & & Rate Max. \\
            \hline
             \cite{qin2025joint} & MEC & Non-Reciprocal & Hybrid & No. of Completed Task Bits Max. \\
		\hline
		\end{tabular}
\end{table*}

\subsubsection{BD-RIS in Terahertz Communications and Massive MIMO}
The need for more spectrum resources and higher frequency bands, such as Terahertz frequency bands, to provide higher spectrum efficiency and support emerging applications is one important trend for future networks. However, one bottleneck of Terahertz communications is the severe path loss in wireless channels, which makes the transmission vulnerable to blockages \cite{akyildiz2022terahertz}. BD-RIS has been shown to be a reliable solution to Terahertz communications, which can be useful to bypass blockages, increase coverage and  system rates \cite{mahmood2024enhancing}.

To compensate for the severe path loss in higher frequency bands, serve numerous users in the same time-frequency resource, and support increasing connectivity in 6G, another promising technique is massive MIMO \cite{bjornson2017massive}. However, the high spectrum efficiency of massive MIMO is achieved at the cost of numerous RF chains, leading to energy inefficiency in the network. 
To reduce the required number of RF chains while maintaining satisfactory spectrum efficiency, \cite{mishra2023transmitter,raeisi2025modern} have recently proposed to integrate BD-RIS within the radome of a transmitter as an auxiliary passive array, enabling a small-dimensional active antenna array at the transmitter to provide highly-directional beams reconfigured by BD-RIS. 
While \cite{mishra2023transmitter,raeisi2025modern} focused on single-transmitter scenarios, \cite{li2025beamforming} has considered a more general cell-free network and verified the performance benefits provided by BD-RIS. 

\subsection{BD-RIS for Sensing and ISAC}

Future networks are expected to support an increasing number of spectrum-demanding applications, which will cause increasing radio spectrum congestion. 
This growth motivates the design of shared paradigms, which enable cooperative spectrum and resource sharing among systems. 
One such solution is the emerging ISAC technique \cite{liu2022integrated}, which realizes dual functions with shared hardware, platform, and resources to reduce costs and increase resource utilization.  
However, since wireless sensing relies on strong LoS channels, which can be obstructed by obstacles in complex wireless propagation environments, it might be difficult to achieve a satisfactory ISAC performance in practical scenarios.  
To tackle this challenge, \cite{zhang2024full} has proposed a multi-sector BD-RIS self-sensing system, where an active source controller is installed on one sector such that signals can be scattered toward other sectors to achieve full-space coverage. 
BD-RIS can also perform as a bridge to link active and passive localization \cite{raeisi2024efficient} and effectively increase localization precision. 
Beyond enabling sensing and localization, there have been a few works on BD-RIS-aided ISAC \cite{wang2024dual,esmaeilbeig2024beyond,guang2024power,liu2024enhancing,chen2024transmitter,nguyen2025beyond}. 
Specifically, \cite{wang2024dual} proposed a hybrid BD-RIS-aided ISAC system where targets and users are located at both transmitting and reflecting sectors of BD-RIS, such that the performance of both functions and coverage is improved. 
In addition, deploying BD-RIS in ISAC systems can also help to increase signal-to-noise ratio (SNR) for both functions \cite{esmaeilbeig2024beyond}, reduce transmit power \cite{guang2024power}, enhance network throughput \cite{liu2024enhancing}, achieve better communication-sensing trade-off \cite{chen2024transmitter}, and meet ISAC outage requirements \cite{nguyen2025beyond}.

\subsection{BD-RIS for WPT and SWIPT}
Energy harvesting techniques are fundamental enablers for realizing seamless and green connectivity between low-power devices and supporting IoT in future wireless networks \cite{lopez2021massive}. Among various energy harvesting techniques, WPT is a paradigm making full use of wireless to deliver energy \cite{clerckx2021wireless}. Beyond WPT, a shared paradigm has been proposed, namely SWIPT \cite{clerckx2021wireless,perera2017simultaneous}, which enables using wireless to not only deliver energy but also convey information to best use the RF spectrum. To further enlarge the effective distance of WPT/SWIPT and improve the power transfer efficiency, \cite{azarbahram2025beyond,azarbahram2025beamforming} have deployed BD-RIS in the wireless system, showing that BD-RIS can flexibly shape the wireless channel to facilitate WPT. 
In addition, \cite{hua2024cell} has focused on a SWIPT system and demonstrated that BD-RIS is beneficial in improving both harvested energy of energy users and spectral efficiency of information users.

\subsection{BD-RIS with Other Techniques and Systems}
\subsubsection{BD-RIS and Stacked Intelligent Metasurface (SIM)}
Stacked intelligent surface (SIM) \cite{an2023stacked} consists in stacking multiple layers of RIS or metasurfaces to provide more flexible wave manipulations. 
While SIM layers have been implemented using D-RIS \cite{an2023stacked}, \cite{nerini2024physically} has recently shown that SIM layers can be implemented using BD-RIS.
Theoretical and numerical results have demonstrated that 1-layer SIM implemented with BD-RIS (tree-connected) is sufficient to achieve performance upper-bound in SISO systems, while multi-layer SIM implemented with D-RIS is suboptimal.

\subsubsection{BD-RIS with Advanced Multiple Access}
Multiple access refers to techniques that make use of resource dimensions, such as time, frequency, space, and power, to serve multiple users. 
Non-orthogonal multiple access (NOMA) has been actively studied recently due to its benefits in exploiting the available resources more efficiently with the aid of superposition coding at the transmitter and successive interference cancellation (SIC) at the receiver \cite{liu2017nonorthogonal}. To further boost system performance, the interaction between NOMA and BD-RIS has been studied in \cite{zhang2024beyond,agarwal2025fairness}, showing that better user fairness can be achieved. 
In addition, rate-splitting multiple access (RSMA) is a novel and general framework for non-orthogonal transmissions, softly including NOMA as an extreme case \cite{mao2022rate} and moving beyond that to provide enhanced performance. Given that channel acquisition of RIS-aided wireless systems is generally challenging while RSMA is robust to CSI imperfections, the integration of the two has emerged \cite{fang2022fully,kim2023group,soleymani2023optimization,khisa2024gradient,li2024synergizing}. Specifically, \cite{fang2022fully,kim2023group,soleymani2023optimization,khisa2024gradient} have focused mainly on perfect CSI and have shown the benefits of such an integration in increasing spectral efficiency, fairness rate, and fairness energy efficiency. 
\cite{li2024synergizing} has further shown that the integration of the two can still increase system sum-rate when only imperfect CSI is available at the transmitter. 

\subsubsection{BD-RIS Aiding Unmanned Aerial Vehicle (UAV) and Non-Terrestrial Networks (NTN)}
Unmanned Aerial Vehicle (UAV) has gained attention for 5G and 6G networks due to its advantages in extending coverage, easing the deployment, and supporting controllable mobility \cite{gupta2015survey}. These advantages of UAVs can be further strengthened by cooperating with BD-RIS. 
\cite{huroon2023optimized} and \cite{lin2024securing} studied the transmission strategy and user scheduling schemes of a ground-based BD-RIS in a network consisting of multiple UAVs, showing that the system performance can be much improved with the aid of BD-RIS.
\cite{mahmood2023joint} proposed to colocate BD-RIS and UAVs to minimize the computation latency and UAV hovering time. 
Not limited to UAV-enabled systems, BD-RIS can also provide performance benefits in general non-terrestrial networks (NTN) typically realized by satellites, high-altitude platform stations (HAPS), and UAVs \cite{azari2022evolution}. 
Results in \cite{khan2025enhancing,khan2025beyondnew} have shown that having a BD-RIS mounted UAV \cite{khan2025enhancing} or HAPS \cite{khan2025beyondnew} as a secondary transmitter in a cognitive radio enabled NTN helps increase secrecy rate and spectral efficiency.

\subsubsection{BD-RIS Aiding Mobile Edge Computing (MEC)}
Mobile edge computing (MEC) \cite{chen2022irs} is a platform that pushes mobile computing, network control, and storage to the network edges, such as base stations and access points. In this way, MEC enables resource-limited mobile devices to complete computational-intensive tasks with much reduced latency. 
To further improve the scalability and performance of MEC to meet the growing user demands in 5G and future 6G networks, \cite{qin2025joint} has proposed deploying BD-RIS in the wireless propagation environment as an enabler to significantly improve the number of completed task bits.

\section{Challenges and Future Research Directions of BD-RIS}
\label{sec:challenges}

Despite the significant benefits of BD-RIS architectures and modes, the study of BD-RIS is still in its infancy. There exist technical challenges in designing and implementing BD-RIS for practical wireless systems, which motivate meaningful future research directions. 
In this section, we discuss key challenges in BD-RIS from the perspectives of implementation, circuit design, optimization, and transmission protocol, each of which is followed by potential research directions.

\subsection{BD-RIS Implementation}

\begin{figure}
    \centering
    \subfigure[Inner and side views of hybrid BD-RIS]{\includegraphics[width=0.42\textwidth]{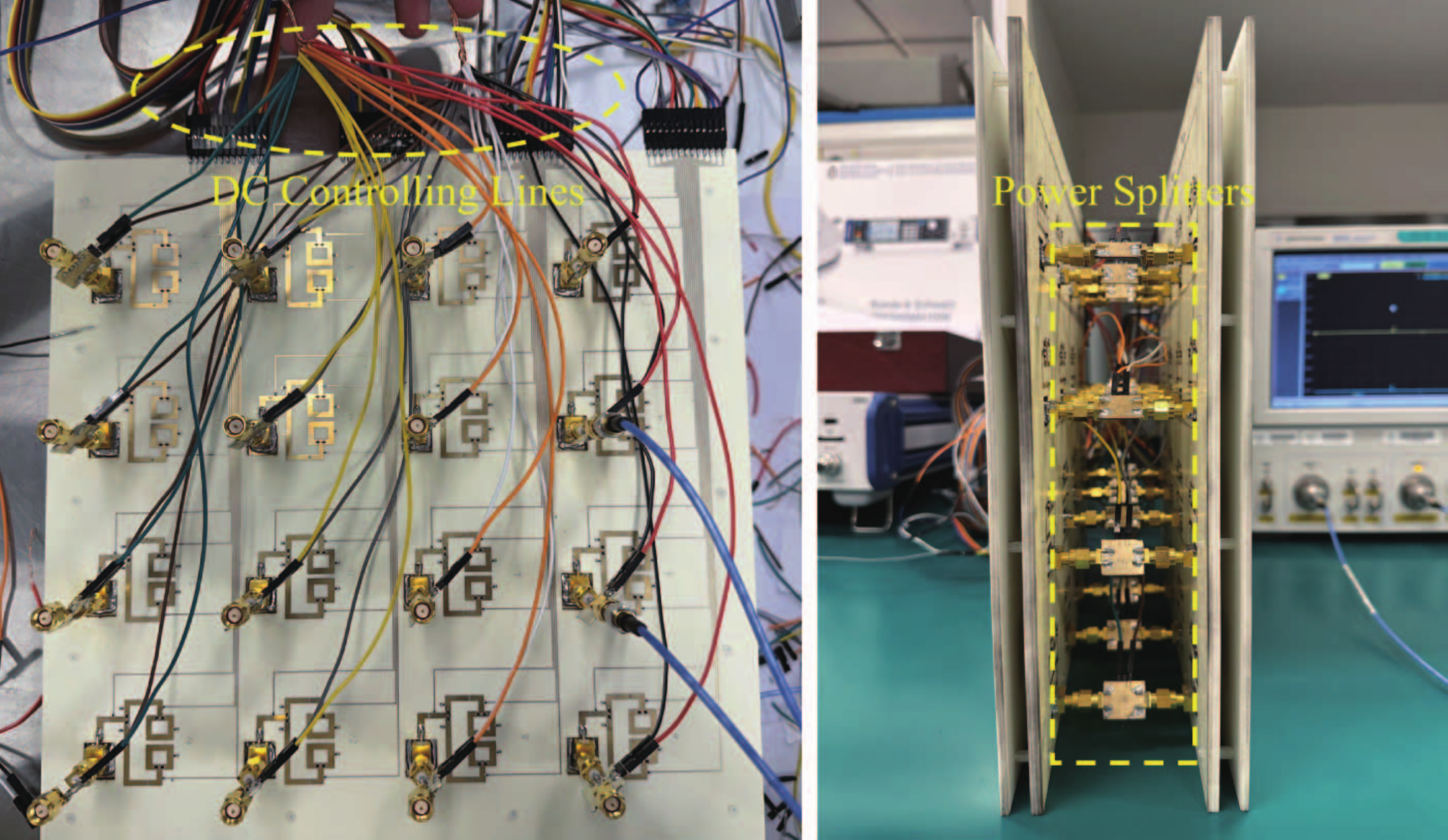}}
    \subfigure[An overall view of hybrid BD-RIS]{\includegraphics[width=0.42\textwidth]{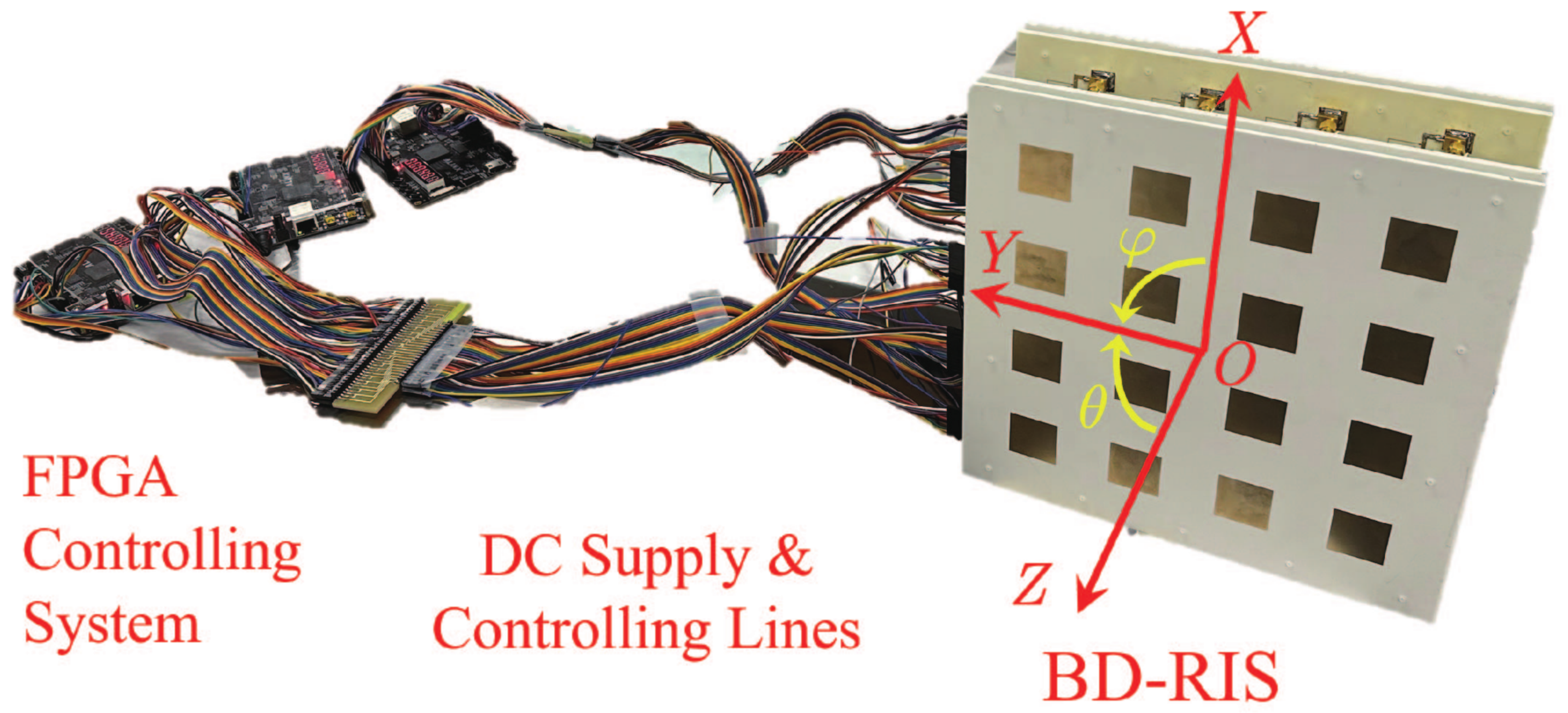}}
    \caption{Photograph of the prototype of the hybrid BD-RIS with $4\times4$ cells and the controlling systems \cite{ming2025hybrid}.}\vspace{-0.2 cm}
    \label{fig:prototype}
\end{figure}

\begin{figure}
    \centering
    \subfigure[Photo of the tridiagonal BD-RIS showing an antenna array (top) connected to a tunable load network (bottom)]{\includegraphics[width=0.42\textwidth]{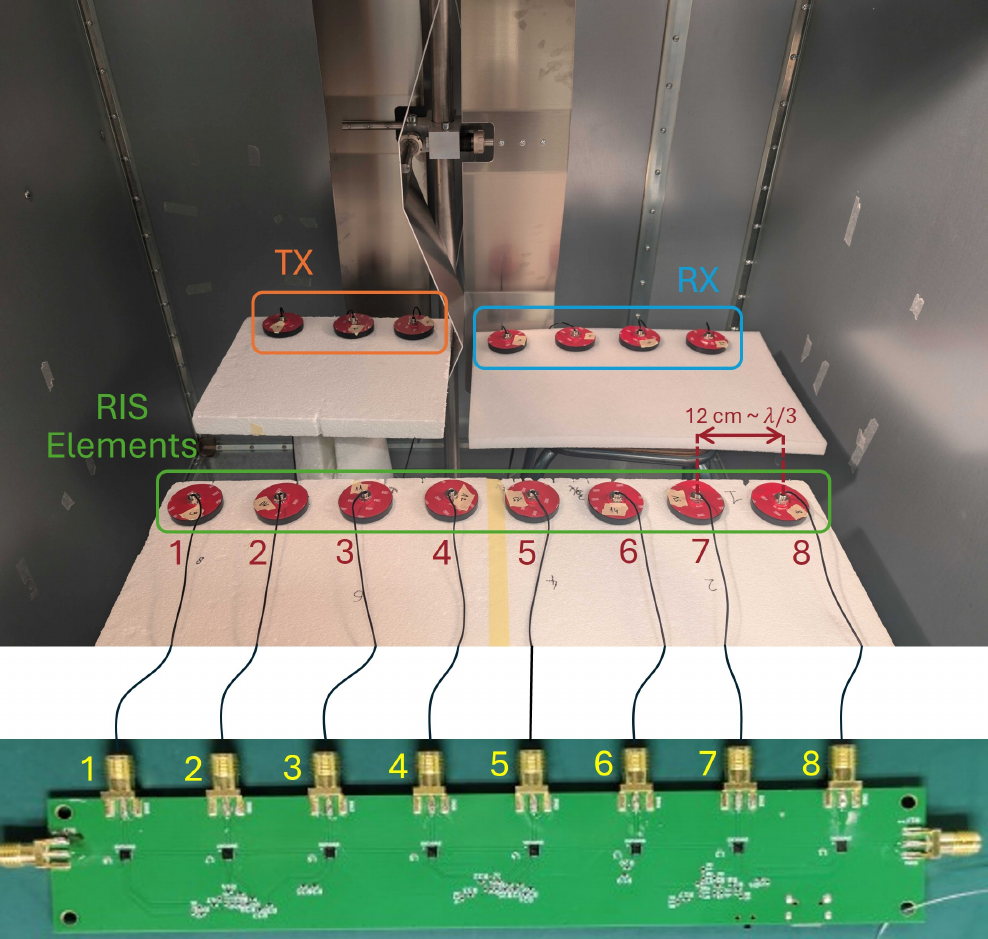}}
    \subfigure[Schematic of the tridiagonal BD-RIS and its matrix $\mathbf{Y}_I$ determined by the switch configuration in the schematic]{\includegraphics[width=0.42\textwidth]{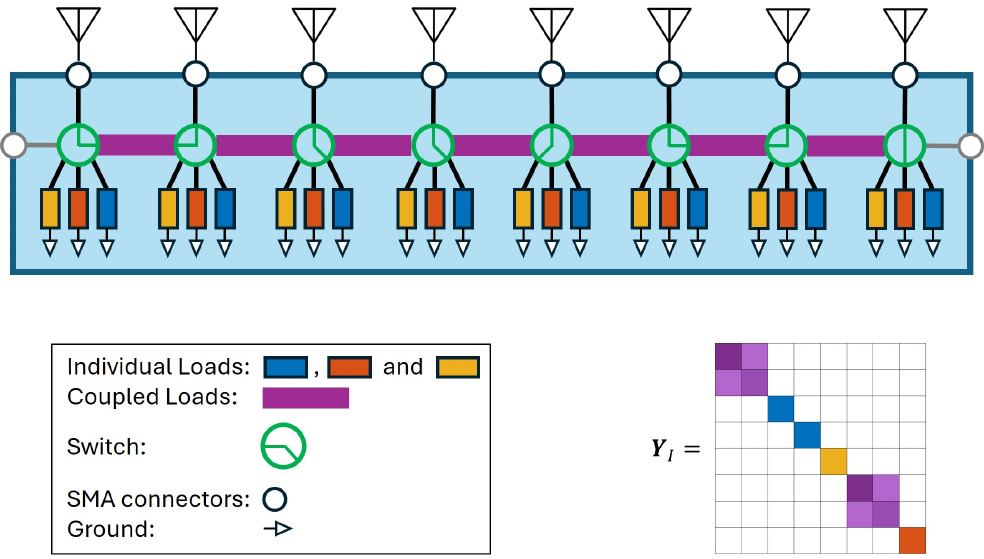}}
    \caption{Photograph and schematic of the prototype of the tridiagonal BD-RIS with $8$ elements \cite{tap2025}.}
    \label{fig:prototype2}\vspace{-0.2 cm}
\end{figure}

The hardware implementation of BD-RIS is a fundamental yet challenging issue. Until now, there has been the implementation of one special type of BD-RIS, namely STAR-RIS (or IOS, STARS) \cite{zhang2022intelligent}. In addition, a hybrid mode BD-RIS has been recently implemented using power splitters as shown in Fig. \ref{fig:prototype}, enabling independent beam control of reflected and transmitted waves \cite{ming2025hybrid}.
Furthermore, a reflective mode BD-RIS, specifically a tridiagonal BD-RIS, has been implemented through a tunable load network allowing interconnections among adjacent RIS elements as shown in Fig. \ref{fig:prototype2} \cite{tap2025}.
These results show that it is physically feasible to implement BD-RIS using existing RF techniques, designs, and devices. However, the comprehensive implementation of various BD-RIS modes and architectures is still on its way. Based on the illustration in Section \ref{sec:model}, an $M$-element BD-RIS consists of an $M$-antenna array and an $M$-port reconfigurable impedance network. 
\begin{itemize}
    \item The implementation of the $M$-antenna array varies with operation modes. For the reflective mode, the conventional uniform linear or planar array can be directly used. For the hybrid mode, each pair of antennas with uni-directional radiation pattern (such as patch antennas) should be back to back placed to construct a cell, and all cells should be arranged in a uniform array. For the multi-sector mode, every $L$ antennas with corresponding beamwidth should be placed as a polygon and all cells should be arranged in a uniform array. 
    \item The implementation of the $M$-port reconfigurable impedance network varies with architectures. For fully/group-, tree/forest-, stem/band-connected architectures, as per the circuit topology, varactors consisting of inductance and tunable capacitance can be used to implement continuous-value reconfigurable admittance components; PIN diodes can be used to implement discrete-value reconfigurable admittance components. For dynamically connected architectures, additional switches are required between every two elements. For non-reciprocal architectures, non-reciprocal devices, such as circulators, isolators, and gyators, are required and the number of them depends on specific circuit designs. 
\end{itemize}

\subsubsection{Challenge} The unique challenge of implementing BD-RIS compared to D-RIS lies in the increasing circuit complexity, cost, and power losses of the reconfigurable impedance network. Especially when the number of elements and/or groups increases, the required number of reconfigurable admittance components significantly increases. 
This will cause a heavy burden on circuit design and, more importantly, accumulated power losses, and power consumption of drive circuits and elements induced by more reconfigurable admittance components \cite{wang2024power}, which may weaken the performance benefits of BD-RIS architectures. In addition, for future system-level experiments, the theoretical channel estimation and beamforming optimization methods might require too much time such that an effective data transmission cannot be guaranteed.

\subsubsection{Future Research Direction} 
Although the tree/forest-connected architecture \cite{nerini2024beyond} and the stem/band-connected architecture \cite{wu2025beyond} have been proposed to achieve the best performance-complexity trade-off in, respectively, MISO and multi-user MIMO systems, these conclusions still stop in the analytical layer. The practical limitation of these architectures taking into account the circuit design feasibility, cost, and power losses still remains unexplored. 
In addition, it is important to establish a fair power consumption model for BD-RIS architectures and validate its feasibility by practical measurements.
How to balance the circuit complexity, computational complexity for optimization, and system performance for a practical BD-RIS-assisted test bed is also a crucial problem.
Therefore, it is challenging but worthwhile to implement and prototype different BD-RIS architectures to have a practical picture on which kind of architecture could best balance the performance, hardware considerations, and computational complexity.

\subsection{Active BD-RIS}
The existing literature of BD-RIS has focused purely on its passive (at best lossless) form, in the sense that the power scattered by the surface is no larger than the incident power. This assumption comes from D-RIS literature and brings many advantages to both D-RIS and BD-RIS, such as low cost, low power consumption, and negligible thermal noise \cite{di2020smart}. However, since the array gain provided by numerous elements of RIS cannot fully compensate for the huge multiplicative fading induced by RIS, the performance improvement of passive RIS is very limited. For D-RIS, the gain in some scenarios is even not visible when the direct link between the transmitter and receiver exists. For BD-RIS, in SISO systems, the fully-connected architecture achieves a maximum of 62\% gain over D-RIS \cite{shen2021} with the absence of the direct link, which is not sufficiently high, especially when the direct link is strong. To break the fundamental performance bottleneck caused by multiplicative fading, the concept of active D-RIS has been proposed in \cite{zhang2022active}. The key feature of such D-RIS is its ability to reflect signals with amplified powers, thanks to the introduction of power amplifiers in RIS elements. The benefits of active D-RIS has also been shown in many aspects, such as achieving a substantial sum-rate gain of 130\% \cite{zhang2022active}, achieving higher energy efficiency than passive D-RIS \cite{zhi2022active}, and further enhancing both communication quality and sensing performance \cite{zhu2023joint}. This motivates the consideration of active BD-RIS: \textit{Will integrating active power amplifiers in BD-RIS architectures provide orders of gains?}

\subsubsection{Challenge}
To answer the above question, one fundamental yet challenging issue is to establish a physics-consistent model for active BD-RIS. Different from passive BD-RIS which is modeled as an antenna array connected to a multi-port reconfigurable impedance network, the active BD-RIS, following the modeling principle in \cite{zhang2022active}, should be modeled as a series connection of an antenna array, a multi-port reconfigurable impedance network, and a reflection power amplifier network. In this sense, the admittance/impedance/scattering matrices themselves of the reconfigurable impedance network are not sufficient to characterize the active BD-RIS, and the impact of power amplifiers should be carefully taken into account.

\subsubsection{Future Research Direction}
The consideration of making BD-RIS active opens many interesting and meaningful research directions, such as theoretically deriving the power scaling law of active BD-RIS for different channel fading conditions; revisiting the optimal performance-circuit complexity Pareto frontier or exploring the optimal performance-power consumption Pareto frontier for active BD-RIS; and fully making use of the active property to reduce channel estimation error and overhead. Beyond these examples, other fundamental directions in passive BD-RIS, such as the impact of hardware impairments, will also be worth studying in active BD-RIS.  In this case, hardware impairments will not be limited to those summarized in Section \ref{sec:hardware}, but will also include the imperfections of power amplifiers.

\subsection{Artificial Intelligence (AI)-Driven Beamforming Solutions}

From the beamforming design perspective, most existing literature on BD-RIS uses conventional optimization methods to design BD-RIS beamforming, aiming at either finding closed-form solutions with a guaranteed performance or finding iterative solutions with a guaranteed convergence. However, these optimization methods have the following drawbacks: 1) They depend heavily on accurate physics modeling of wireless channels and thus are not easy to be used in practical scenarios where theoretical models cannot fully reflect the true characteristics of wireless environments; 2) The computational complexity can be extremely high when considering large-dimensional scenarios, which are unfortunately quite common in the modern world. Although BD-RIS theoretically has appealing benefits supported by these optimization methods, it is important to improve the scalability and feasibility of BD-RIS design methods to prepare for the possible future commercialization of BD-RIS. This thus motivates the consideration of using artificial intelligence (AI)-driven solutions, since they are not sensitive to models and suitable for large-scale problems \cite{eldar2022machine,jiang2016machine}.

\subsubsection{Challenge}
Although adopting AI-driven solutions in BD-RIS-aided wireless systems presents significant opportunities for optimization, automation, and efficiency, several challenges must be addressed. For example, in BD-RIS-aided wireless communication systems, it is expected to flexibly tune BD-RIS to adapt to the time-varing dynamic environments, which requires fast real-time decisions and is thus challenging due to the practical limitations of resources, costs, and hardware.  
For another example, 
AI-driven solutions heavily depend on high-quality training data, while it is challenging to acquire such data in BD-RIS-aided systems due to practical hardware limitations, channel dynamics, and measurement overhead. 

\subsubsection{Future Research Direction}
Although \cite{sobhi2024joint,loli2024meta} have adopted deep reinforcement learning (DRL) and meta-learning to design BD-RIS-aided wireless communication systems and shown the benefits of such AI-driven solutions in being friendly to large-scale wireless systems, the study in this direction is still in its infancy. Future research avenues include, but are not limited to, designing more efficient online adaptation schemes to enable highly reliable instantaneous transmissions and establishing proper AI models that can properly adapt to various BD-RIS architectures, that are robust to real hardware constraints of BD-RIS, and that have good training overhead-performance trade-offs.

\vspace{-0.35 cm}

\subsection{CSI-Free Protocols}

Channel estimation$\rightarrow$feedback and beamforming$\rightarrow$data transmission is a protocol typically used in wireless communication systems \cite{molisch2012wireless}. Most studies on RIS-aided wireless communication systems also adopts this protocol, in which the data transmission performance depends highly on the channel estimation accuracy. 
However, as discussed in Section \ref{subsec:estimation}, due to the passive property of RIS, the channel estimation of BD-RIS-aided systems suffers high channel estimation error and/or high training overhead which grow with the circuit complexity of architectures.  
This motivates the consideration on making the protocol itself less dependent on accurate channel acquisition. 

\subsubsection{Challenge}
In D-RIS-aided systems, protocols based on beam training schemes have been proposed to reduce overhead and avoid the requirement for accurate CSI acquisition \cite{mu2024reconfigurable,wang2022beam,liu2023low}. Specifically, a codebook consisting of possible scattering matrices of RIS is predefined in an offline stage. During the online stage, the transmitter consecutively sends pilot signals while the RIS switches between codewords in the codebook. The receiver collects the received signal powers corresponding to all codewords, finds the index of the codeword which leads to the best received power, and feeds this index back to the transmitter. In this way, the channel estimation and beamforming phases in the typically used protocol are replaced with a beam training phase supported by a predefined codebook, such that the RIS can be optimized without explicitly knowing the CSI. The main difficulty in this protocol lies in the construction of RIS codebook. Especially when it comes to BD-RIS with various architectures, it is challenging to construct codebooks whose codewords perfectly follow the constraints of corresponding architectures and have sufficiently large diversity to reflect the flexibility of BD-RIS architectures compared to D-RIS. 

\subsubsection{Future Research Direction}
Based on the above discussions, before going to the design of the codebook, it remains unexplored but is important to establish suitable and rigorous criteria to evaluate if a codebook for a specific BD-RIS architecture is ``good''.  With these criteria, it is interesting to explore the relationship between the weights in a neural network and codewords in a codebook, such that learning-based methods can be adopted to facilitate the codebook design. Beyond this, the idea of end-to-end learning for communication systems \cite{ait2019model} can be potentially adopted in practical BD-RIS-aided systems without even knowing the channel model. 

\subsection{New BD-RIS Architectures}

BD-RIS represents a general class of RIS architectures that are more flexible than D-RIS as they include additional tunable components interconnecting the RIS elements with each other.
Thanks to this additional flexibility, BD-RIS can further improve the channel shaping capabilities of D-RIS, leading to superior performance in wireless systems, e.g., in terms of channel gain in single-user systems \cite{shen2021} or sum-rate in multi-user systems \cite{li2022beyond}.
Since this performance gain comes at the cost of increased hardware complexity due to the additional impedance components, the fundamental question that arises when considering the deployment of BD-RIS is ``\textit{Is the additional hardware complexity justified by the resulting performance improvements?}''

\subsubsection{Challenge}
To answer this question, the limits of the trade-off between performance and hardware complexity enabled by BD-RIS need to be studied.
In this direction, a model of BD-RIS architectures based on graph theory has been used in previous literature to derive the tree-, forest-, stem-, and band-connected BD-RIS architectures as efficient BD-RIS architectures with reduced complexity \cite{nerini2024beyond,zhou2024novel,wu2025beyond}.
In addition, the set of BD-RIS architectures achieving the Pareto frontier of this performance-complexity trade-off has been characterized for SISO systems, uni-polarized \cite{nerini2023pareto} as well as dual-polarized \cite{nerini2025dual}.
However, only a limited number of BD-RIS architectures have been explored so far, which are proven to achieve a good trade-off between performance and complexity only under ideal assumptions, e.g., arbitrarily reconfigurable  components, lossless RIS, and absence of mutual coupling.

\subsubsection{Future Research Direction}
From the graph theoretical model of BD-RIS, we find that the number of possible $M$-element BD-RIS architectures is $2^{M(M-1)/2}$, as there are $M(M-1)/2$ possible interconnections between the $M$ elements.
This huge number of potential BD-RIS architectures offers a vast design space that remains largely unexplored.
A promising direction for future research is the development of novel BD-RIS architectures that offer favorable performance-complexity trade-offs, particularly under practical constraints such as hardware impairments and real-world non-idealities.

\section{Conclusion}
\label{sec:conclusion}

The existing literature has shown that the capability of BD-RIS in reconfiguring the coupling between elements with inter-element admittance components provides additional flexibility to manipulate signals and waves in the analog domain. This capability of BD-RIS has generated multidimensional benefits to enable higher performance, larger coverage, and denser connectivity for future wireless networks.
In this paper, we have provided the first holistic tutorial on BD-RIS as a brand-new advance in the RIS technique. We have thoroughly shared all the technical tools and strategies to model, design, and optimize BD-RIS. In support of these fundamentals, we have highlighted key benefits, emerging applications, challenges, and possible future research directions of BD-RIS. The scope of this tutorial spanned from fundamental physics-consistent modeling using multi-port network analysis, mode analysis, reciprocal and non-reciprocal architecture designs; representative signal processing strategies for BD-RIS beamforming and channel estimation; and modeling and analysis of key hardware impairments of BD-RIS, to higher-level summaries of benefits supported by flexible modes and architectures; applications in wireless communications, sensing, and power transfer; and challenges from perspectives of hardware implementation and signal processing which triggered directions crucial for promoting the possible commercialization of BD-RIS. 
We hope that this paper will serve as a useful resource to teach and inspire readers in the wireless society, and provide fresh blood to activate interesting and meaningful future research on BD-RIS.


\bibliographystyle{IEEEtran}
\bibliography{refs}
\end{document}